\documentclass[12pt, onecolumn, a4paper, fleqn]{article}

\usepackage{STY/PACKAGES}
\usepackage{STY/DEFINITIONS}
\usepackage{STY/SETTINGS}
\usepackage{STY/TIKZ}
\usepackage{STY/TITLEPAGE}

\usepackage{mdframed} 

\begin{document}
\edef\myindent{\the\parindent}
\author{Kronberg, Vì C.E. \and Anthonissen, Martijn J.H. \and ten Thije Boonkkamp, Jan H.M. \and IJzerman, Wilbert L.}
\title{Modelling Surface Light Scattering in the Context of Freeform Optical Design}

\thispagestyle{empty} 
\pagenumbering{gobble} 

\begin{figure}[t]
	\centering
	\includegraphics[width=0.4\textwidth]{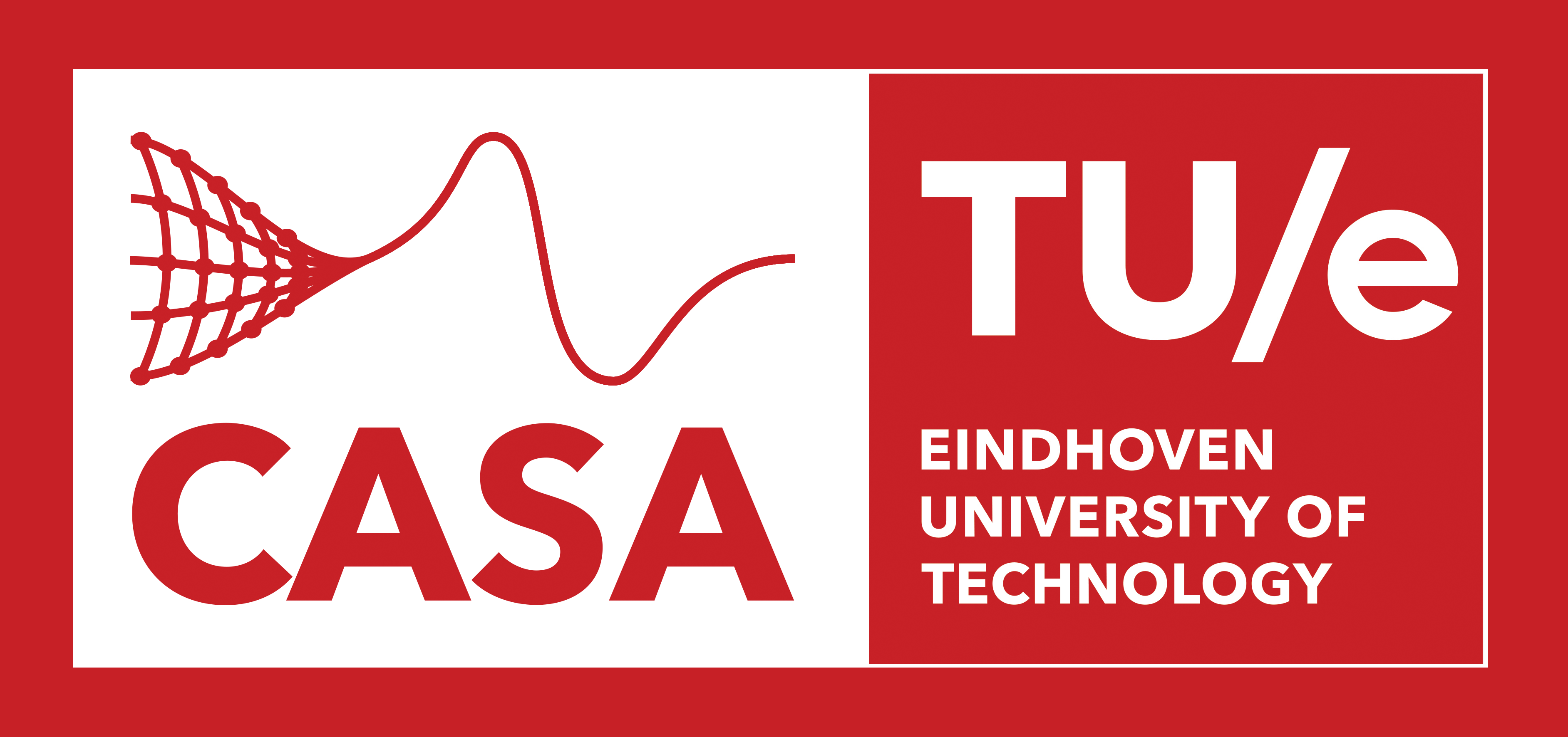}
\end{figure}

\vspace*{-0.7cm}
\begin{center}
	\LARGE \textbf{Modelling Surface Light Scattering in the Context of Freeform Optical Design}\\[0.5cm]
\end{center}

\begin{center}
	\large
	{\itshape V\`{i} C.E.}~\textsc{Kronberg},\textsuperscript{1,*} {\itshape Martijn J.H.}~\textsc{Anthonissen},\textsuperscript{1} {\itshape Jan H.M.}~\textsc{ten Thije Boonkkamp},\textsuperscript{1} and {\itshape Wilbert L.}~\textsc{IJzerman}\textsuperscript{1,2}
\end{center}

\begin{changemargin}{1cm}{1cm}
\noindent
\textsuperscript{1}{\itshape Deparment of Mathematics and Computer Science, Eindhoven University of Technology}, PO Box 513, 5600 MB Eindhoven, The Netherlands\\
\textsuperscript{2}{\itshape Signify Research}, High Tech Campus 7, 5656 AE Eindhoven, The Netherlands\\
\textsuperscript{*}{\color{blue}\href{mailto:s.kronberg@tue.nl}{\texttt{s.kronberg@tue.nl}}}\\
\url{https://www.win.tue.nl/~martijna/Optics/}\\
\end{changemargin}

\hrule
\begin{changemargin}{1cm}{1cm}
{\textbf{\textsc{PACS:}} 02.30.Z $\cdot$ 42.15.−i $\cdot$ z42.25.Fx $\cdot$ 42.79.Fm \hfill | \hfill \textbf{\textsc{AMS:}} 78A05 $\cdot$ 78A45 $\cdot$ 78A46}
\end{changemargin}
\hrule
\begin{changemargin}{1cm}{1cm}
	\textsc{\textbf{Abstract:}}
	We present a novel approach of modelling surface light scattering in the context of freeform optical design.
	The model relies on energy conservation and optimal transport theory.
	For isotropic scattering in cylindrically or rotationally symmetric systems with in-plane scattering, the scattered light distribution can be expressed as a convolution between a scattering function, which characterises the optical properties of the surface, and a specular light distribution.
	Deconvolving this expression allows for traditional specular reflector design procedures to be used, whilst accounting for scattering.
\end{changemargin}
\hrule

\clearpage

\pagestyle{main}
\pagenumbering{arabic}
\setcounter{page}{1}

\section{Introduction}
The engineering field of optical design is often restricted to working within the confines of the so-called geometrical optics (GO) approximation, where light propagation is modelled using light rays --- lines collinear with the Poynting vector.
Within the GO approximation, phenomena such as diffraction and interference are typically not accounted for \cite[p.~159]{Hecht2017Optics}.
The absence of such phenomena in the design procedure of an optical element can result in discrepancies between the raytraced distributions and the ones measured using the actual component.
Additionally, it places restrictions on what kind of optical systems one can design, as scattering cannot be utilised or accounted for in the design process.

There are a myriad of approaches one could consider to include scattering.
The most drastic, but in some ways most natural, approach is that of solving Maxwell's equations.
This would constitute a substantial departure in terms of strategy, and would in theory work for most realistic systems, but it is often impractical due to the enormous complexity of the task.
As such, a multitude of approximations based on Maxwell's equations have been formulated which are applicable to problems within several regimes of parameter values, such as surface roughness or incident angle.
Some highlights include the rigorous vector perturbation theory published by Lord Rayleigh in 1907 \cite{Rayleigh1907}, and later expanded by Rice (1951) \cite{Rice1951}, and the rather different Kirchhoff approach, based on random phase variations due to microtopographic surface features, most commonly attributed to Beckmann and Spizzichino (1963) \cite{Beckmann1987}.
These approaches are valid in different regimes.
In particular, Rayleigh-Rice vector perturbation theory agrees well with experimental measurements of wide-angle scattering (up to approximately $50^\circ$ of the polar angle of detector/source) for scattering from optically smooth surfaces.
Here, ``optically smooth'' refers to the root mean square (RMS) surface roughness $\sigma_s$ divided by the wavelength $\lambda$ being much less than unity, i.e., $\sigma_s/\lambda \ll 1$ \cite[p.~49]{Harvey2019}.
The Beckmann-Kirchhoff theory, on the other hand, is valid for rougher surfaces, but due to a moderate-angle assumption as part of its derivation, it is not suitable for use with wide scattering angles and/or large angles of incidence.

There have been numerous developments since these early theories were formulated.
Here, we highlight the work of Church, who published numerous papers during the 1970s on Rayleigh-Rice theory in the context of surface scattering from optically smooth surfaces \cite{Church1975}.
According to Harvey, his contributions were instrumental in shaping the applied optics community at the time \cite[p.~50]{Harvey2019}.
The last work from this era we want to highlight is that of Harvey and Shack from 1976, where they developed a linear systems formulation of surface scattering based on a surface transfer function \cite{HarveyPhD}.
This approach allows for the use of the Fourier transform of the surface transfer function to compute a scattered radiance function ``closely related to the bidirectional reflectance distribution function (BRDF)'' \cite[p.~50]{Harvey2019}.
This was later extended to the generalised Harvey-Shack surface scattering theory, which is able to treat arbitrarily rough surfaces with arbitrary incident and scattering angles \cite{GHS}.

Rather than solving Maxwell's equations, with or without approximations, one could alter the GO rays in some manner such that they can be used to compute scattering phenomena.
One example of such an approach, that still still retains many of the computational benefits of GO, is to modify the GO rays such that they carry information regarding the phase of the light, which can form the basis of diffraction calculations.
A good overview of this approach can be found in McNamara \cite{mcnamara1990introduction}.
Whilst computationally efficient, such an approach would still require substantial modifications to contemporary reflector design procedures.

In our approach, we remain in the domain of traditional GO, and as an alternative to carrying phase information, we propose a surface scattering model inspired by optimal transport theory \cite{Villani2003Topics}, which leads to a convolution integral for cylindrically and rotationally symmetric problems with isotropic in-plane scattering.
This convolution integral yields the scattered light distribution, given a scattering function and a specular target distribution, where the former characterises the optical properties of the surface.
This is a forward problem, but it can also be cast in terms of an inverse problem --- given a desired target distribution and a scattering function, one may perform deconvolution to find an appropriate intermediate specular target distribution, which can in turn be used to design the optical element using traditional specular design methods.
In contrast to the phase-carrying rays formulation, the specular reflector design procedures do not need to be modified, as the effect of scattering can be considered a pre-processing step.
In this paper, we utilise Maes's work on specular reflector design, and in particular the two-dimensional procedure applicable to cylindrically and rotationally symmetric problems outlined in \cite[Ch.~3]{Maes1997ReflectorDesign}.
For three-dimensional freeform specular optical design, there have been several recent developments \cite{Prins2014, Prins2014AMonge, Prins2015, Lith2017, Yadav2018, Beltman2018, Yadav2019, Romijn2020, Romijn2020A}.
The advantage of our approach is that we may separate the scattering calculations from the reflector design step, allowing us to greatly benefit from the maturity of specular reflector design procedures, whilst still accounting for scattering.

The structure of the paper is as follows.
Cylindrically symmetric problems with in-plane scattering are covered first, together with a few words about deconvolution in Sec.~\ref{sec:2D-refl}, followed by rotationally symmetric problems with in-plane scattering in Sec.~\ref{sec:3D-refl}.
Next, two-dimensional specular reflector design is briefly discussed in Sec.~\ref{sec:reflectorDesign}, followed by some results in the form of reflectors and ray traced distributions for validation of both the cylindrically and rotationally symmetric systems in Sec.~\ref{sec:results}.
Finally, conclusions with some proposals for expansions of the model are presented in Sec.~\ref{sec:conclusion}.

\section{Cylindrically Symmetric Problems}\label{sec:2D-refl}
\begin{figure}[htbp]
	\centering
	\begin{minipage}{0.5\linewidth}
		\centering
		\includegraphics[width=0.9\linewidth]{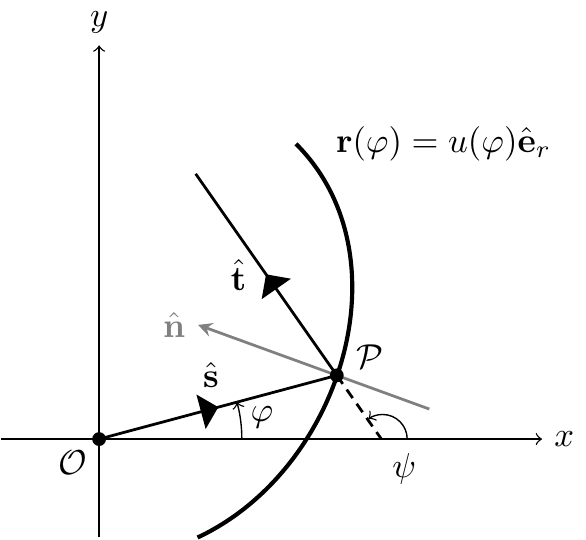}
		\captionsetup{width=0.9\linewidth}
		\caption{Specular reflector.}
		\label{fig:2D_LoR}
	\end{minipage}\hfill
	\begin{minipage}{0.5\linewidth}
		\centering
		\includegraphics[width=0.9\linewidth]{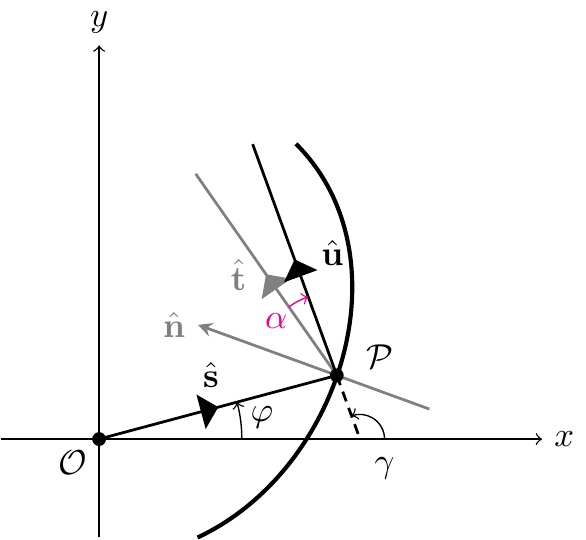}
		\captionsetup{width=0.9\linewidth}
		\caption{Diffuse reflector.}
		\label{fig:2D_uu}
    \end{minipage}
\end{figure}
\noindent Starting with cylindrically symmetric problems, consider the situation depicted in Fig.~\ref{fig:2D_LoR}.
This specular problem can be viewed as a cross-section of a translationally invariant problem, such as an extruded optical element, with a line-source along the suppressed $z$-axis.
For such a system, the specular problem may be analysed in two-dimensions \cite[Ch.~3]{Maes1997ReflectorDesign}.
Thus, the intensities and reflector surfaces are independent of $z$ and we may study a cross-section in the plane of incidence, which is spanned by the source and reflected rays, and which contains the unit normal.
The reflector is parametrised by $\mathbf{r}(\varphi) = u(\varphi) \hat{\mathbf{e}}_r$, where $u(\varphi)\to \mathbb{R}^+,\ \varphi \in [\varphi_1, \varphi_2]$ is at least twice continuously differentiable and $\hat{\mathbf{e}}_r = \big(\!\cos(\varphi),\sin(\varphi)\big)^\intercal$ is the radial unit vector in polar coordinates.
The angle $\varphi$ is measured counter clockwise from the positive $x$-axis, and it fully characterises the source ray along $\hat{\mathbf{s}} = \mathbf{\hat{e}}_r$, emitted form the line source at the origin $\mathcal{O}$.
The hat ($\ \hat{}\ $) indicates unit vectors throughout this paper.
The source ray along $\hat{\mathbf{s}}$ intersects the reflector at some point $\mathcal{P}$, where the unit normal of the reflector is given by $\hat{\mathbf{n}}$.
We take the convention $\hat{\mathbf{s}}\cdot\hat{\mathbf{n}}<0$, i.e., the normal is chosen directed towards the light source.
From the specular law of reflection (LoR), we get an expression for the reflected direction $\hat{\mathbf{t}} = \big(\!\cos(\psi),\sin(\psi)\big)^\intercal$, i.e.,
\begin{equation}\label{eq:LoR}
	\hat{\mathbf{t}} = \hat{\mathbf{s}} - 2(\hat{\mathbf{s}} \cdot \hat{\mathbf{n}}) \hat{\mathbf{n}},
\end{equation}
where we have denoted the angle between the positive $x$-axis and $\hat{\mathbf{t}}$ by $\psi$.

To introduce scattering, consider the situation depicted in Fig.~\ref{fig:2D_uu}.
Inherent in this description is that we have assumed the scattering is limited to the plane of incidence, such that we can again study a cross-section of the translationally invariant problem.
Here, the source ray along $\hat{\mathbf{s}}$ gets mapped to a scattered ray along $\hat{\mathbf{u}} = \big(\!\cos(\gamma), \sin(\gamma)\big)^\intercal$, where $\gamma$ is measured counter-clockwise from the positive $x$-axis.
The scattered direction $\mathbf{\hat{u}}$ can be described as a rotation of $\hat{\mathbf{t}}$ by a stochastic parameter $\alpha$ around the axis parallel to the $z$-axis passing through $\mathcal{P}$, i.e.,
\begin{equation}\label{eq:2D_uu}
	\hat{\mathbf{u}} = \mathbf{R}(\alpha) \hat{\mathbf{t}},
	\quad
	\mathbf{R}(\alpha) =
	\begin{pmatrix}
		\cos(\alpha) & {-}\!\sin(\alpha)\\
		\sin(\alpha) & \cos(\alpha)
	\end{pmatrix}.
\end{equation}
The stochastic parameter $\alpha$ is related to the scattering characteristics of the surface.
We note that $\alpha$ depends on $\psi$, both in the sense that it will almost certainly have a different stochastic value for a given $\psi$ --- in fact, since $\alpha$ is sampled from a probability distribution, it has multiple values for all $\psi$, and in the sense that the probability distribution from which it is sampled may be different for different values of $\psi$.
We shall return to the meaning of this, both mathematically and physically, in Sec.~\ref{sec:2D-energyBalance}.
Finally, we also note that $\alpha$ can be negative, which is the case in Fig.~\ref{fig:2D_uu}.

\subsection{Mappings}
\begin{figure}[H]
	\centering
	\includegraphics[width=0.8\linewidth]{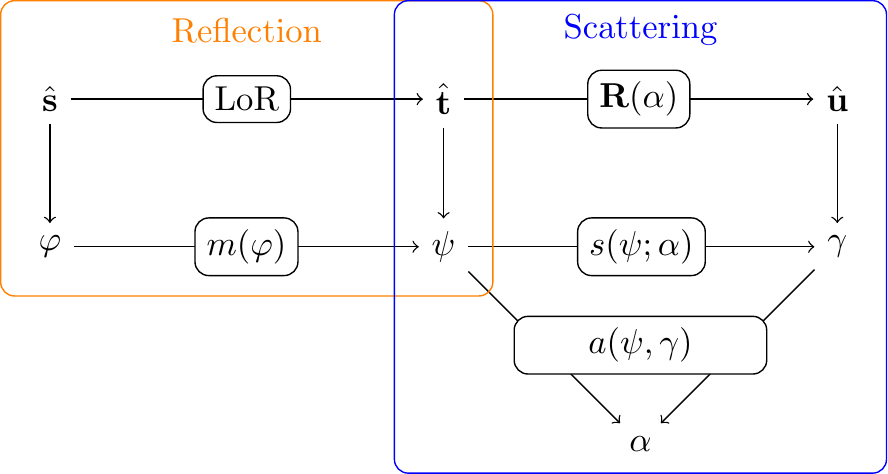}
	\captionsetup{width=0.8\linewidth}
	\caption{Relations between the unit vectors and angles.}
	\label{fig:2D_coordinateRelations}
\end{figure}

To formulate the above in terms of angles, let us introduce two mappings which give the reflected and scattered directions, i.e.,
\begin{equation}
	m(\varphi) = \psi \quad \text{and} \quad s(\psi; \alpha) = \gamma,
\end{equation}
where the former is the law of reflection, and the latter represents the scattering.
This might seem superfluous, but it will simplify the discussion later, especially in the general three-dimensional case, which we intend to treat in a future publication.
In addition to these maps, we require their inverses to exist,
\begin{equation}
	m^{-1}(\psi) = \varphi \quad \text{and} \quad s^{-1}(\gamma; \alpha) = \psi.
\end{equation}
Additionally, we define a mapping yielding $\alpha$, for fixed $\psi$ and $\gamma$, i.e.,
\begin{equation}
	a(\psi, \gamma) = \alpha.
\end{equation}
For a schematic summary, see Fig.~\ref{fig:2D_coordinateRelations}.

\noindent In total, the maps are
\begin{equation}\label{eq:2D_mappings}
	\begin{split}
		m(\varphi) &= \varphi + \arccos(1-2\big(\hat{\mathbf{s}}\cdot\hat{\mathbf{n}}\big)^2) \equiv \psi,\\
		m^{-1}(\psi) &= \psi - \arccos(1-2\big(\hat{\mathbf{s}}\cdot\hat{\mathbf{n}}\big)^2) \equiv \varphi,\\
		s(\psi; \alpha) &= \psi + \alpha \equiv \gamma,\\
		s^{-1}(\gamma; \alpha) &= \gamma - \alpha \equiv \psi,\\
		a(\psi,\gamma) &= \gamma - \psi \equiv \alpha,
	\end{split}
\end{equation}
where the first two relations follow from $\cos(\psi - \varphi) = \hat{\mathbf{s}} \cdot \hat{\mathbf{t}}$, see Fig.~\ref{fig:2D_LoR}, and the LoR, Eq.~\eqref{eq:LoR}.
The existence of inverse mappings is not \textit{a priori} guaranteed for all situations, but we shall restrict our attention to problems where they do exist.

\subsection{Energy Balances}\label{sec:2D-energyBalance}
Having presented the mappings for the angles, we are now ready to formulate the energy balances.
To start, fix the angles
\begin{itemize}\setlength\itemsep{0.1em}
	\item $\varphi_1, \varphi_2 \in (-\pi,\pi),\ \varphi_1 < \varphi_2$,
	\item $\psi_1, \psi_2 \in(-\pi,\pi),\ \psi_1 < \psi_2$,
	\item $\gamma_1, \gamma_2 \in (-\pi,\pi),\ \gamma_1 < \gamma_2$,
\end{itemize}
and introduce the intensity distributions (illuminance) [lm/rad]:
\begin{itemize}\setlength\itemsep{0.1em}
	\item source intensity distribution $f(\varphi) \to \mathbb{R}^+,\ \varphi \in [\varphi_1, \varphi_2]$,
	\item intermediate specular intensity distribution $g(\psi) \to \mathbb{R}^+,\ \psi \in [\psi_1,\psi_2]$,
	\item diffuse target intensity distribution $h(\gamma) \to \mathbb{R}^+,\ \gamma \in [\gamma_1,\gamma_2]$.
\end{itemize}
In the design procedure outlined in Sec.~\ref{sec:reflectorDesign}, the source and diffuse target distributions, $f$ and $h$, are given, and the intermediate specular intensity distribution $g$ is computed, and used in the design of the reflector.
Assuming no light is lost along the way from source to target, we may formulate the global energy balances as
\begin{equation}\label{eq:2D_globalEnergyBalance}
	\uint_{\varphi_1}^{\varphi_2} f(\varphi) \, \text{d}\varphi =
	\uint_{\psi_1}^{\psi_2} g(\psi) \, \text{d}\psi =
	\uint_{\gamma_1}^{\gamma_2} h(\gamma) \, \text{d}\gamma.
\end{equation}

\begin{figure}[htbp]
	\centering
	\begin{minipage}{0.5\linewidth}
		\centering
		\includegraphics[width=0.95\linewidth]{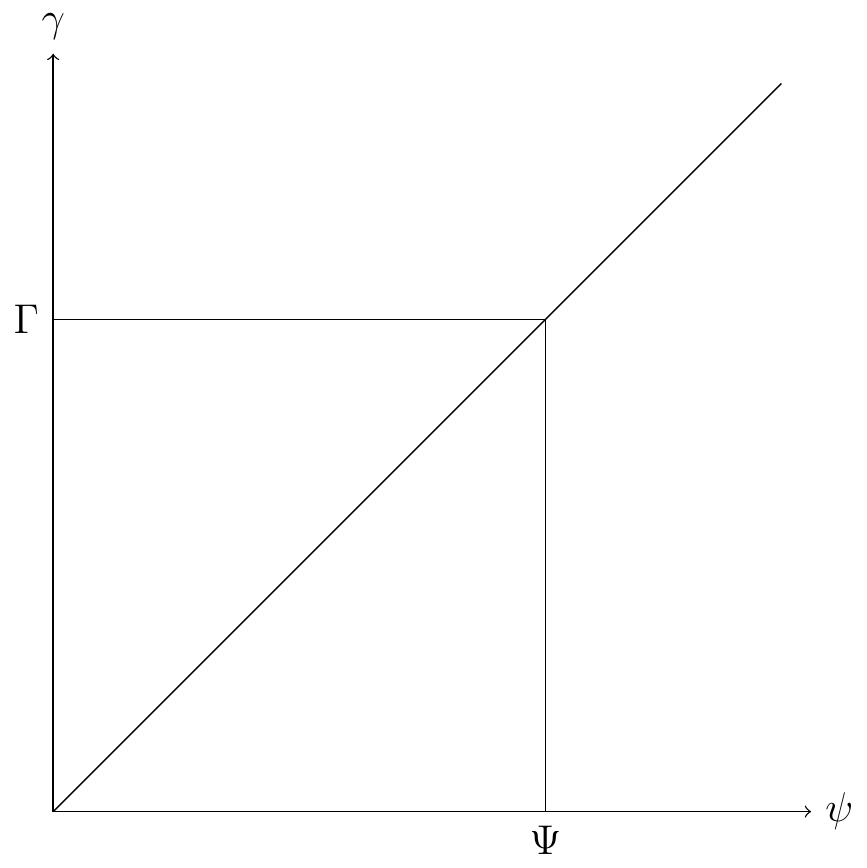}
		\captionsetup{width=0.95\linewidth}
		\caption{Specular map $\Psi \to \Gamma$.}
		\label{fig:psigammaSpecular}
	\end{minipage}\hfill
	\begin{minipage}{0.5\linewidth}
		\centering
		\includegraphics[width=0.95\linewidth]{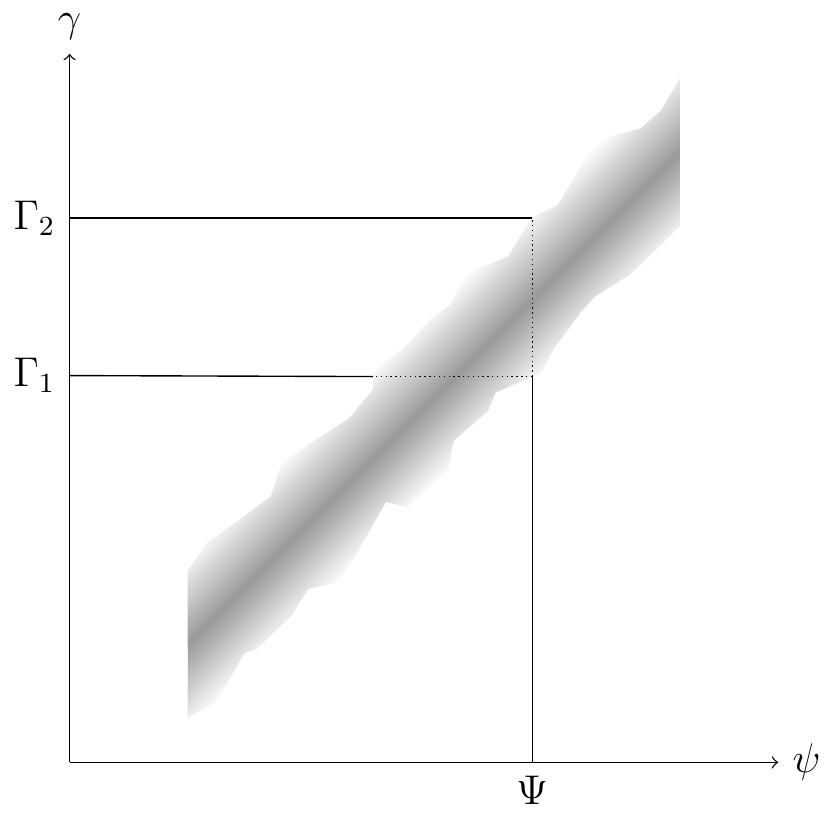}
		\captionsetup{width=0.95\linewidth}
		\caption{Diffuse map $\Psi \to \Gamma$.}
		\label{fig:psigammaDIFFERENT}
    \end{minipage}
\end{figure}

\noindent Consider next the relationship between $\psi$ and $\gamma$ for a fixed $\psi = \Psi$.
Suppose we have a perfect specular reflector (i.e., a mirror).
Then, $\alpha$ always vanishes, such that $\psi \equiv \gamma$ and for fixed $\psi = \Psi$, we simply get a fixed $\gamma = \Gamma$.
This is depicted schematically in Fig.~\ref{fig:psigammaSpecular}.
Consider now the case where we have nonzero scattering, and indeed where the scattering may vary depending on the incident angle.
This yields a situation like the one depicted schematically in Fig.~\ref{fig:psigammaDIFFERENT}.
Here, fixing $\psi = \Psi$ and tracking where all the light emerges, we see that it falls within the interval $[\Gamma_1, \Gamma_2] \subset [\gamma_1, \gamma_2]$.

\noindent Motivated by the more fundamental concept of optimal transport theory, and in particular Monge-Kantorovich problems \cite[Ch.~1]{Villani2003Topics}, let us introduce the \textit{density} $\rho(\psi,\gamma) \to \mathbb{R}^+,\ \psi \in [\psi_1,\psi_2], \gamma \in [\gamma_1,\gamma_2]$, with the properties
\begin{subequations}\label{eq:2D_angRhoIntegral}
	\renewcommand{\theequation}{(\arabic{parentequation}\alph{equation})}
	\begin{align}
		\uint_{\gamma_1}^{\gamma_2} \rho(\psi,\gamma) \, \text{d}\gamma &= g(\psi),\label{eq:2D_angRhoIntegral_g}\\
		\uint_{\psi_1}^{\psi_2} \rho(\psi,\gamma) \, \text{d}\psi &= h(\gamma).\label{eq:2D_angRhoIntegral_h}
	\end{align}
\end{subequations}
There are several natural requirements on $\rho$, including positivity and a compact support.
In Fig.~\ref{fig:psigammaDIFFERENT}, its support may be considered the shading, where darker values represent a higher density, and the support is clearly a function of the angles.
Notice that the second energy balance in Eq.~\eqref{eq:2D_globalEnergyBalance} is trivially fulfilled by direct substitution:
\begin{subequations}
	\renewcommand{\theequation}{(\arabic{parentequation}\alph{equation})}
	\begin{align}
		\uint_{\gamma_1}^{\gamma_2} h(\gamma) \, \text{d}\gamma &=
		\uint_{\gamma_1}^{\gamma_2} \uint_{\psi_1}^{\psi_2} \rho(\psi,\gamma) \, \text{d}\psi \text{d}\gamma,\\
		\uint_{\psi_1}^{\psi_2} g(\psi) \, \text{d}\psi &=
		\uint_{\psi_1}^{\psi_2} \uint_{\gamma_1}^{\gamma_2} \rho(\psi,\gamma) \, \text{d}\gamma \text{d}\psi,
	\end{align}
\end{subequations}
which are the same after a change of integration order.

\noindent Let us now attempt to find a suitable choice of $\rho(\psi,\gamma)$.
In particular, consider
\begin{equation}\label{eq:2D_rhoPsiGamma}
	\rho(\psi,\gamma) = p\big(a(\psi,\gamma);\psi\big) g(\psi),
\end{equation}
where $p$ is a function describing the redistribution of light, subject to an energy constraint we shall formulate momentarily.
Physically, this choice can be motivated as follows.
In the specular case, i.e., Fig.~\ref{fig:psigammaSpecular}, $p\big(a(\psi,\gamma); \psi\big) = \delta(\gamma-\psi)$, where $\delta$ represents the Dirac delta function, meaning the light will be scattered in exactly one direction $\gamma \equiv \psi$.
When $p$ is some other appropriate function, light in direction $\psi$ is scattered over multiple angles and we have a situation similar to that in Fig.~\ref{fig:psigammaDIFFERENT}, i.e., this choice of $\rho$ represents the physical properties of light scattering.
Note, however, that $p$ includes the parameter $\psi$, which highlights the possibility of unique $p$ functions for each specular ray, which is what is schematically shown in Fig.~\ref{fig:psigammaDIFFERENT}.
We now insert this density in Eq.~\eqref{eq:2D_angRhoIntegral_g} to get
\begin{equation}
	\uint_{\gamma_1}^{\gamma_2} p\big(a(\psi,\gamma);\psi\big) \, \text{d}\gamma = 1.
\end{equation}
Transforming the integration variable $\gamma$ to $\alpha$, recalling that $\gamma=s(\psi;\alpha)$ and $a(\psi,\gamma) = \alpha$, yields
\begin{subequations}\label{eq:2D_alphaLimits}
	\renewcommand{\theequation}{(\arabic{parentequation}\alph{equation})}
	\begin{align}
		&\uint_{\alpha_1}^{\alpha_2} p(\alpha;\psi) \, \left|\frac{\partial s(\psi;\alpha)}{\partial\alpha}\right| \, \text{d}\alpha = 1,\\
		\alpha_1 = &\min\Big\{ a(\psi,\gamma)\ \big|\ \psi \in [\psi_1,\psi_2],\ \gamma \in [\gamma_1,\gamma_2] \Big\},\\
		\alpha_2 = &\max\Big\{ a(\psi,\gamma)\ \big|\ \psi \in [\psi_1,\psi_2],\ \gamma \in [\gamma_1,\gamma_2] \Big\}.
	\end{align}
\end{subequations}
Note that with our choice $s(\psi;\alpha) = \psi + \alpha$ in Eq.~\eqref{eq:2D_mappings}, the Jacobian $\abs{\partial s/\partial \alpha} = 1$, so that
\begin{equation}
	\uint_{\alpha_1}^{\alpha_2} p(\alpha;\psi) \, \text{d}\alpha = 1, \quad \forall \psi \in [\psi_1, \psi_2].
\end{equation}
With one additional obvious requirement that $p(\alpha;\psi) \geq 0$, it is clear that with our choice of $s(\psi;\alpha)$, $p$ becomes a probability density function (PDF).

\subsection{Integral Equation}
Let us now focus on Eq.~\eqref{eq:2D_angRhoIntegral_h}.
Substituting our choice of $\rho(\psi,\gamma)$ from Eq.~\eqref{eq:2D_rhoPsiGamma} yields
\begin{equation}\label{eq:2D_hgammaGeneralPsi}
	h(\gamma) =
	\uint_{\psi_1}^{\psi_2} p\big(a(\psi,\gamma); \psi\big) g(\psi) \, \text{d}\psi.
\end{equation}

\noindent We once again utilise $a(\psi,\gamma)$ to change the integration variable from $\psi$ to $\alpha$, together with $\psi = s^{-1}(\gamma; \alpha)$, to get
\begin{equation}\label{eq:2D_hgammaGeneralAlpha}
	h(\gamma) =
	\uint_{\alpha_1}^{\alpha_2} p\big(\alpha; s^{-1}(\gamma; \alpha)\big) g\big(s^{-1}(\gamma; \alpha)\big) \, \left|\frac{\partial s^{-1}(\gamma; \alpha)}{\partial \alpha}\right| \, \text{d}\alpha,
\end{equation}
where $\alpha_1$ and $\alpha_2$ were defined in Eq.~\eqref{eq:2D_alphaLimits}.
Here, we note that this is a Fredholm integral equation for $g$, as $p$ depends on both $\gamma$ (via $\psi = s^{-1}(\gamma;\alpha)$) and $\alpha$.
That is, $p$ is a spatially varying kernel function, $h$ is the prescribed target and $g$ is to be determined.

\begin{figure}[H]
	\centering
	\includegraphics[width=0.522\linewidth]{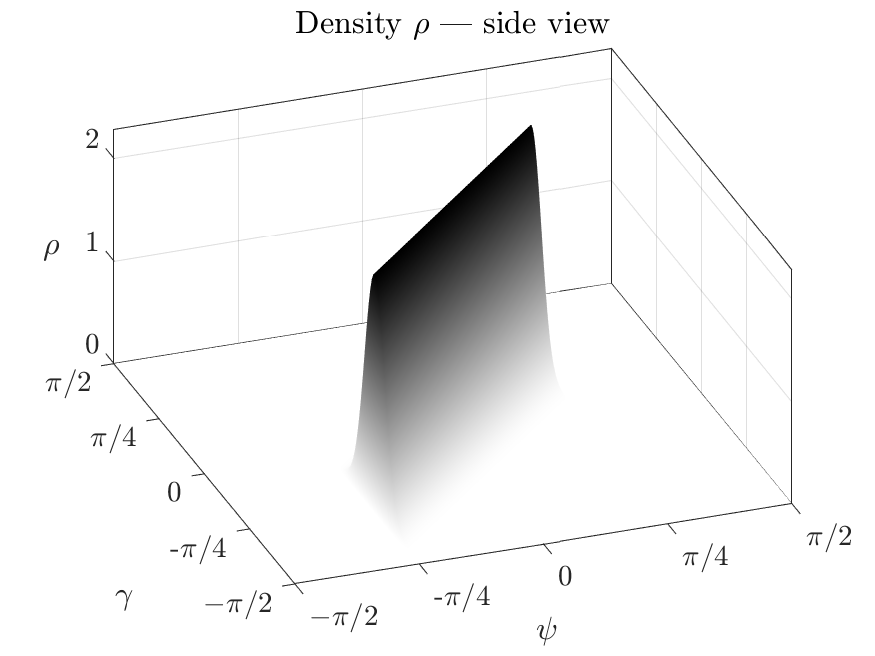}%
	\includegraphics[width=0.478\linewidth]{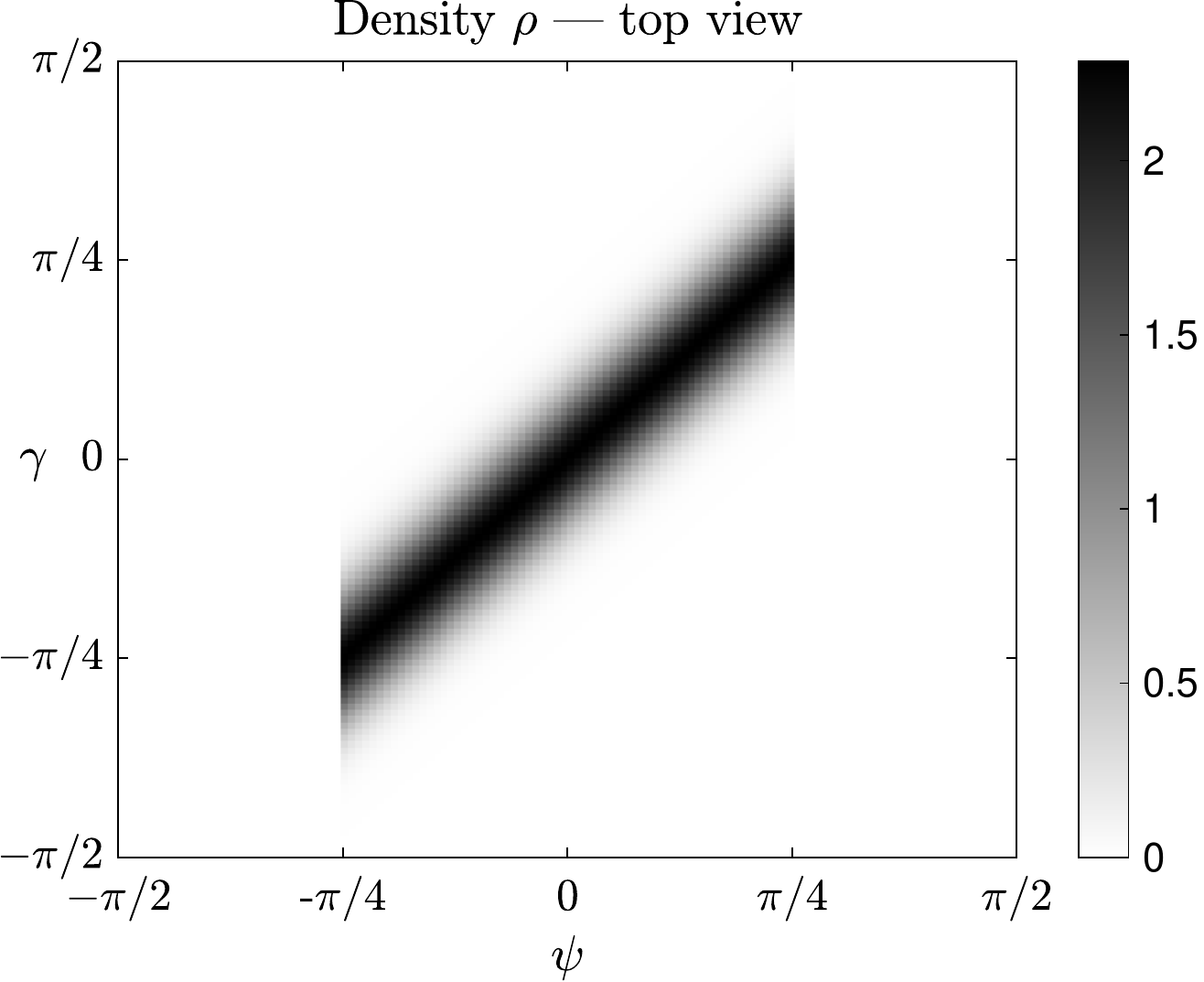}
	\captionsetup{width=\linewidth}
	\caption{Diffuse map $\psi \to \gamma$ (isotropic scattering; example \#$2$ in Sec.~\ref{sec:results}).}
	\label{fig:psigamma}
\end{figure}

\noindent In the case of isotropic scattering, the explicit $\psi$-dependence in $p$ is omitted, meaning we get $p(a(\psi,\gamma))$, or simply $p(\alpha)$.
The $\psi$ vs.~$\gamma$ plot for such a situation is shown in Fig.~\ref{fig:psigamma}.
In contrast to Fig.~\ref{fig:psigammaDIFFERENT}, the support of $\rho$ is now a band of constant width, and the data represent that of Example \#$2$ in Sec.~\ref{sec:results}.
Inserting $a(\psi,\gamma)$ and $s^{-1}(\gamma;\alpha)$ from Eq.~\eqref{eq:2D_mappings} into Eqs.~\eqref{eq:2D_hgammaGeneralPsi} and \eqref{eq:2D_hgammaGeneralAlpha} yields
\begin{subequations}\label{eq:2D_hgammaFinal}
	\renewcommand{\theequation}{(\arabic{parentequation}\alph{equation})}
	\begin{eqnarray}
		h(\gamma) = \uint_{\psi_1}^{\psi_2} p(\gamma-\psi) g(\psi) \, \text{d}\psi,\label{eq:2D_hgammaFinalPsi}\\
		h(\gamma) = \uint_{\alpha_1}^{\alpha_2} p(\alpha) g(\gamma-\alpha) \, \text{d}\alpha,\label{eq:2D_hgammaFinalAlpha}
	\end{eqnarray}
\end{subequations}
which are convolution integrals.
We shall use the common notation of $h(\gamma) = (p*g)(\gamma)$ for the convolution in Eq.~\eqref{eq:2D_hgammaFinalPsi}.
Due to the commutativity property of convolution integrals, an equivalent definition is $h(\gamma) = (g*p)(\gamma)$ in Eq.~\eqref{eq:2D_hgammaFinalAlpha} \cite[p.~309]{Rade2004}.
Obtaining $g$ is now a matter of deconvolving Eq.~\eqref{eq:2D_hgammaFinalPsi} or \eqref{eq:2D_hgammaFinalAlpha}.
There are a large number of different deconvolution methods, but not all are equally suitable for our purposes.
In particular, $g$ must be nonnegative.
This can be achieved using an iterative ratio method, such as Gold's method, or the more common Richardson-Lucy method of deconvolution \cite{Jansson1997Deconvolution}.
We shall return to this topic in Sec.~\ref{sec:results}.

\clearpage
\section{Rotationally Symmetric Reflectors}\label{sec:3D-refl}
\hspace{1pt}\vspace{-11pt}
\begin{wrapfigure}{r}{0.39\textwidth}
	\vspace{-30pt}
	\includegraphics[width=\linewidth]{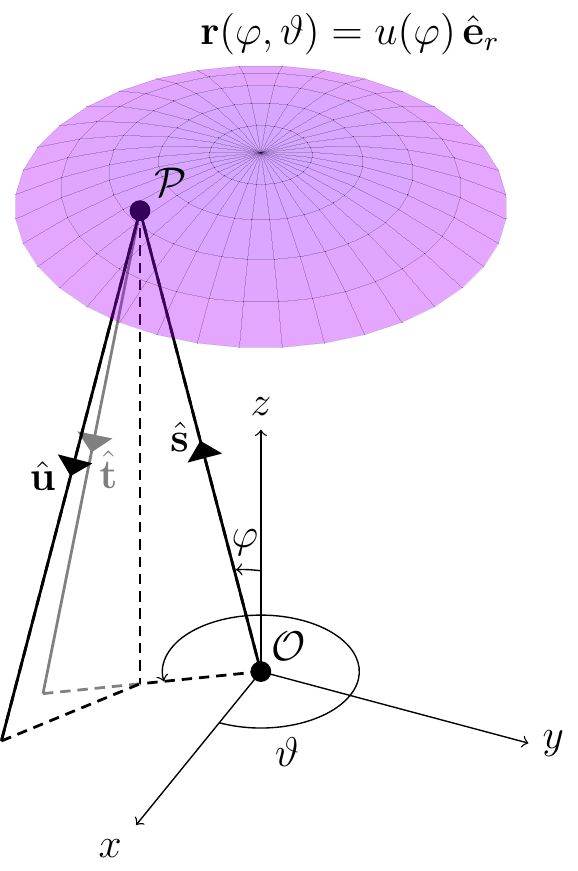}
	\caption{A rotationally symmetric reflector with arbitrary scattering.}
	\label{fig:3D_uu_sym}
	\vspace{-30pt}
\end{wrapfigure}
\noindent To introduce the rotationally symmetric three-dimensional reflectors, consider Fig.~\ref{fig:3D_uu_sym}.
The situation is as follows.
A point-source is located at the origin $\mathcal{O}$.
The reflector is parametrised by $\mathbf{r}(\varphi,\vartheta) = u(\varphi)\hat{\mathbf{e}}_r$, where $u(\varphi) \to \mathbb{R}^+,\ \varphi \in [\varphi_1,\varphi_2]$ is at least twice continuously differentiable and $\hat{\mathbf{e}}_r = \big(\! \sin(\varphi)\cos(\vartheta),\sin(\varphi)\sin(\vartheta),\cos(\varphi)\big)^\intercal$ is the radial unit vector in spherical coordinates.
The angle $\varphi \in [0,\pi)$ is measured from the positive $z$-axis and the angle $\vartheta \in (-\pi, \pi)$ is measured from the positive $x$-axis.
Consider now a ray in direction $\hat{\mathbf{s}}$ emitted from the point source at $\mathcal{O}$.
Following its trajectory, it strikes the reflector at a point $\mathcal{P}$, where the unit normal (not shown) is given by $\hat{\mathbf{n}}$.
Just like in the two-dimensional case, we adopt the convention $\hat{\mathbf{s}} \cdot \hat{\mathbf{n}} < 0$, i.e., the normal points towards the light source.
From the LoR, Eq.~\eqref{eq:LoR}, we get an expression for the reflected ray $\hat{\mathbf{t}}$.
Finally, the scattered ray $\hat{\mathbf{u}}$ may in general be computed by picking a new direction in a cone coaxial with $\hat{\mathbf{t}}$.
In general, we would thus require two stochastic parameters to fix $\hat{\mathbf{u}}$.

\vspace*{10pt}
\subsection{In-plane Scattering}
\hspace{1pt}\vspace{-25pt}
\begin{wrapfigure}{r}{0.39\textwidth}
	\includegraphics[width=\linewidth]{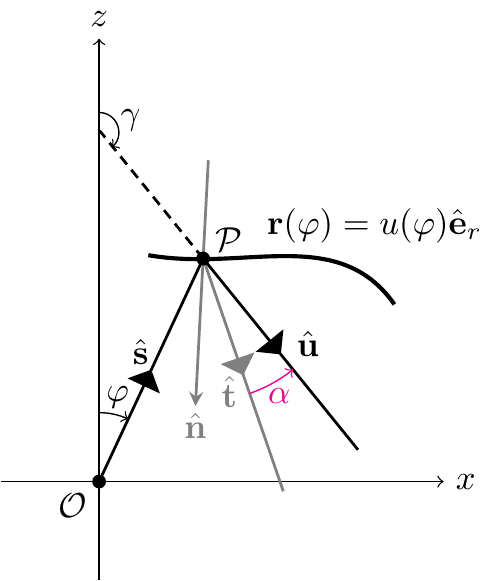}
	\caption{Cross-section in the plane of incidence under the assumption of in-plane scattering.}
	\label{fig:3D_rot}
\end{wrapfigure}

\noindent If, in addition to the reflector being rotationally symmetric, the scattered direction is in the plane of incidence, the problem may be analysed in two dimensions.
The fact that the specular problem reduces to two dimensions is shown in \cite[Ch.~3]{Maes1997ReflectorDesign}, and in-plane scattering preserves this symmetry.
The situation is depicted in Fig.~\ref{fig:3D_rot}, depicting the cross-section in the plane of incidence.
The two-dimensional reflector is parametrised by $\mathbf{r}(\varphi) = u(\varphi) \hat{\mathbf{e}}_r$, where $\hat{\mathbf{e}}_r = \big(\!\sin(\varphi),\cos(\varphi)\big)^\intercal$, $\varphi \in [0,\pi)$ measured from the positive $z$-axis.
	The source ray $\hat{\mathbf{s}} = \big(\!\sin(\varphi),\cos(\varphi)\big)^\intercal$ intersects the reflector at a point $\mathcal{P}$ with unit normal $\hat{\mathbf{n}}$, and the specular direction $\hat{\mathbf{t}} = \big(\!\sin(\psi),\cos(\psi)\big)^\intercal$, where $\psi \in [0,\pi)$ is measured from the positive $z$-axis, is given by the LoR, Eq.~\eqref{eq:LoR}.
	The scattered direction $\mathbf{\hat{u}} = \big(\!\sin(\gamma),\cos(\gamma)\big)^\intercal$, where $\psi \in [0,\pi)$ is measured from the positive $z$-axis, is given by a rotation by a stochastic parameter $\alpha$ in accordance with Eq.~\eqref{eq:2D_uu}.

Before proceeding, let us briefly discuss when such an approximation is valid.
The rotationally symmetric reflector is self-explanatory, but the in-plane scattering is less straight-forward.
The proposed situation where this may hold is as follows.
A rotationally symmetric reflector was machined in a manner which left the surface chiseled.
Specifically, the chisel-marks follow a tightly-wound spiral along the reflector.
In this situation, we postulate that we may study the cross-section in the plane of incidence.
Within these restrictions, i.e., a rotationally symmetric reflector with chisel-induced scattering, and under our postulate of in-plane scattering only, we may readily use the formulae in the prior sections, with a few changes that shall be highlighted shortly.
We furthermore assume that the intensity distributions are rotationally symmetric, such that the following description will suffice [lm/sr]:
\begin{itemize}\setlength\itemsep{0.1em}
	\item source intensity distribution $f(\varphi) \to \mathbb{R}^+,\ \varphi \in [\varphi_1, \varphi_2]$,
	\item intermediate specular intensity distribution $g(\psi) \to \mathbb{R}^+, \psi \in [\psi_1,\psi_2]$,
	\item diffuse target intensity distribution $h(\gamma) \to \mathbb{R}^+, \gamma \in [\gamma_1,\gamma_2]$,
\end{itemize}
where $\varphi_1 < \varphi_2$, $\psi_1 < \psi_2$ and $\gamma_1 < \gamma_2$ are the supports of the distributions.
The energy balances in Eq.~\eqref{eq:2D_globalEnergyBalance} become
\begin{equation}\label{eq:3D_globalEnergyBalance}
		\uint_{\varphi_1}^{\varphi_2} f(\varphi) \sin(\varphi) \, \dd\varphi =
		\uint_{\psi_1}^{\psi_2} g(\psi) \sin(\psi) \, \dd\psi =
		\uint_{\gamma_1}^{\gamma_2} h(\gamma) \sin(\gamma) \, \dd\gamma,
\end{equation}
where the sine terms come from integration over the unit sphere.
Following the procedure in the previous section, let us introduce the density $\rho(\psi,\gamma) \to \mathbb{R}^+,\ \psi \in[\psi_1, \psi_2],\ \gamma [\gamma_1,\gamma_2]$, such that
\begin{subequations}\label{eq:rot_angRhoIntegral}
	\renewcommand{\theequation}{(\arabic{parentequation}\alph{equation})}
	\begin{align}
		\uint_{\gamma_1}^{\gamma_2} \rho(\psi,\gamma) \sin(\gamma) \, \text{d}\gamma &= g(\psi),\label{eq:rot_angRhoIntegral_g}\\
		\uint_{\psi_1}^{\psi_2} \rho(\psi,\gamma) \sin(\psi) \, \text{d}\psi &= h(\gamma).\label{eq:rot_angRhoIntegral_h}
	\end{align}
\end{subequations}
Comparing these equations to Eq.~\eqref{eq:2D_angRhoIntegral}, notice that they are the same up to the sine terms from the spherical area elements.
As such, the natural choice for $\rho$ becomes (recall Eq.~\eqref{eq:2D_rhoPsiGamma}; assuming isotropic scattering, i.e., no explicit $\psi$-dependence)
\begin{equation}\label{eq:rot_rho}
	\rho(\psi,\gamma) \sin(\gamma) = p\big(a(\psi,\gamma)\big)g(\psi).
\end{equation}
We now substitute this $\rho$ into Eq.~\eqref{eq:rot_angRhoIntegral_g} to get
\begin{equation}
	\uint_{\gamma_1}^{\gamma_2} p\big(a(\psi,\gamma)\big) \, \dd \gamma = 1.
\end{equation}
Transforming $\gamma$ to $\alpha$ yields
\begin{equation}
	\uint_{\alpha_1}^{\alpha_2} p(\alpha) \abs{\pdv{s(\psi;\alpha)}{\alpha}} \, \dd \alpha = 1,
\end{equation}
where $\alpha_1$ and $\alpha_2$ were defined in Eq.~\eqref{eq:2D_alphaLimits}.
The Jacobian will become unity via the mappings in Eq.~\eqref{eq:2D_mappings}.
Whence, the normalisation of $p$ is
\begin{equation}
	\uint_{\alpha_1}^{\alpha_2} p(\alpha) \, \dd \alpha = 1.
\end{equation}
Substituting $\rho$, defined in Eq.~\eqref{eq:rot_rho}, into Eq.~\eqref{eq:rot_angRhoIntegral_h} yields
\begin{equation}
	h(\gamma)\sin(\gamma) = \uint_{\psi_1}^{\psi_2} p\big(a(\psi,\gamma)\big)g(\psi) \sin(\psi) \, \dd\psi.
\end{equation}
Absorbing the sine terms in the intensity distributions by defining
\begin{equation}
	\tilde{h}(\gamma) := h(\gamma) \sin(\gamma), \quad \tilde{g}(\psi) := g(\psi)\sin(\psi),
\end{equation}
and inserting $a(\psi,\gamma)$ from Eq.~\eqref{eq:2D_mappings}, yields
\begin{subequations}\label{eq:rot_hgammaFinal}
	\renewcommand{\theequation}{(\arabic{parentequation}\alph{equation})}
	\begin{align}
		\tilde{h}(\gamma) = \uint_{\psi_1}^{\psi_2} p(\gamma - \psi) \tilde{g}(\psi) \, \dd\psi,\\
		\tilde{h}(\gamma) = \uint_{\alpha_1}^{\alpha_2} p(\alpha) \tilde{g}(\gamma-\alpha) \, \dd\alpha,
	\end{align}
\end{subequations}
where the second equation is obtained by transforming $\psi$ to $\alpha$.
Comparing these to Eq.~\eqref{eq:2D_hgammaFinal}, it is clear that they are the same, up to the sine terms from the spherical area elements in the modified distributions.
As such, deconvolution is still a vital tool to obtain the specular target distribution $g$, used in the reflector design procedure.

\section{Specular Reflector Design}\label{sec:reflectorDesign}
Our goal is as follows.
Determine a specular reflector which transforms the given source distribution into the given target distribution.
The approach we have chosen involves solving two ordinary differential equations (ODEs) for the radius function $u(\varphi)$ and the mapping $m(\varphi)$, which together fully characterise the reflector.
This is similar to the approach outlined in \cite[Ch.~3.3]{Maes1997ReflectorDesign}.

\subsection{Cylindrically Symmetric Reflectors}
Recall that the reflector is parametrised by $\mathbf{r}(\varphi) = u(\varphi) \hat{\mathbf{e}}_r$, and that $\varphi$ is measured counter clockwise from the positive $x$-axis (refer to Fig.~\ref{fig:2D_LoR}).
To start, note that a tangent vector to the reflector is
\begin{equation}
	\boldsymbol{\tau} = \mathbf{r}'(\varphi) = u'(\varphi) \hat{\mathbf{e}}_r + u(\varphi) \hat{\mathbf{e}}_\varphi,
\end{equation}
where $\hat{\mathbf{e}}_r = \big(\!\cos(\varphi),\sin(\varphi)\big)^\intercal$ and $\hat{\mathbf{e}}_\varphi = \big({-}\!\sin(\varphi),\cos(\varphi)\big)^\intercal$ are the standard unit vectors in polar coordinates.
The corresponding normal vector can be constructed by rotating this vector counter clockwise, i.e.,
\begin{equation}
	\mathbf{n} = \mathbf{R}(\pi/2)\boldsymbol{\tau}, \quad \mathbf{R}(\pi/2) = \begin{pmatrix} 0 & -1\\ 1 & 0 \end{pmatrix},
\end{equation}
where the rotation matrix $\mathbf{R}$ was initially defined in Eq.~\eqref{eq:2D_uu}.
The associated unit normal is
\begin{equation}
	\hat{\mathbf{n}} = \frac{\mathbf{n}}{|\mathbf{n}|} = \frac{\mathbf{n}}{|\boldsymbol{\tau}|} = \frac{-u(\varphi) \hat{\mathbf{e}}_r + u'(\varphi) \hat{\mathbf{e}}_\varphi}{\sqrt{u(\varphi)^2+u'(\varphi)^2}},
\end{equation}
where we made use of the fact that $\mathbf{R}(\pi/2) \hat{\mathbf{e}}_r = \hat{\mathbf{e}}_{\varphi}$ and $\mathbf{R}(\pi/2) \hat{\mathbf{e}}_{\varphi} = - \hat{\mathbf{e}}_r$.
Let $v(\varphi) := \ln\!\big(u(\varphi)\big)$, so that
\begin{equation}
	\hat{\mathbf{n}} = \frac{- \hat{\mathbf{e}}_r + v'(\varphi) \hat{\mathbf{e}}_\varphi}{\sqrt{1+v'(\varphi)^2}}.
\end{equation}
Let us compute
\begin{equation}\label{eq:2D_sn}
	\hat{\mathbf{s}} \cdot \hat{\mathbf{n}} = \frac{-1}{\sqrt{1+v'(\varphi)^2}},
\end{equation}
which we shall use momentarily.
In doing so, we made use of the relation $\hat{\mathbf{s}} = \hat{\mathbf{e}}_r$.
Note that $\hat{\mathbf{s}} \cdot \hat{\mathbf{n}} < 0$, indicating we rotated $\boldsymbol{\tau}$ the correct way to get $\mathbf{n}$.
From the LoR, Eq.~\eqref{eq:LoR}, we get
\begin{equation}
	\hat{\mathbf{s}} \cdot \hat{\mathbf{t}} = 1 - 2(\hat{\mathbf{s}} \cdot \hat{\mathbf{n}})^2.
\end{equation}
By geometrical arguments --- see Fig.~\ref{fig:2D_LoR}, it is clear that $\hat{\mathbf{s}} \cdot \hat{\mathbf{t}} = -\hat{\mathbf{s}} \cdot - \hat{\mathbf{t}} = \cos(\psi - \varphi)$, such that together with Eq.~\eqref{eq:2D_sn}, we get
\begin{equation}
	\cos(\psi - \varphi) = 1 - \frac{2}{1+v'(\varphi)^2},
\end{equation}
or, equivalently
\begin{equation}
	v'(\varphi) = \pm \sqrt{\frac{1 + \cos(\psi - \varphi)}{1 - \cos(\psi- \varphi)}} = \cot(\frac{m(\varphi)-\varphi}{2}),
\end{equation}
where we used the tangent half-angle relation \cite[p.~127]{Rade2004} and switched from $\psi$ to $m(\varphi)$ in the last step to highlight that this is indeed the specular map, which has an explicit $\varphi$-dependence.
To solve this ODE, we shall make use of the arbitrary boundary condition $v(\varphi_1) = 0$.
We thus have the initial value problem (IVP)
\begin{equation}\label{eq:2D_u_ODE}
	\begin{cases}
		\mathlarger{v'(\varphi) = \cot(\frac{m(\varphi)-\varphi}{2})},\\
		v(\varphi_1) = 0.
	\end{cases},\quad
	\varphi_1 < \varphi < \varphi_2.
\end{equation}

\noindent Next, we consider a monotonically increasing or decreasing optical map $m(\varphi)$.
Suppose we have a monotonically increasing function $m(\varphi) =: m_\mathrm{div}(\varphi)$, together with the boundary condition $m_\mathrm{div}(\varphi_1) = \psi_1$, such that the reflected rays do not intersect, i.e., the ray bundle is divergent.
Since $m_\mathrm{div}$ is by construction a valid solution, i.e., it achieves $g$, given $f$, the following must hold for all $\varphi \in [\varphi_1, \varphi_2]$ (recall that $\psi_1 < \psi < \psi_2$)
\begin{equation}\label{eq:reflDesign_psiPlus}
	\uint_{\varphi_1}^{\varphi} f(\tilde{\varphi}) \, \text{d}\tilde{\varphi} = \uint_{\psi_1}^{m_\mathrm{div}(\varphi)} g(\tilde{\psi}) \, \text{d}\tilde{\psi}.
\end{equation}
Differentiation with respect to $\varphi$ immediately yields the IVP
\begin{equation}\label{eq:2D_psiPlusODE}
	\begin{cases}
		\mathlarger{m_\mathrm{div}'(\varphi) = \frac{f(\varphi)}{g\big(m_\mathrm{div}(\varphi)\big)}},\\
		m_\mathrm{div}(\varphi_1) = \psi_1.
	\end{cases}
\end{equation}
Analogous considerations with a monotonically decreasing function $m(\varphi) := m_\mathrm{conv}(\varphi)$, where we instead have a convergent ray bundle, yield, for all $\varphi$,
\begin{equation}\label{eq:refldesign_psiMinus}
	\uint_{\varphi_1}^{\varphi} f(\tilde{\varphi}) \, \text{d}\tilde{\varphi} = \uint_{m_\mathrm{conv}(\varphi)}^{\psi_2} g(\tilde{\psi}) \, \text{d}\tilde{\psi},
\end{equation}
or in terms of the equivalent IVP,
\begin{equation}\label{eq:2D_psiMinusODE}
	\begin{cases}
		\mathlarger{m_\mathrm{conv}'(\varphi) = -\frac{f(\varphi)}{g\big(m_\mathrm{conv}(\varphi)\big)}},\\
		m_\mathrm{conv}(\varphi_1) = \psi_2.
	\end{cases}
\end{equation}
Once the desired $m(\varphi)$ has been obtained, it is substituted into Eq.~\eqref{eq:2D_u_ODE}, which is then solved, yielding $v(\varphi)$ and consequently $u(\varphi)=\mathrm{e}^{v(\varphi)}$, which fully characterises the reflector.
Throughout this paper, we have utilised Matlab's \texttt{ode15s} routine to solve the IPVs.

\subsection{Rotationally Symmetric Reflectors}
In the case of rotationally symmetric reflectors, we measure $\varphi$ clockwise from the positive $z$-axis, in the plane of incidence.
Since $\vartheta$ is constant, the reflector is now parametrised by $\mathbf{r}(\varphi) = u(\varphi) \hat{\mathbf{e}}_r$, where $\hat{\mathbf{e}}_r = \big(\! \sin(\varphi),\cos(\varphi)\big)^\intercal$ is the radial unit vector in this particular polar coordinate system.
Thus, the tangent vector becomes
\begin{equation}
	\boldsymbol{\tau} = \mathbf{r}'(\varphi) = u'(\varphi) \hat{\mathbf{e}}_r + u(\varphi) \hat{\mathbf{e}}_{\varphi},
\end{equation}
where $\hat{\mathbf{e}}_{\varphi} = \big(\! \cos(\varphi),-\sin(\varphi)\big)^\intercal$ is the angular unit vector in this polar coordinate system.
To obtain the unit normal pointing towards the source, we rotate the tangent vector clockwise, i.e.,
\begin{equation}
	\mathbf{n} = \mathbf{R}(-\pi/2)\boldsymbol{\tau}, \quad \mathbf{R}(-\pi/2) = \begin{pmatrix} 0 & 1\\ -1 & 0 \end{pmatrix}.
\end{equation}
Following the procedure from the previous section, we end up with a unit normal
\begin{equation}
	\mathbf{\hat{n}} = \frac{-\mathbf{\hat{e}}_r+v'(\varphi)\mathbf{\hat{e}}_\varphi}{\sqrt{1+v'(\varphi)^2}},
\end{equation}
after introducing $v(\varphi) := \ln\!\big(u(\varphi)\big)$.
Finally, we consider the geometry of the situation (see Fig.~\ref{fig:3D_rot}; $\psi$ is analogous to $\gamma$) to conclude that $\hat{\mathbf{s}} \cdot \hat{\mathbf{t}} = -\hat{\mathbf{s}} \cdot -\hat{\mathbf{t}} = \cos(\psi - \varphi)$, so that together with the law of reflection, Eq.~\eqref{eq:LoR}, the boundary condition $v(\varphi_1) = 0$, and the tangent half-angle relation, we once again arrive at the IVP in Eq.~\eqref{eq:2D_u_ODE}.

As in the cylindrically symmetric case, suppose we have a monotonically increasing specular map $m(\varphi) = m_\mathrm{div}(\varphi)$.
Then, for any $\varphi \in [\varphi_1, \varphi_2]$,
\begin{equation}
	\uint_{\varphi_1}^{\varphi} f(\tilde{\varphi}) \sin(\tilde{\varphi}) \, \text{d}\tilde{\varphi} = \uint_{\psi_1}^{m_\mathrm{div}(\varphi)} g(\tilde{\psi}) \sin(\tilde{\psi}) \, \text{d}\tilde{\psi},
\end{equation}
or, formulated as an IVP
\begin{equation}
	\begin{cases}
		\mathlarger{m_\mathrm{div}'(\varphi) = \frac{f(\varphi)\sin(\varphi)}{g\big(m_\mathrm{div}(\varphi)\big)\sin\big(m_\mathrm{div}(\varphi)\big)}},\label{eq:3D_psiPlusODE}\\
		m_\mathrm{div}(\varphi_1)=\psi_1.
	\end{cases}
\end{equation}
Similarly, for a monotonically decreasing $m(\varphi) = m_\mathrm{conv}(\varphi)$,
\begin{equation}
	\uint_{\varphi_1}^{\varphi} f(\tilde{\varphi}) \sin(\tilde{\psi}) \, \text{d}\tilde{\varphi} = \uint_{m_\mathrm{conv}(\varphi)}^{\psi_2} g(\tilde{\psi}) \sin (\tilde{\psi}) \, \text{d}\tilde{\psi},
\end{equation}
and the IVP becomes
\begin{equation}
	\begin{cases}
		\mathlarger{m_\mathrm{conv}'(\varphi) = - \frac{f(\varphi)\sin(\varphi)}{g\big(m_\mathrm{conv}(\varphi)\big)\sin\big(m_\mathrm{conv}(\varphi)\big)}},\label{eq:3D_psiMinusODE}\\
		m_\mathrm{conv}(\varphi_1)=\psi_2.
	\end{cases}
\end{equation}
With these changes in mind, and the knowledge that the IVP for $v$ remains the same, we can conclude that the design procedure outlined previously may be used without further modifications.

\subsection{Raytracing}
Irrespective of how the reflector is computed, and whether it is cylindrically or rotationally symmetric, a validation method is required.
The natural choice is raytracing, and we have written our own two-dimensional raytracer, which includes the effects of scattering in accordance with our model.
In particular, source, specular and diffuse rays are all collected.
The source rays are generated from the appropriate source distribution using Matlab's \texttt{rand} routine, followed by an intersection computation.
When computing the reflector, we are left with discrete data points, and these form so-called reflector bins.
All rays that fall within a reflector bin will result in the normal of the piecewise-linear interpolation between the data points constituting the reflector being used for the computation of the reflected direction.
The intersection is computed by checking which centre-angle of the reflector bins is closest to the angle of the generated ray using Matlab's \texttt{dsearchn} routine, thus avoiding a relatively expensive intersection computation.
The specular rays are then computed using the law of reflection, Eq.~\eqref{eq:LoR}, whilst the diffuse rays are computed using a rotation matrix with the stochastic parameter $\alpha$, sampled from the probability distribution $p$ --- recall Eq.~\eqref{eq:2D_uu}.
The sampling of $\alpha$ depends on the chosen $p$.
We shall use either a Gau{\ss}ian and Matlab's \texttt{randn} routine or a Lorentzian (Cauchy distribution) and Matlab's \texttt{rand} routine in the appropriate cumulative distribution function.
The ray collection is performed by equidistantly dividing the relevant angular domain ($(-\pi,\pi)$ or $[0,\pi)$, for cylindrically and rotationally symmetric distributions, respectively), thus forming collection bins.
The centres of the collection bins are known, and the appropriate bin for a given ray is then computed via a nearest point search using \texttt{dsearchn}, and the number of rays in the bin is incremented.
After this, the process is repeated until the requested number of rays have been traced though the system.
Finally, the number of rays per collection bin is converted into an intensity by dividing the probability of falling in each bin by the size of the bins and multiplying with the total flux of the source, i.e.,
\begin{equation}\label{eq:2D_I_j}
	I_j = \frac{\text{Pr}(\varphi_{j-1} \leq \varphi < \varphi_j)}{\Delta \varphi} \uint_{\varphi_1}^{\varphi_2} f(\varphi) \, \dd \varphi,
\end{equation}
for the cylindrically reflectors, and for the $j$th bin.
Here, $\text{Pr}(\varphi_{j-1} \leq \varphi < \varphi_j)$ is the number of rays in the $j$th bin divided by the total number of rays traced, and $\Delta \varphi$ is the angular size of the collection bins.
The total flux of the source is given by the integral over $f$.
For the rotationally symmetric reflectors, we have
\begin{equation}\label{eq:3D_I_j}
	I_j = \frac{\text{Pr}(\varphi_{j-1} \leq \varphi < \varphi_j)}{\Delta \varphi} \uint_{\varphi_1}^{\varphi_2} f(\varphi) \sin(\varphi) \, \dd \varphi.
\end{equation}
This is outlined in \cite[p.~34]{Filosa2018}.

\clearpage
\subsubsection{Angle Convention}
\hspace{1pt}\vspace{-11pt}
\begin{wrapfigure}{r}{0.4\textwidth}
	\includegraphics[width=\linewidth]{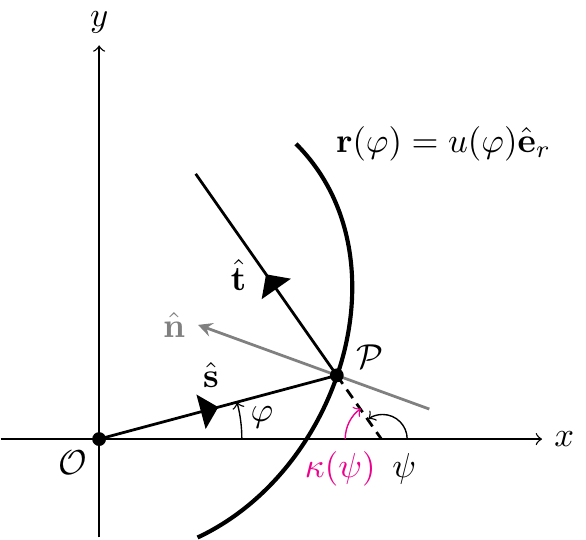}
	\caption{The angle convention in the results section.}
	\label{fig:2D_angConv}
\end{wrapfigure}
We adopt the $(-\pi, \pi)$ angle convention for cylindrically reflectors, so that we can make use of Matlab's \texttt{atan2} function to compute the angles of the rays.
The rotationally symmetric problems will still use the polar $[0, \pi)$ convention.
In addition to this, we shall define our target distributions in terms of $\kappa(\psi)$ and $\kappa(\gamma)$, where
\begin{equation}\label{eq:kappa}
	\kappa(\theta) :=
	\begin{cases}
		- \pi - \theta, & \theta < 0,\\
		\pi - \theta, & \theta > 0.
	\end{cases}
\end{equation}
That is, $\kappa(\psi)$ represents the angle from the negative $x$-axis in the $(-\pi,\pi)$ convention --- see Fig.~\ref{fig:2D_angConv} for $\kappa(\psi)$, and analogously for $\kappa(\gamma)$.
In addition, we note that $\kappa(\theta)$ can be used in the cylindrical angle convention as well, since the polar angle is always positive, and reflection along the $z$-axis results in a change of the azimuthal angle by $\pi$.

\subsubsection{Validation Criteria}
We utilise raytracing as a validation technique, so let us define a criterion to quantify the differences between the raytraced and predicted distributions.
Specifically, we use the root mean square (RMS) defined as follows
\begin{equation}\label{eq:h_RMS}
	\varepsilon(h,h^*) = \sqrt{\frac{1}{N} \sum_{n=1}^N \abs{h_n - h^*_n}^2}\, ,
\end{equation}
where $N$ is the number of collection bins of $h^*$, and the star indicates raytraced distributions.
In most of our examples, we use the \textbf{d}econvolved specular distribution $g_\mathrm{dc}$, obtained by deconvolving Eq.~\eqref{eq:2D_hgammaFinal} or \eqref{eq:rot_hgammaFinal} when designing the reflectors.
In this case, $h$ and $h^*$ will be $h_\mathrm{rc}$ and $h_\mathrm{rc}^*$, respectively, where the `rc' subscript signifies ``\textbf{r}econvolution'', that is $h_\mathrm{rc} := g_\mathrm{dc} * p$.
For the cases where $h$ itself is used, it was averaged over the bins using Eq.~\eqref{eq:2D_I_j} or \eqref{eq:3D_I_j} with $f$ and $\varphi$ replaced by $h$ and $\gamma$, depending on the symmetry of the problem.

\vfill
\section{Results}\label{sec:results}
\noindent This section presents three sample problems: two exhibiting cylindrical symmetry and one with rotational symmetry.
To verify our model, we prescribe the specular target distribution $g$ exactly, and construct the diffuse target distribution $h$ by convolving $g$ and the chosen scattering PDF $p$.
We then compute the deconvolved specular distribution $g_\mathrm{dc}$ and design the reflectors using it.
Finally, we raytrace the system and compare the result to our prediction.
In the rotationally symmetric example, we no longer know the exact $g$, but rather we prescribe $h$ exactly.
This is more similar to how we envision an optical designer working with our model.

\clearpage
\subsection{Example \#1: Smooth Target Distribution}
The specular problem consists of a homogeneous source $f$ being transformed into two partly overlapping Gau{\ss}ians $g$.
As for the choice of $p$, we opted for a  Gau{\ss}ian centred around $\alpha = 0$, with standard deviation $\sigma = 10^\circ$.
This is supposed to represent relatively minor scattering when compared to, e.g., Lambert's cosine law, whilst still being a significant deviation from a specular reflector.
The diffuse distribution $h = p*g$.
Worth noting is that Gau{\ss}ians do not have finite support, meaning we need to truncate the nonzero values outside of $[\alpha_1,\alpha_2]$ when performing the (de-)convolution.
We re-normalised $p$ after truncation to ensure that $\uint_{\alpha_1}^{\alpha_2} p(\alpha) \, \dd \alpha = 1$.
The problem is summarised in the box below, where
\begin{equation}
	\mathcal{N}(\theta; \mu,\sigma) = \frac{1}{\sigma\sqrt{2\pi}} \exp\Bigg({-}\frac{1}{2} \bigg(\frac{\theta - \mu}{\sigma}\bigg)^2\Bigg),
\end{equation}
represents the Gau{\ss}ian, centred at $\mu$ with standard deviation $\sigma$.
The value of $\varphi_2$ was chosen such that energy is conserved up to $10^{-3}$.
\begin{mdframed}
	\textbf{Example \#1: Smooth Target Distribution}
\begin{align*}
	&\text{$\varphi$-range:} 					&[\varphi_1, \varphi_2] 	&= [-\pi/4,29\pi/75]\\
	&\text{$\psi$-range:} 						&[\psi_1, \psi_2] 			&: \text{ see text}\\
	&\text{$\alpha$-range:} 					&[\alpha_1, \alpha_2] 		&= [-\pi,\pi]\\
	&\text{Source distribution:} 				&f(\varphi)					&= \begin{cases} 1, \ \varphi \in [\varphi_1, \varphi_2]\\ 0, \text{ otherwise} \end{cases}\\
	&\text{Specular target distribution:} 		&g(\psi) 					&= \mathcal{N}(\psi; -\pi/8,10^{\circ}) + \mathcal{N}(\psi;\pi/12,12^{\circ})\\
	&\text{Surface scattering function:} 		&p(\alpha) 					&= \mathcal{N}(\alpha; 0,10^{\circ})\\
	&\text{Diffuse distribution prediction:}	&h(\gamma) 					&= (p*g)(\gamma)\\
	&\text{Boundary condition:} 				&u(\varphi_1)				&= 1
\end{align*}
\end{mdframed}

\begin{figure}[htbp]
	\includegraphics[width=0.5\linewidth]{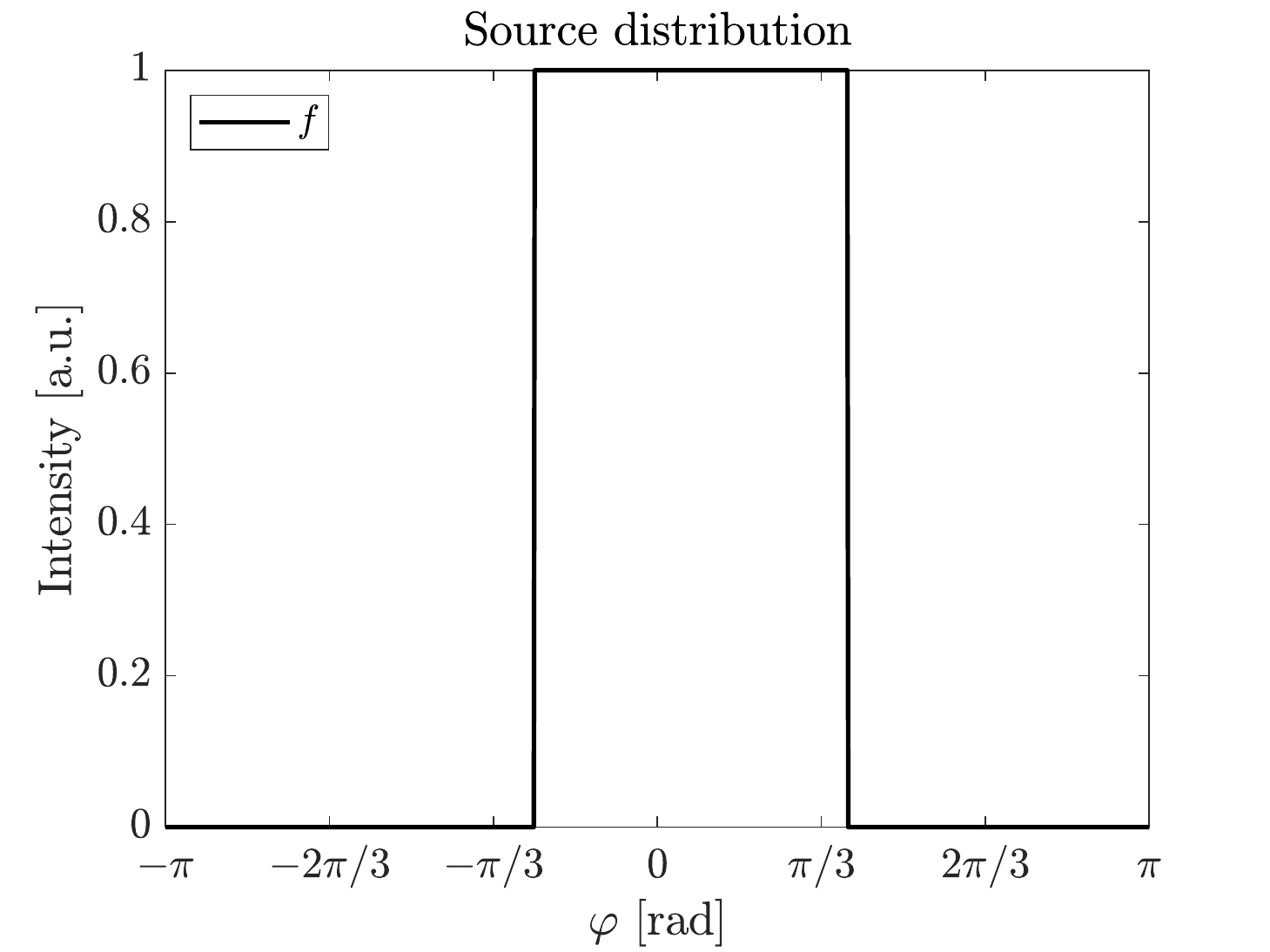}%
	\includegraphics[width=0.5\linewidth]{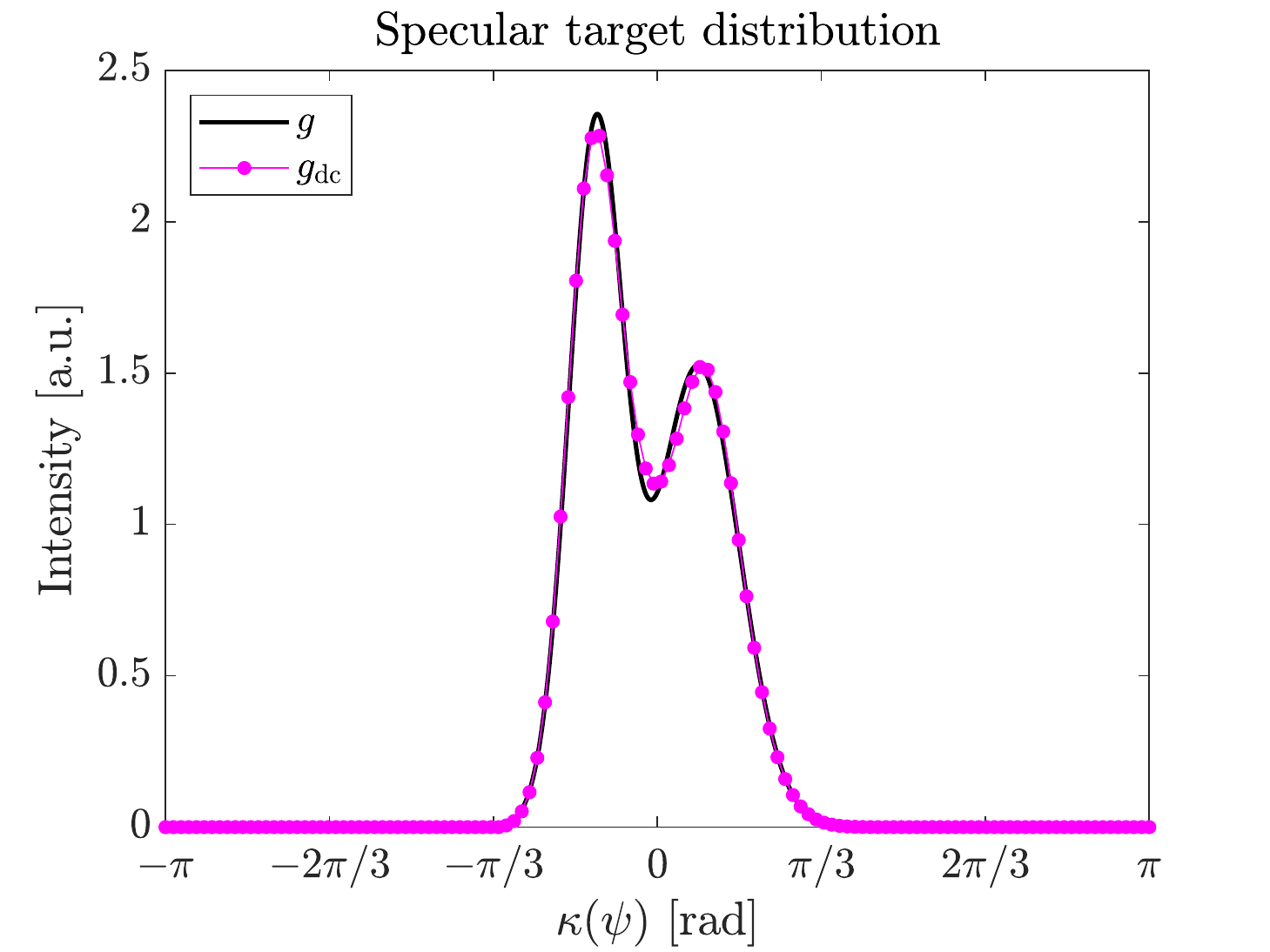}\\[5pt]
	\includegraphics[width=0.5\linewidth]{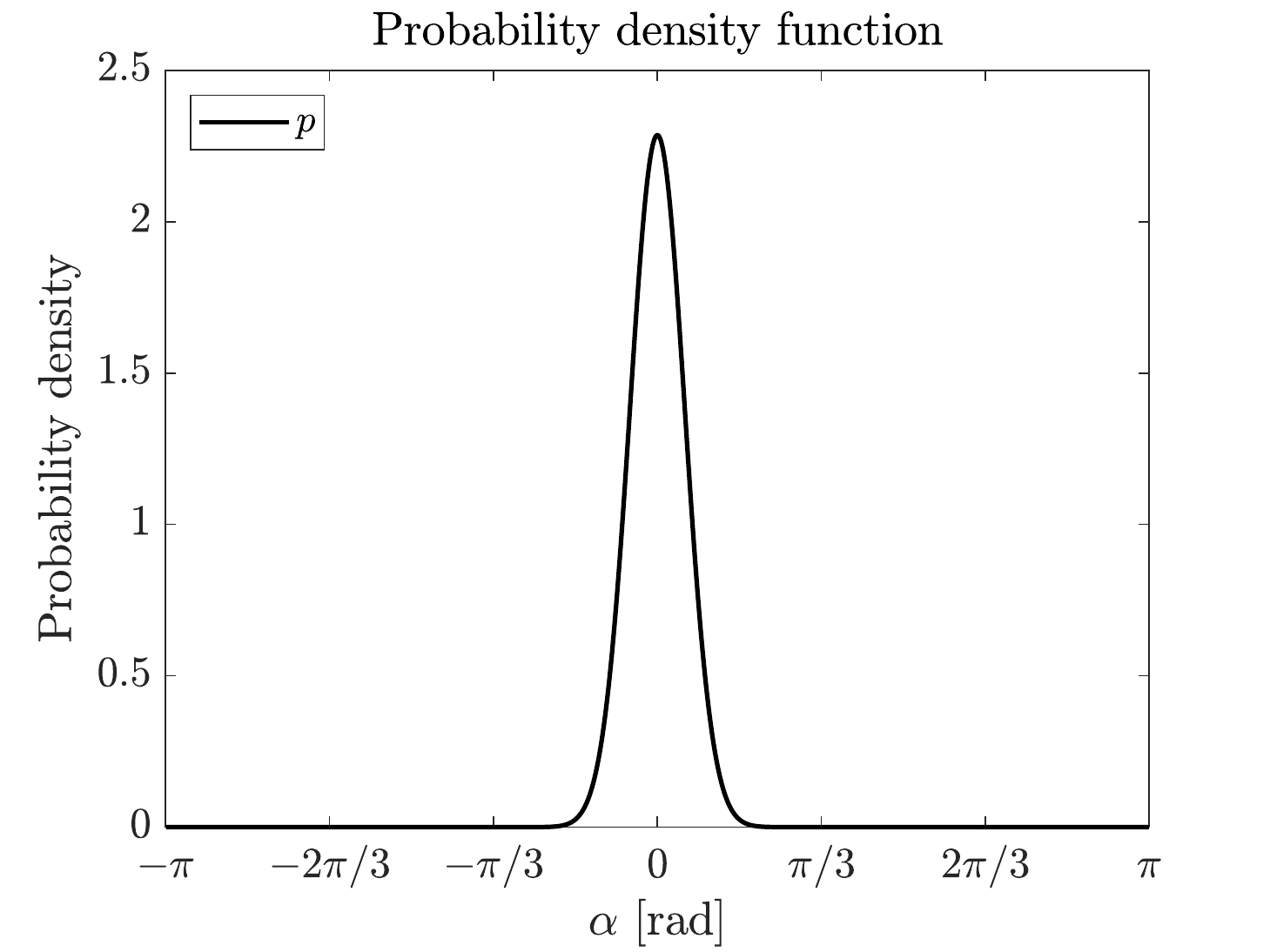}%
	\includegraphics[width=0.5\linewidth]{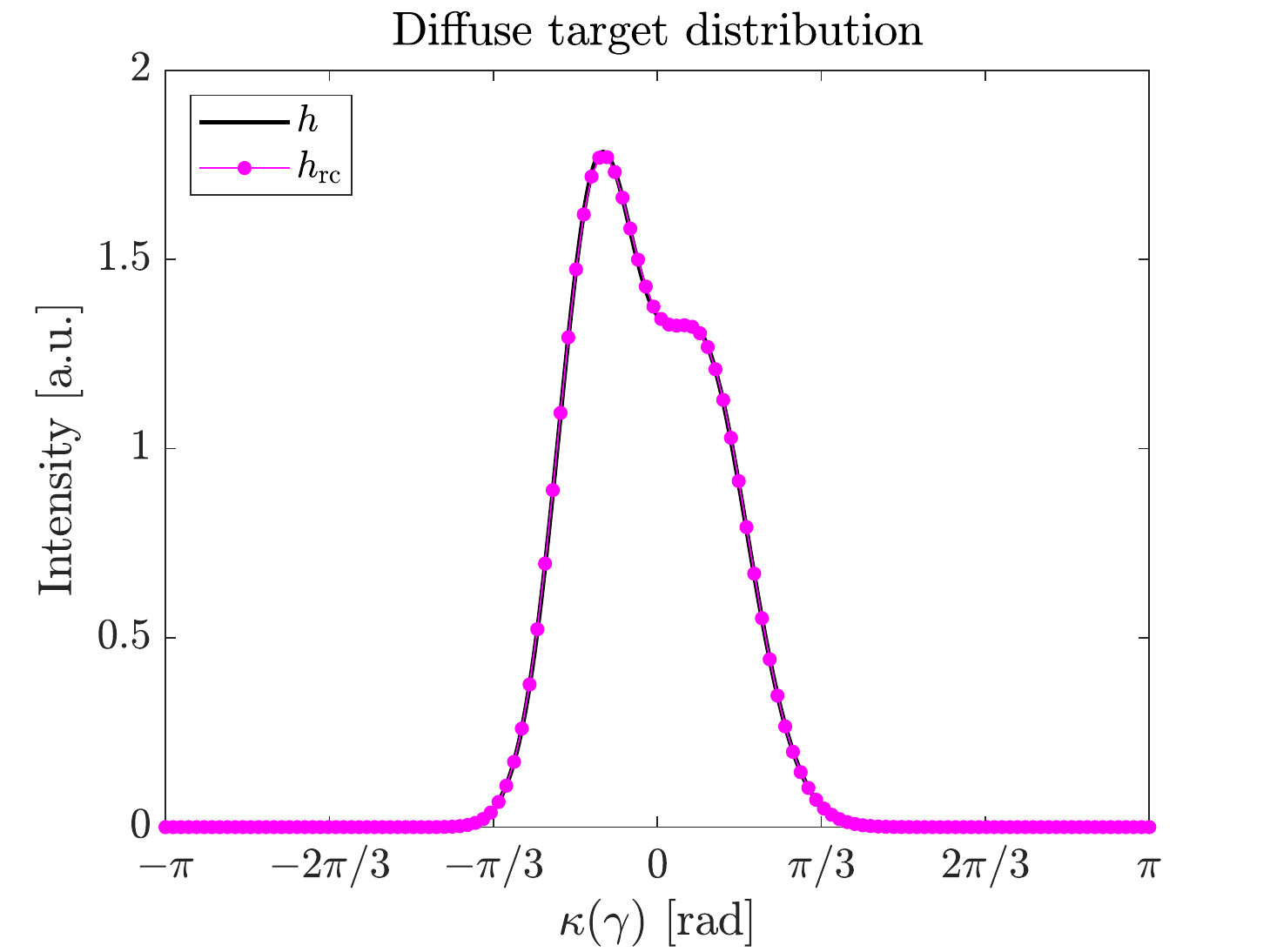}
	\captionsetup{width=\linewidth}
	\caption{Distributions in Example \#1; 128 sample points.}
	\label{fig:example_1-1}
\end{figure}

\noindent Since we are interested in validating the whole proposed solution method, the first step is to find the specular target distribution $g_\mathrm{dc}$ by deconvolving $g$ from $h$.
We shall utilise Richardson-Lucy deconvolution, and in particular Matlab's \texttt{deconvlucy} routine.
This iterative ratio method has numerous benefits compared to direct methods, most crucial for our purposes being guaranteed positivity of the solution.
The functions $f$, $g$, $g_\mathrm{dc}$, $p$, $h$, and $h_\mathrm{rc}$ are shown in Fig.~\ref{fig:example_1-1}, for \texttt{deconvlucy}'s default settings of 10 iterations with 128 sampling points.
Clearly, the recovered $g_\mathrm{dc}$ resembles the original $g$ relatively well, and the reconvolved $h_\mathrm{rc} = g_\mathrm{dc}*p$ is nearly identical to $h$.

\begin{figure}[htbp]
	\centering
	\begin{minipage}{0.5\linewidth}
		\centering
		\includegraphics[width=\linewidth]{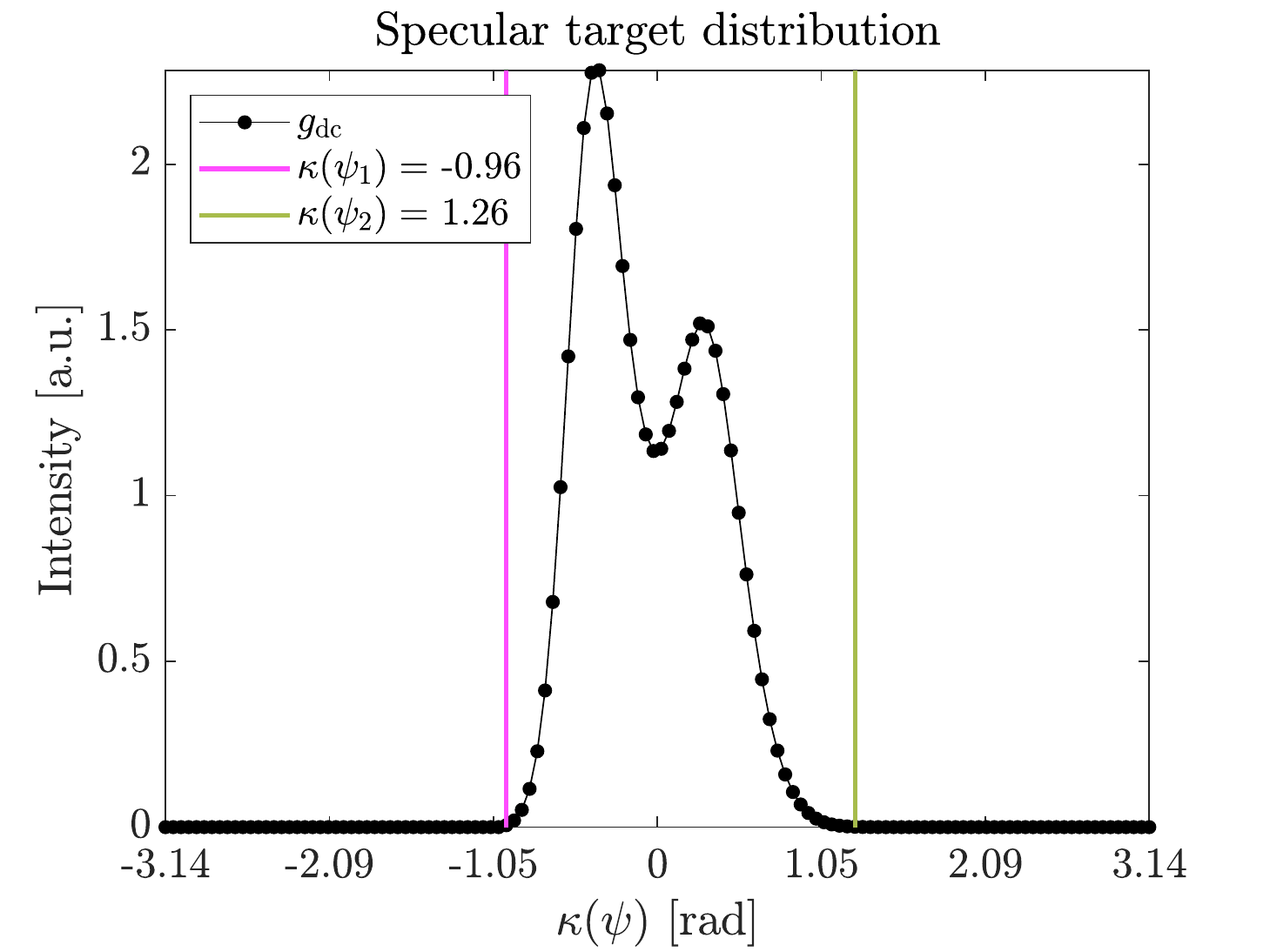}
		\captionsetup{width=0.95\linewidth}
		\caption{The $\kappa(\psi)$-boundaries used as the support of $g_\mathrm{dc}$ in Example \#1.\\\hspace{0pt}}
		\label{fig:example_1-g_deconv-psi_limits}
	\end{minipage}\hfill
	\begin{minipage}{0.5\linewidth}
		\centering
		\includegraphics[width=\linewidth]{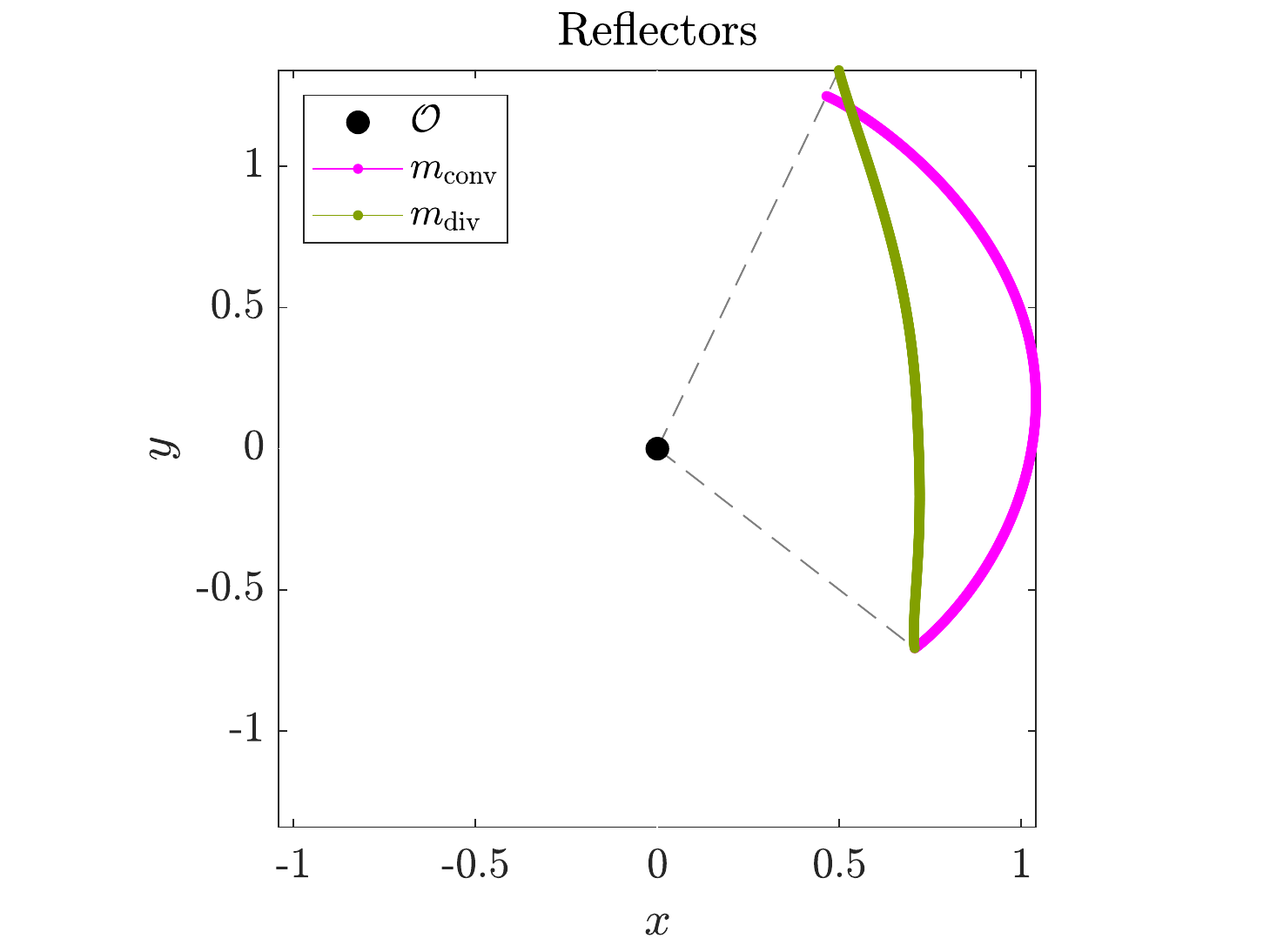}
		\captionsetup{width=0.95\linewidth}
		\caption{Reflectors associated with Example \#1 using $g_\mathrm{dc}$ from Fig.~\ref{fig:example_1-1}; 1024 sample points.\\\hspace{1pt}}
		\label{fig:example_1-reflectors-g_deconv}
	\end{minipage}
\end{figure}

\begin{figure}[htbp]
	\includegraphics[width=0.5\linewidth]{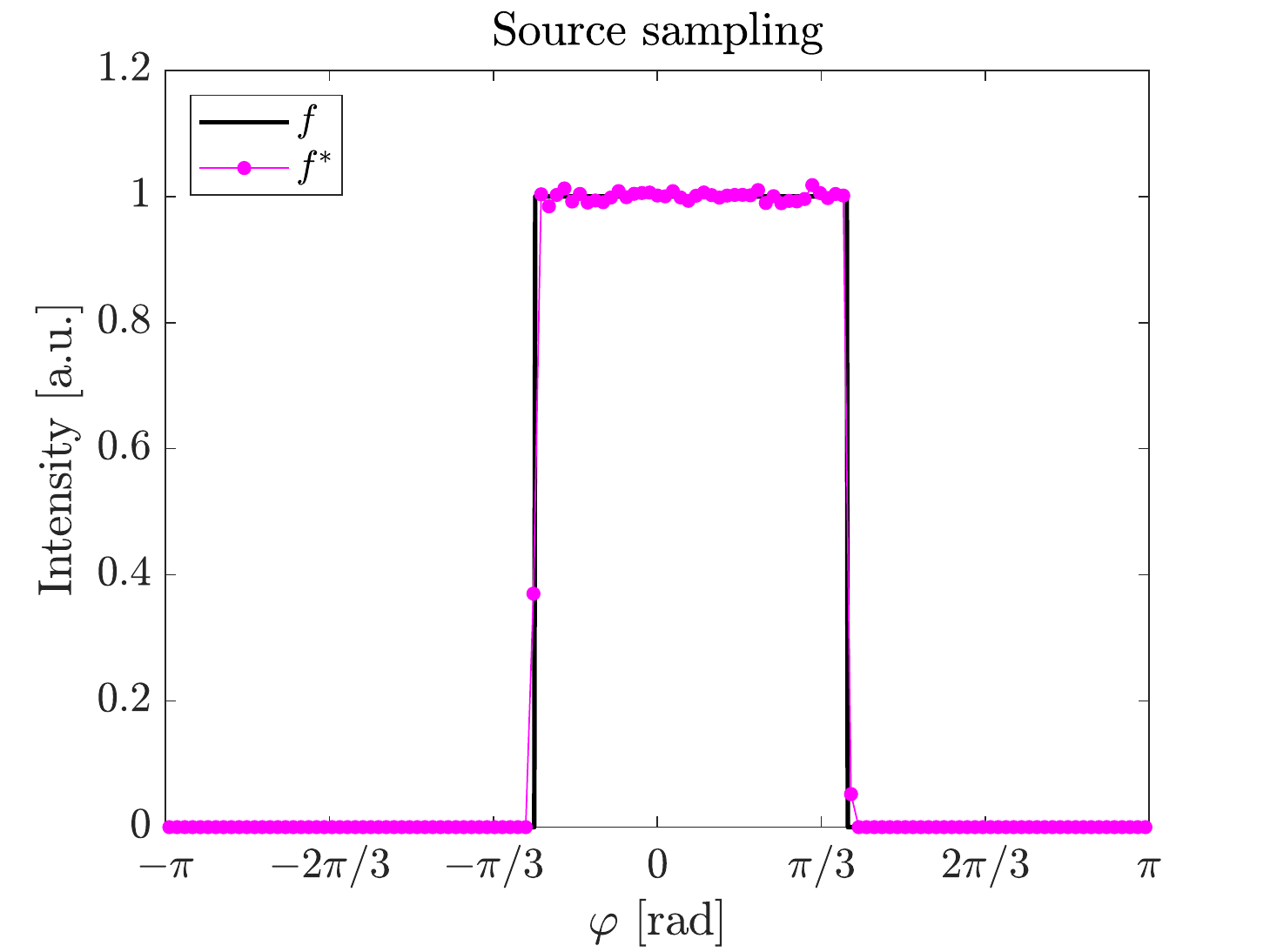}%
	\includegraphics[width=0.5\linewidth]{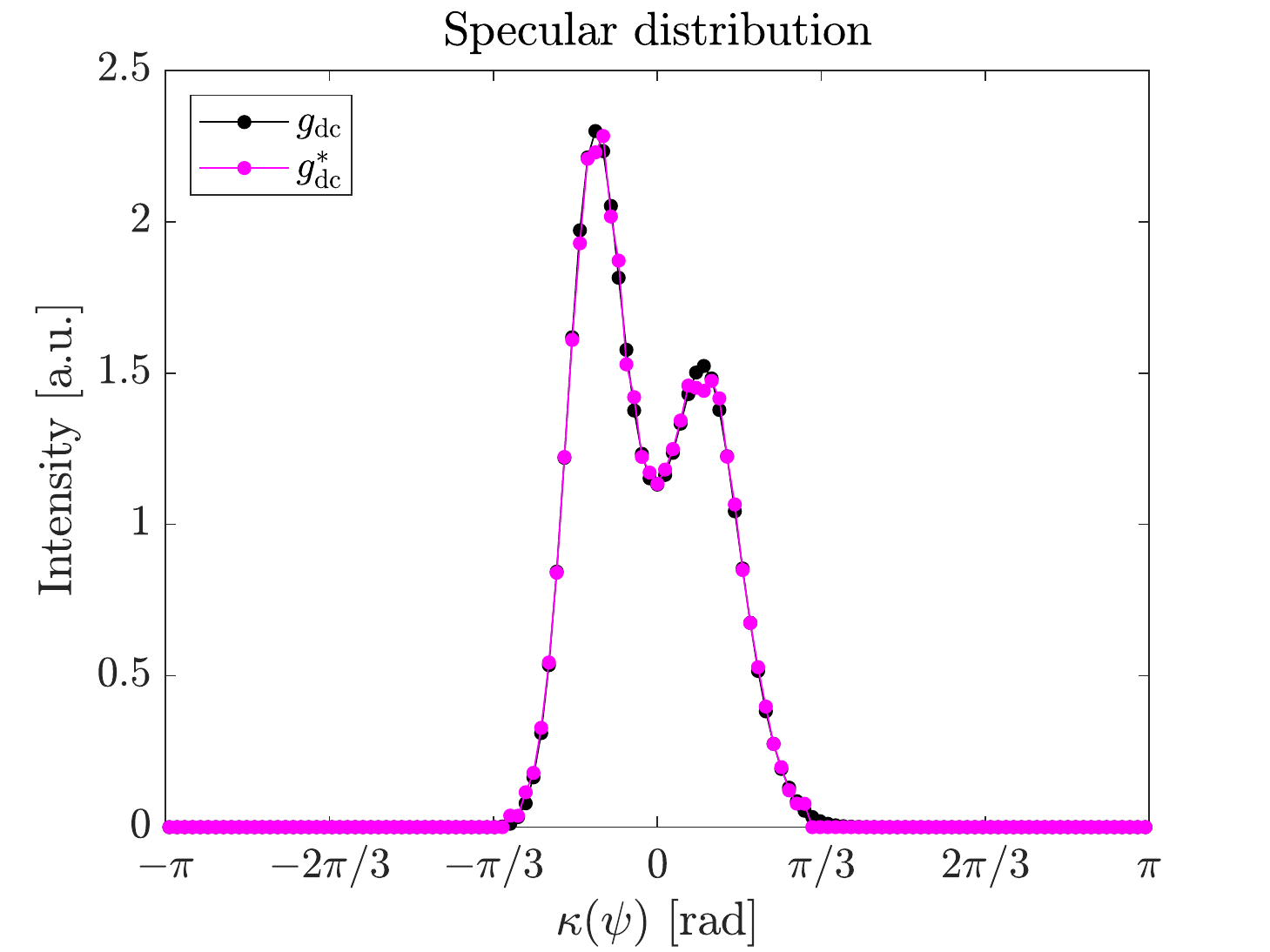}\\[5pt]
	\includegraphics[width=0.5\linewidth]{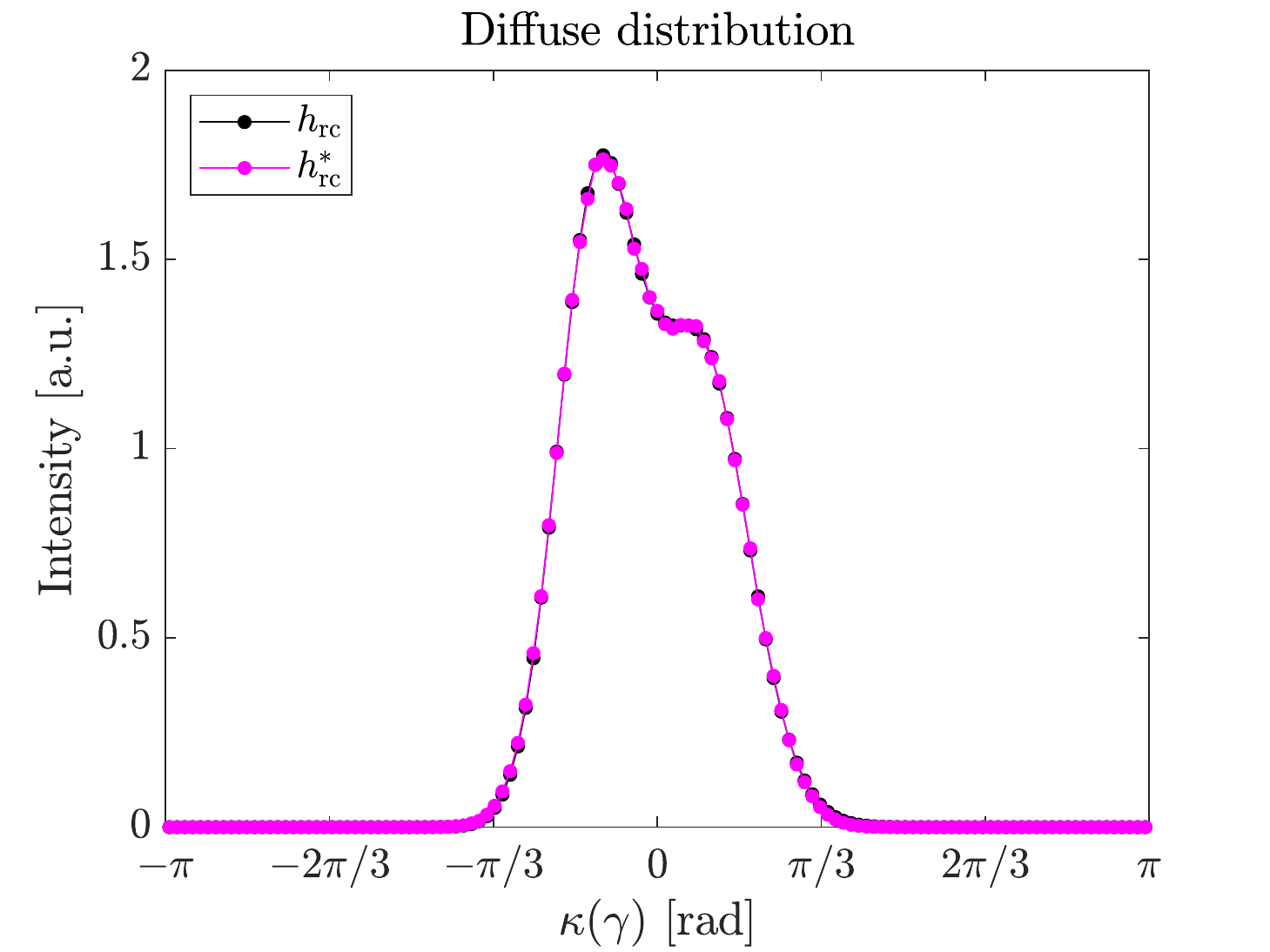}%
	\includegraphics[width=0.5\linewidth]{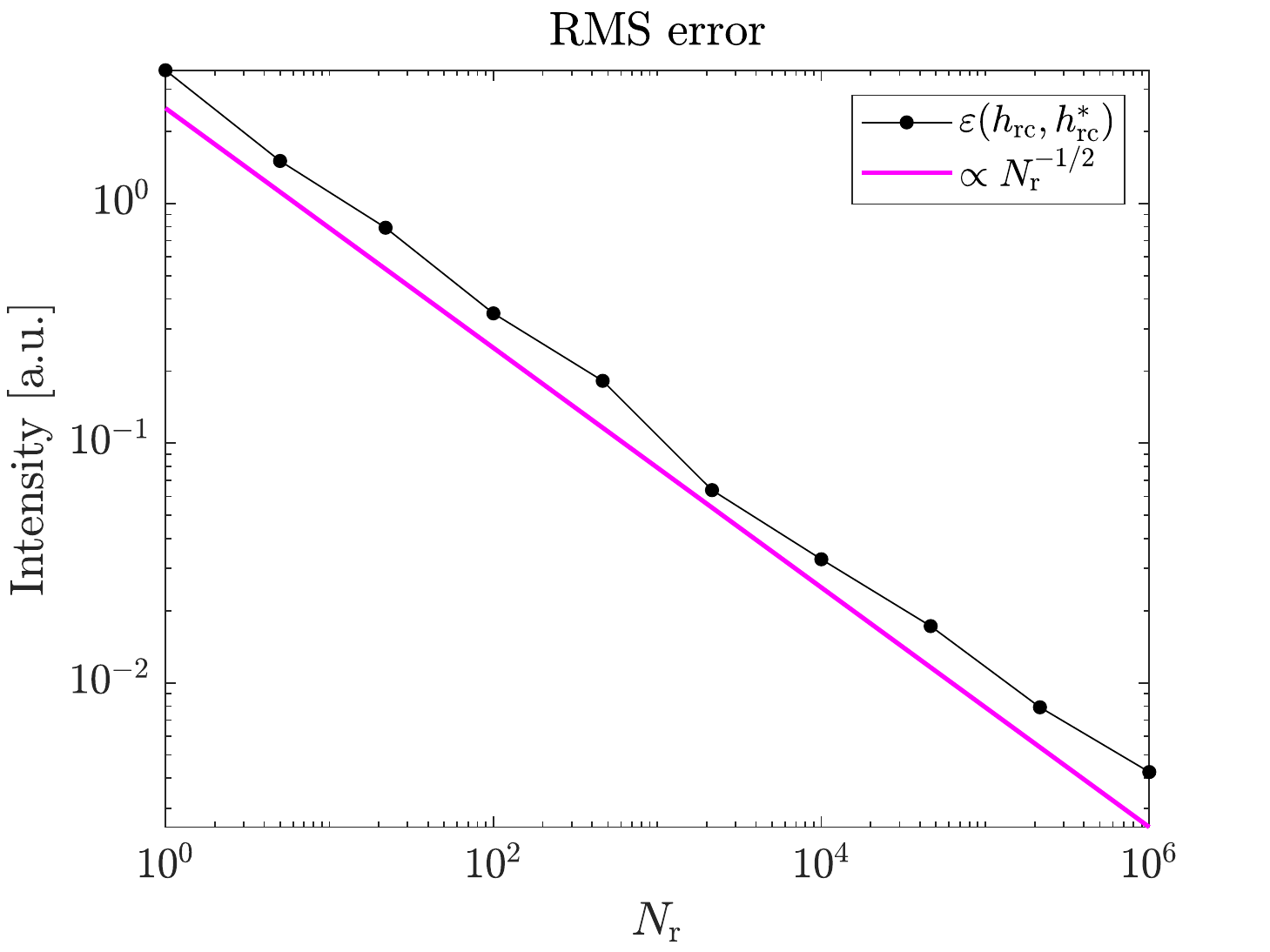}
	\captionsetup{width=\linewidth}
	\caption{Raytraced distributions; Example \#1 with $g_\mathrm{dc}$ and $m_\mathrm{conv}$; $10^6$ rays.}
	\label{fig:example_1-RT-g_deconv}
\end{figure}

The next step is to design the reflectors.
In order to use the procedure outlined in Sec.~\ref{sec:reflectorDesign}, we need the limits $\psi_1$ and $\psi_2$.
Recall that these should represent the support of $g$ (or, rather, $g_\mathrm{dc}$ in this case).
In this example, the limits are ambiguous due to the Gau{\ss}ians.
We computed the limits by fixing a threshold $\eta = 0.001$ and locating the two extrema of $\kappa(\psi)$ where $g_\mathrm{dc}(\psi) = \eta$, using piecewise-linear interpolation between the data points of $g_\mathrm{dc}$.
The limits are shown in Fig.~\ref{fig:example_1-g_deconv-psi_limits}, and the reflectors are shown in Fig.~\ref{fig:example_1-reflectors-g_deconv}.
In this case, we do not know the exact solutions, so we shall not discuss the reflectors further for this example.
We note that one could renormalise $g_\mathrm{dc}$ to ensure more accurate energy conservation, but this has not been done in the data shown.
We have used 1024 sample points for the reflectors in an attempt to minimise discretisation errors due to the reflectors when validating the $m_\mathrm{conv}$ reflector using raytracing.
The sample points are equidistant in the $[\varphi_1, \varphi_2]$-range.
The raytraced distributions and the RMS error from Eq.~\eqref{eq:h_RMS} is shown in Fig.~\ref{fig:example_1-RT-g_deconv}, where we see that the source sampling is appropriate, and the resulting distributions are well predicted by our model.
In addition, the convergence shows the expected $N_\mathrm{r}^{-1/2}$ behaviour of Monte Carlo raytracing \cite[p.~9]{Filosa2018}, where $N_\mathrm{r}$ is the number of rays traced through the system.

There are a couple of regions where the raytraced distributions deviate from our predictions.
Specifically, $f^*$ near $\varphi_1 = - \pi/4$ and $\varphi_2 = 29\pi/75$, $g_\mathrm{dc}^*$ near $\kappa(\psi_2) \approx 1.26$ and near both peaks of the Gau{\ss}ians.
The discrepancies in $f^*$ are due to the binning not aligning perfectly with the support of $f$, such that part of a collection bin may cross the $\varphi_1$- and $\varphi_2$-boundaries.
The discrepancies of $g^*_\mathrm{dc}$ are presumably partly due to energy not being perfectly conserved, and partly from the discretisation of the reflector surface, in addition to the aforementioned binning issue.
Additionally, the very peaks of the Gau{\ss}ians are only one or two data points wide, and achieving that level of precision is no easy feat, using a numerical scheme.
Keeping all of these factors in mind, the results are promising, and from the RMS error plot, we see that increasing the number of rays is likely to improve the result further.

\clearpage
\subsection{Example \#2: Block Function as Target Distribution}
\noindent We now move on to our second example, which at first appears much simpler, but will prove to be quite a challenge for our numerical scheme.
The specular problem consists of homogeneous illumination of a circular disk within $[\psi_1,\psi_2]$ and $f$ is homogeneous on $[\varphi_1,\varphi_2]$.
The scattering PDF $p$ is still a Gau{\ss}ian, this time with a standard deviation of $\sigma = 5^\circ$.
We employ a similar approach to the first example, i.e., prescribe $g$, compute $h = g * p$ and attempt to recover $g$ via deconvolution, then validate the reflectors we design using $g_\mathrm{dc}$ with raytracing.
The example is outlined in the box below.
We wish to highlight that the density $\rho$ is shown explicitly for this example in Fig.~\ref{fig:psigamma}.

\begin{mdframed}
	\textbf{Example \#2: Block Function as Target Distribution}
\begin{align*}
	&\text{$\varphi$-range:} 					&[\varphi_1, \varphi_2] 				&= [-\pi/4,\pi/4]\\
	&\text{$\psi$-range:} 						&[\kappa(\psi_1), \kappa(\psi_2)] 		&= [-\pi/4,\pi/4]\\
	&\text{$\alpha$-range:} 					&[\alpha_1, \alpha_2] 					&= [-\pi,\pi]\\
	&\text{Source distribution:} 				&f(\varphi)								&= \begin{cases} 1, \ \varphi \in [\varphi_1, \varphi_2]\\ 0, \text{ otherwise} \end{cases}\\
	&\text{Specular target distribution:} 		&g(\psi) 								&= \begin{cases} 1, \ \psi \in [\psi_1, \psi_2]\\ 0, \text{ otherwise} \end{cases}\\
	&\text{Surface scattering function:} 		&p(\alpha) 								&= \mathcal{N}(\alpha; 0,5^{\circ})\\
	&\text{Diffuse distribution prediction:} 	&h(\gamma) 								&= (p*g)(\gamma)\\
	&\text{Boundary condition:} 				&u(\varphi_1)							&= 1
\end{align*}
\end{mdframed}

\begin{figure}[htbp]
	\includegraphics[width=0.5\linewidth]{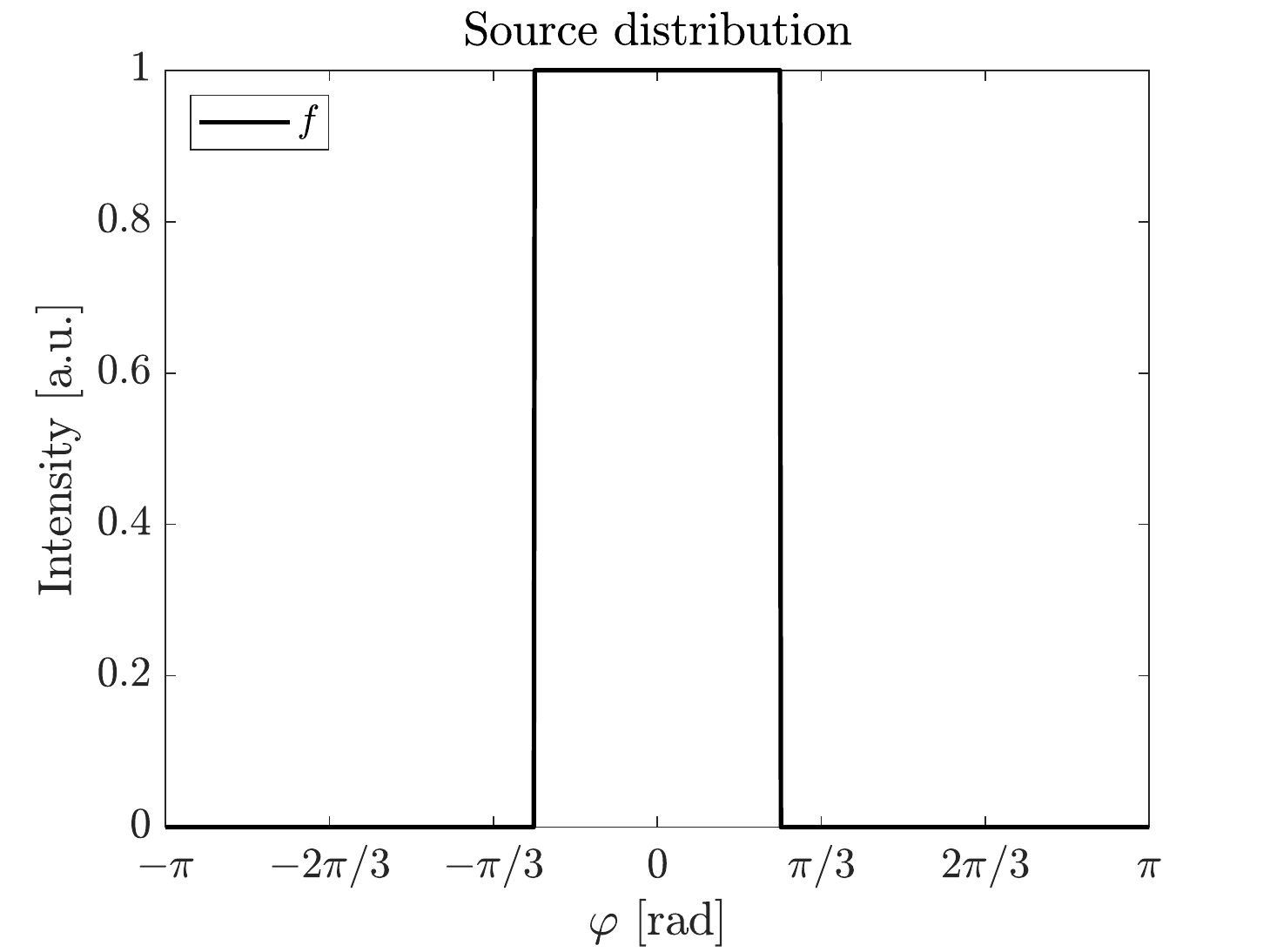}%
	\includegraphics[width=0.5\linewidth]{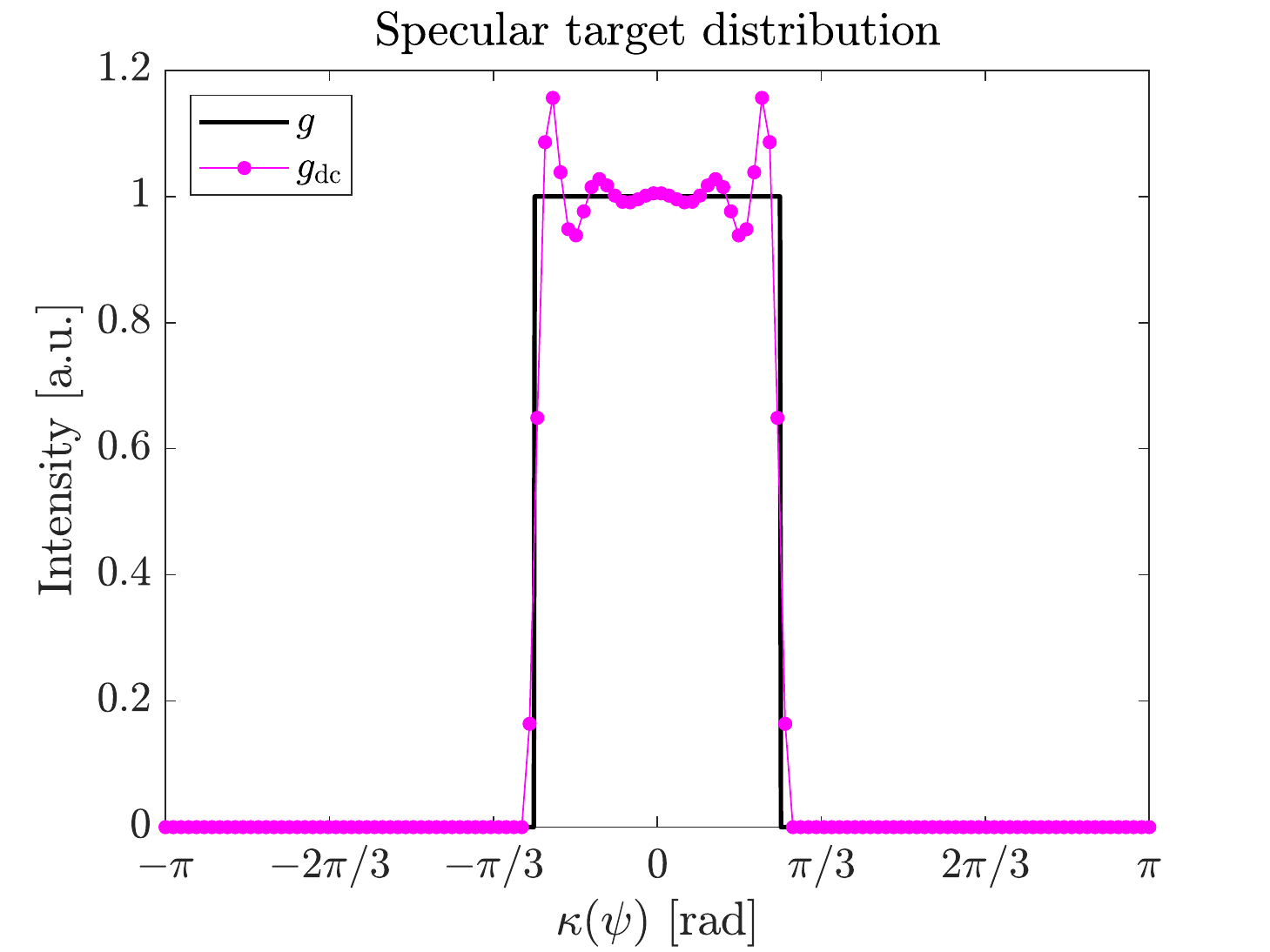}\\[5pt]
	\includegraphics[width=0.5\linewidth]{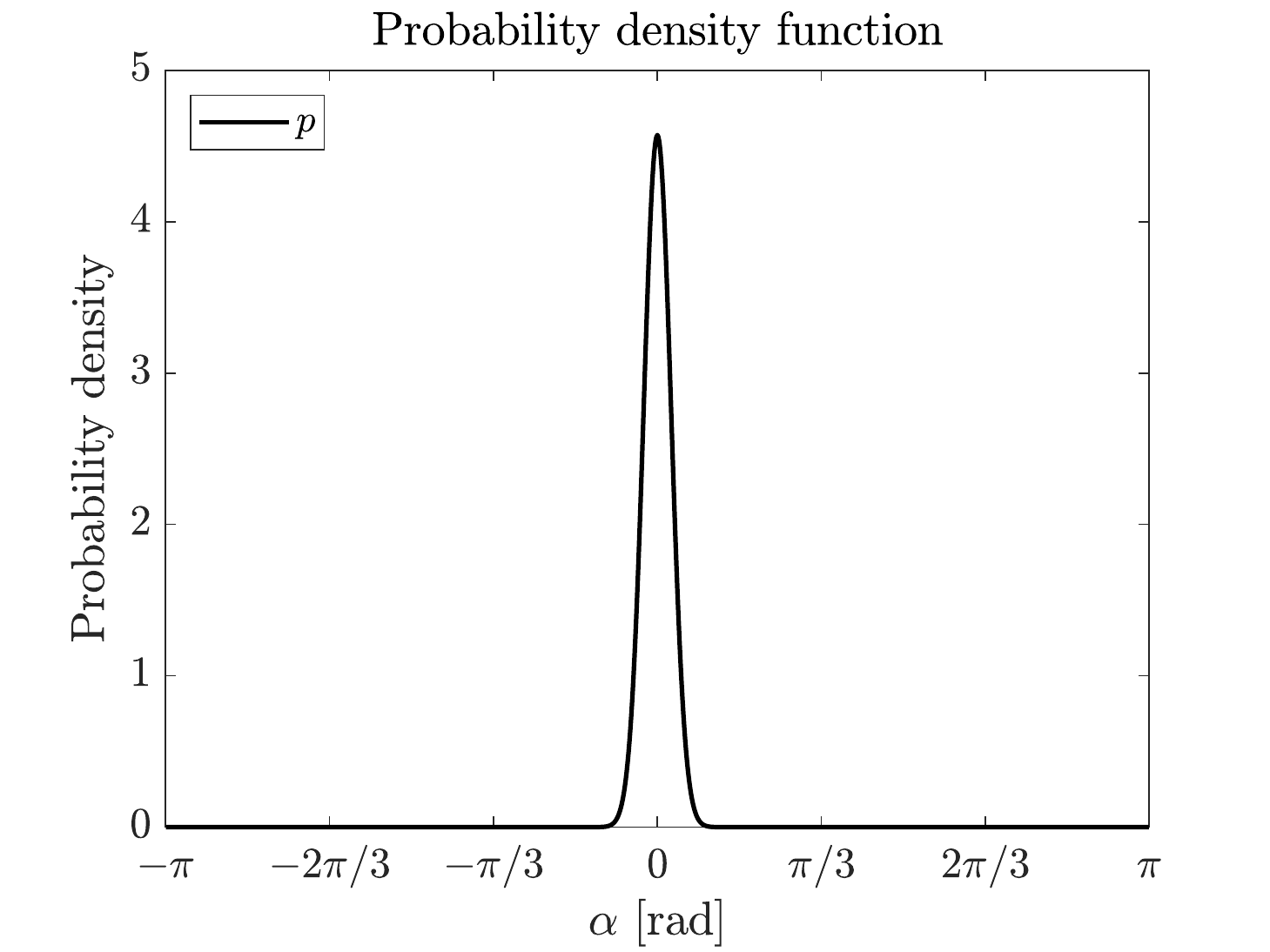}%
	\includegraphics[width=0.5\linewidth]{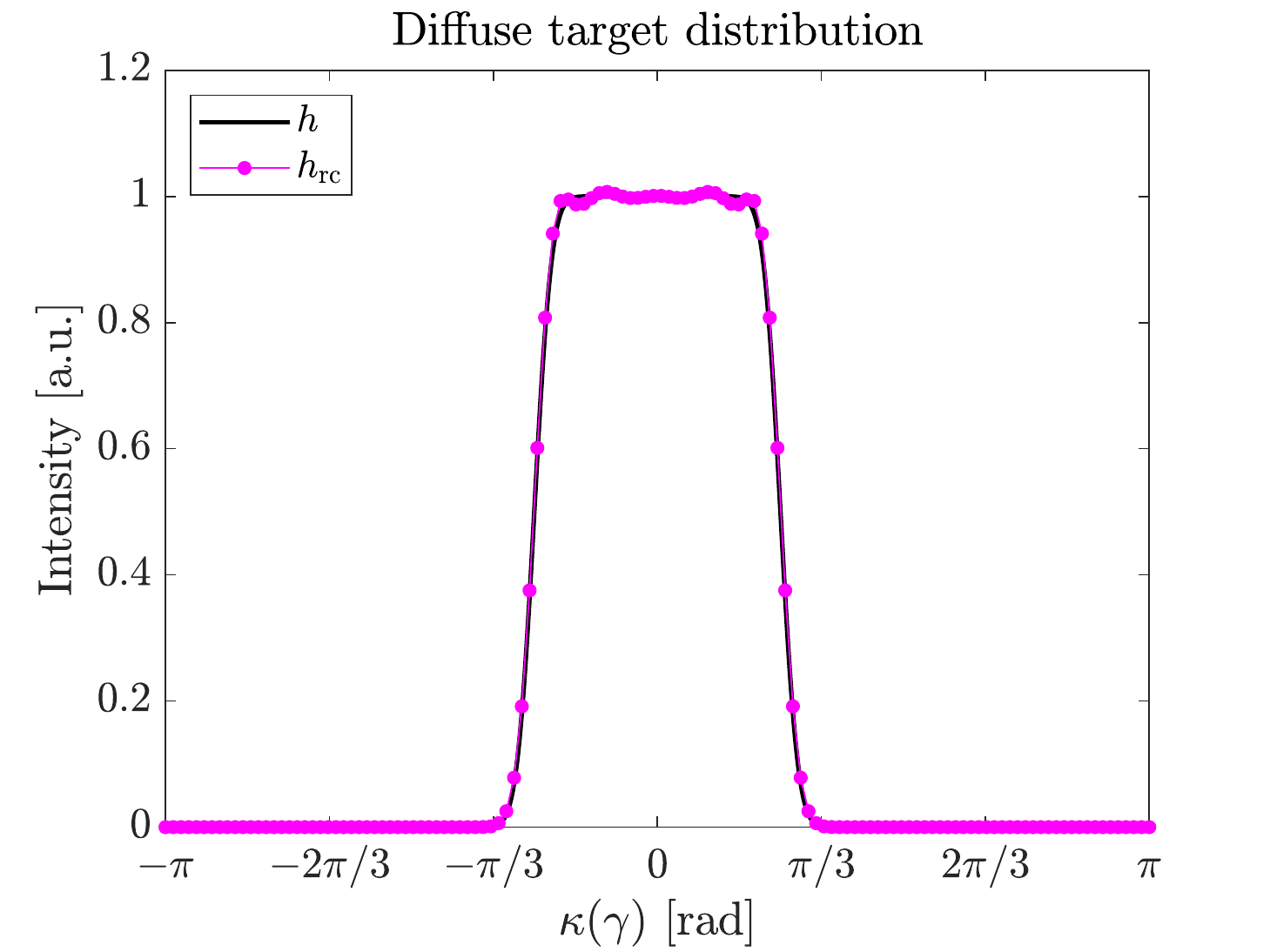}
	\captionsetup{width=\linewidth}
	\caption{Initial distributions in Example \#2; 128 sample points.}
	\label{fig:example_2-1}
\end{figure}

\begin{figure}[htbp]
	\centering
	\begin{minipage}{0.5\linewidth}
		\centering
		\includegraphics[width=\linewidth]{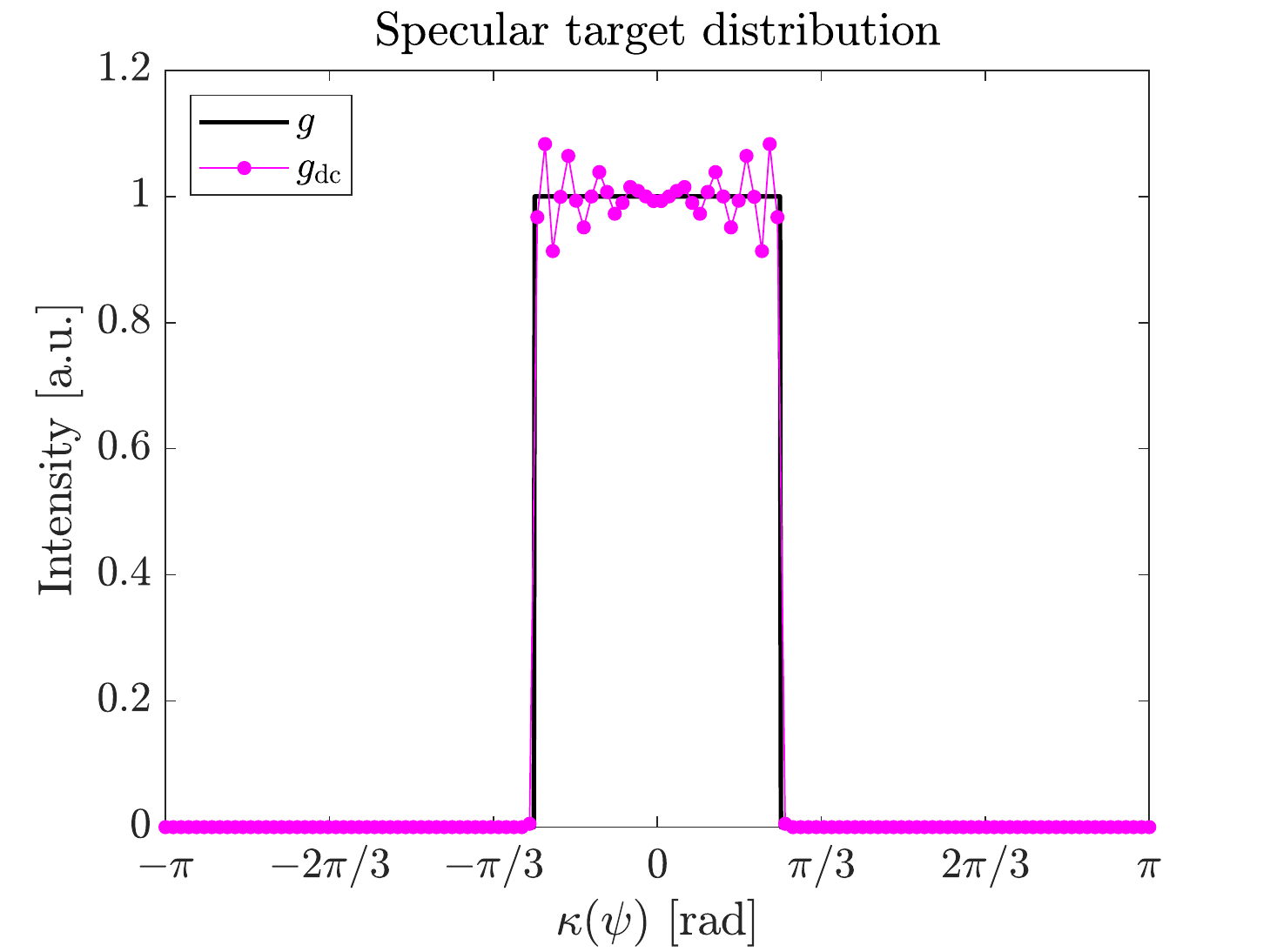}
		\captionsetup{width=0.95\linewidth}
		\caption{Example \#2 with 100 \texttt{deconvlucy} iterations.}
		\label{fig:example_2-g_deconv_100}
	\end{minipage}\hfill
	\begin{minipage}{0.5\linewidth}
		\centering
		\includegraphics[width=\linewidth]{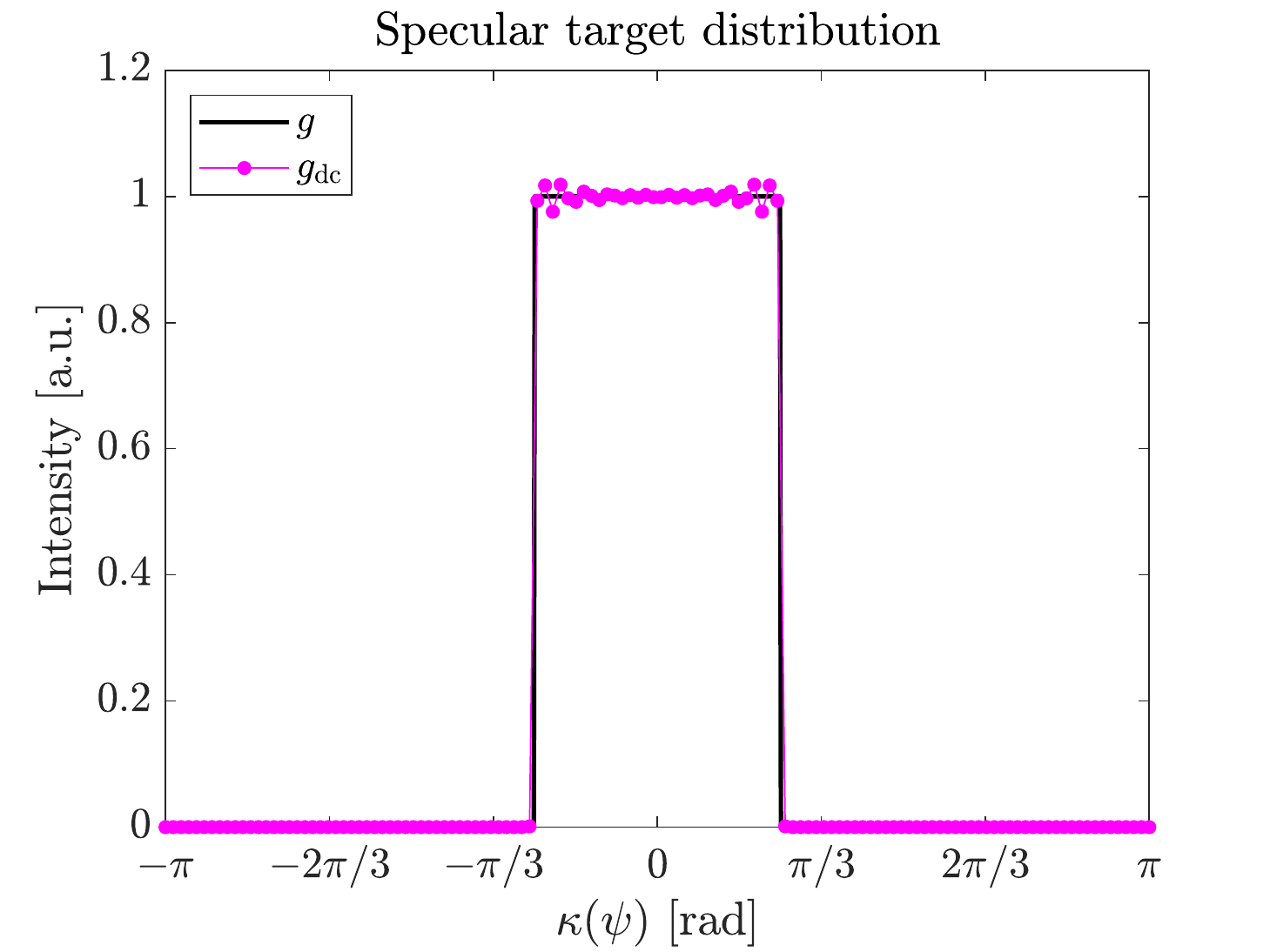}
		\captionsetup{width=0.95\linewidth}
		\caption{Example \#2 with $10^4$ \texttt{deconvlucy} iterations.}
		\label{fig:example_2-g_deconv_10k}
	\end{minipage}
\end{figure}

\noindent Using the default settings of \texttt{deconvlucy} (10 iterations) yields $g_\mathrm{dc}$ in Fig.~\ref{fig:example_2-1}, where we immediately see that it deviates significantly from the original $g$.
Readers who are familiar with signal theory are likely not surprised by this, as representing a block function in Fourier space requires an infinite number of frequencies.
Let us attempt to increase the number of deconvolution iterations by an order of magnitude --- see Fig.~\ref{fig:example_2-g_deconv_100}.
This shows a slight improvement, but we are still quite far from the original $g$.
As such, let us further increase the number of iterations by two orders of magnitude to get the result in Fig.~\ref{fig:example_2-g_deconv_10k}, which is certainly a lot closer to the original $g$.
The RMS error $\varepsilon(g,g_\mathrm{dc})$, recall Eq.~\eqref{eq:h_RMS}, decreased from $0.055$ to $0.020$ and $0.005$, for 10, $10^2$ and $10^4$ iterations, respectively.
In a real problem, this metric would not be available to us, so we would have to rely on the RMS error $\varepsilon(h,h_\mathrm{rc})$, which decreased from $10^{-3}$ to $10^{-4}$ and $10^{-6}$.
Based solely on $\varepsilon(h,h_\mathrm{rc})$, it is not unreasonable that one might design the reflector using the first $g_\mathrm{dc}$, so we shall include it as a worst-case scenario, as well as the best $g_\mathrm{dc}$, in the sense that it has the lowest RMS error.

\begin{figure}[htbp]
	\centering
	\begin{minipage}{0.5\linewidth}
		\centering
		\includegraphics[width=\linewidth]{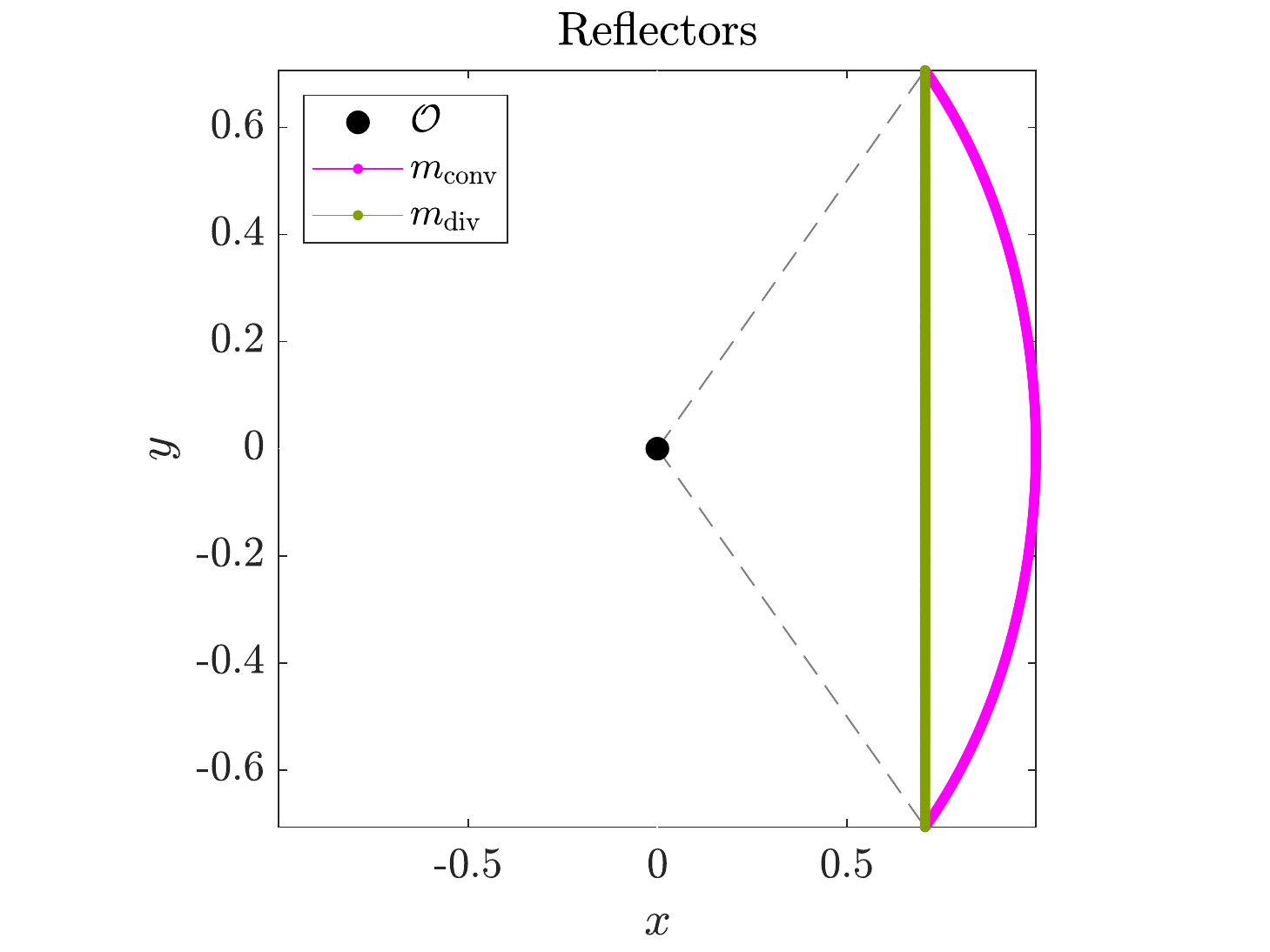}
		\captionsetup{width=0.95\linewidth}
		\caption{Example \#2; reflectors designed using $g$ in Fig.~\ref{fig:example_2-1}; 1024 sample points.}
		\label{fig:example_2-refls-g}
	\end{minipage}\hfill
	\begin{minipage}{0.5\linewidth}
		\centering
		\includegraphics[width=\linewidth]{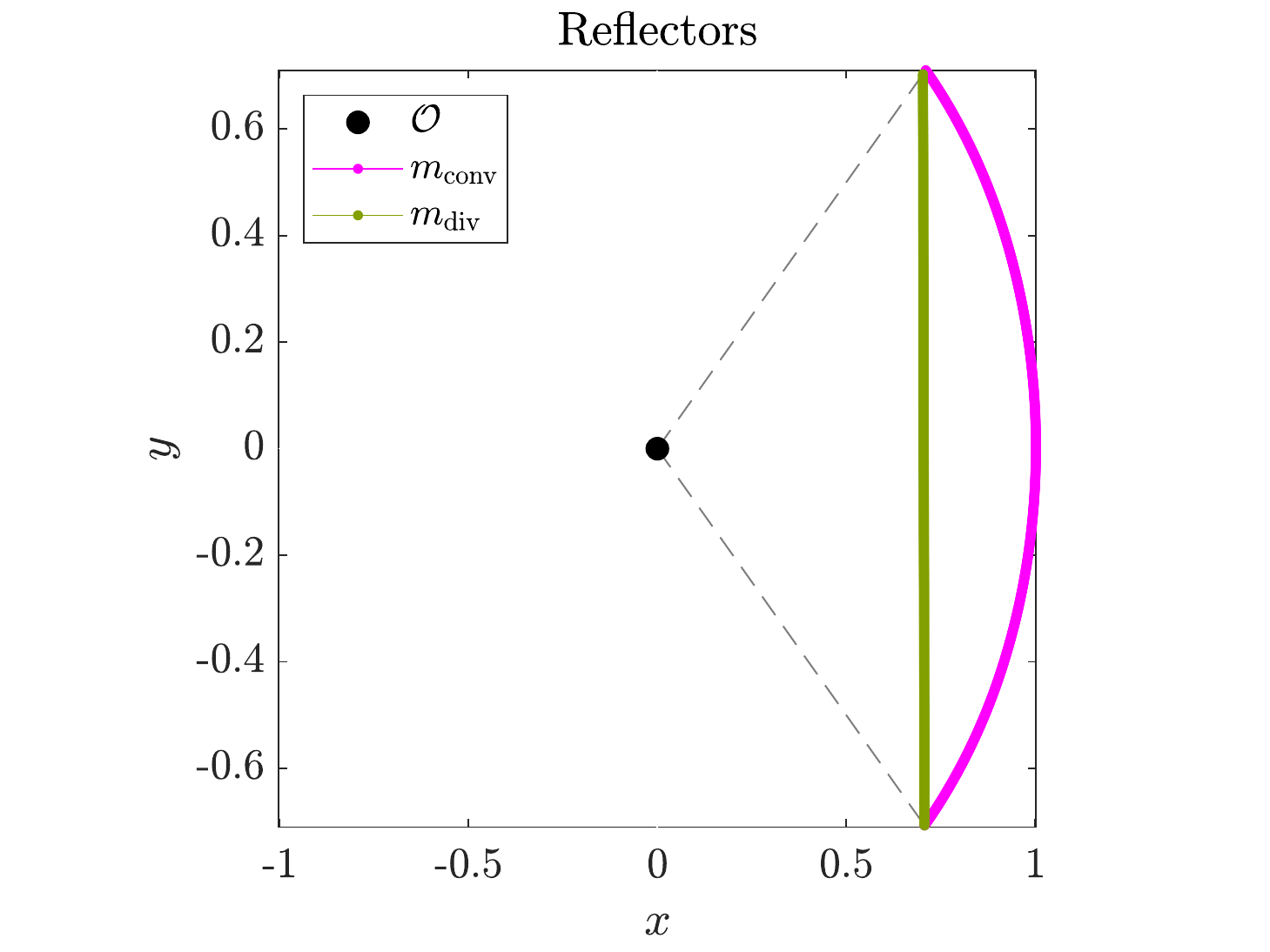}
		\captionsetup{width=0.95\linewidth}
		\caption{Example \#2; reflectors designed using the default $g_\mathrm{dc}$ in Fig.~\ref{fig:example_2-1}; 1024 sample points.}
		\label{fig:example_2-refls-g_deconv_10}
	\end{minipage}
\end{figure}

\begin{figure}[H]
	\includegraphics[width=0.5\linewidth]{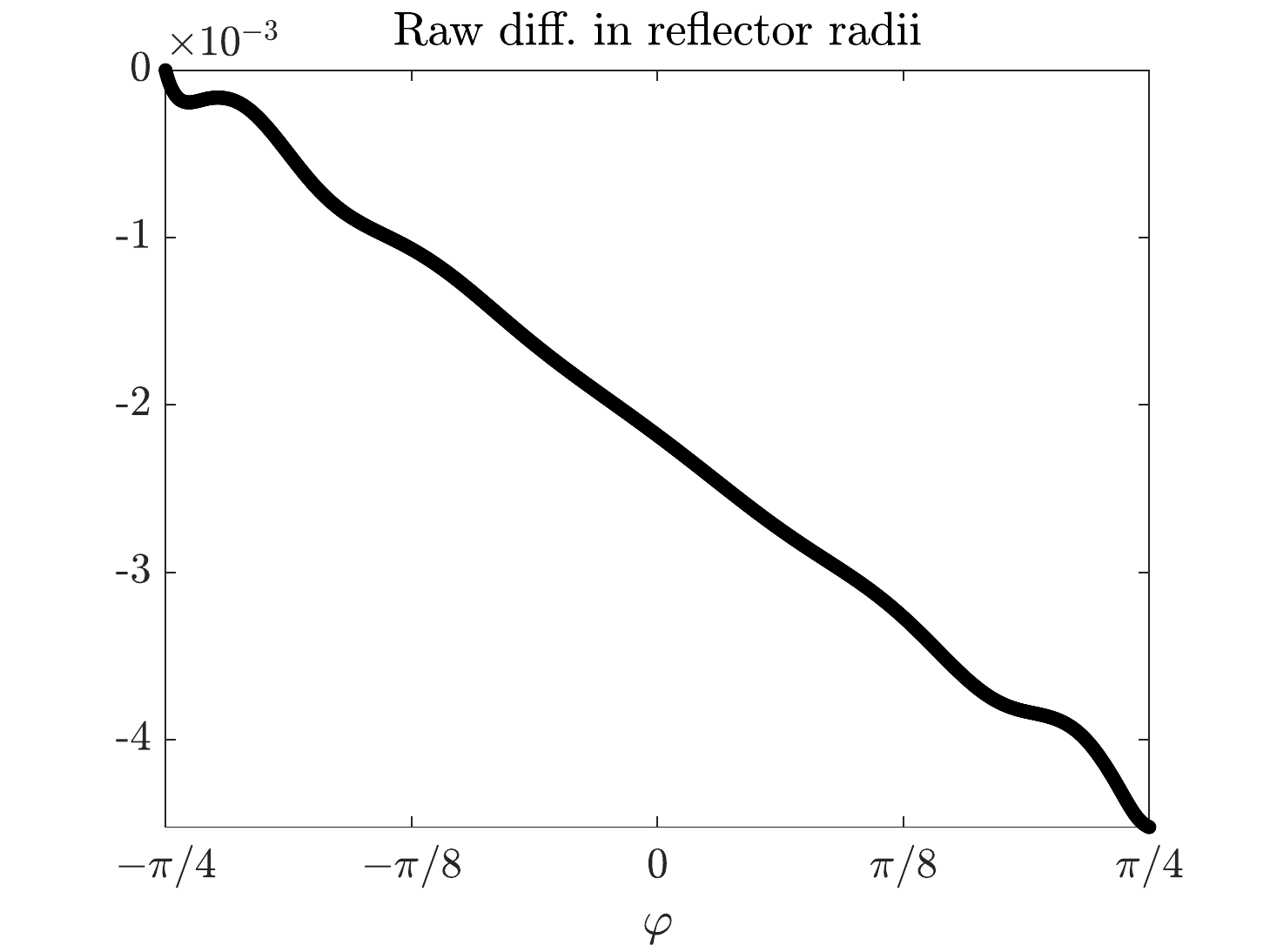}%
	\includegraphics[width=0.5\linewidth]{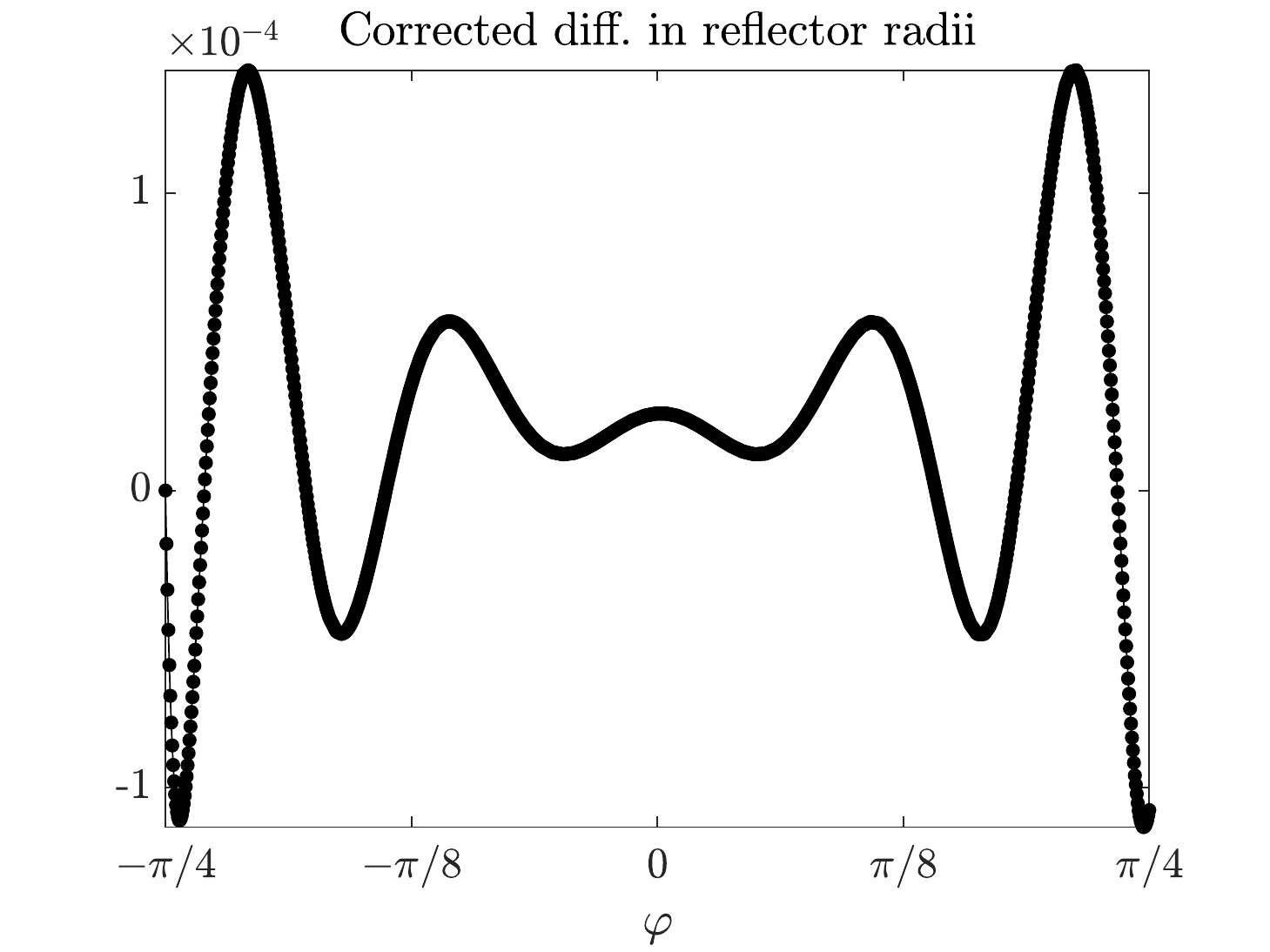}
	\captionsetup{width=0.8\linewidth}
	\caption{Difference in reflector radii of the $m_\mathrm{conv}$ reflectors in Figs.~\ref{fig:example_2-refls-g} and \ref{fig:example_2-refls-g_deconv_10}; slope of the subtracted linear correction term was $4.31\cdot 10^{-6}$.}
	\label{fig:example_2-relfs-radii}
\end{figure}

Turning to the topic of reflector design, consider the reflectors in Figs.~\ref{fig:example_2-refls-g} and \ref{fig:example_2-refls-g_deconv_10}, designed using the original $g$ and the deconvolved $g_\mathrm{dc}$ in Fig.~\ref{fig:example_2-1}, respectively.
We note that the exact solutions to this problem are given in \cite[p.~28]{Maes1997ReflectorDesign} as a circle segment and a straight line, i.e., we recover them using our numerical scheme.
To the naked eye, the two figures appear nigh identical, and it is only when we plot the difference in radii of the $m_\mathrm{conv}$ reflectors in Fig.~\ref{fig:example_2-relfs-radii}, that we can appreciate the differences.
From the raw (or unaltered) graph, we postulate that the deviations can be decomposed into a sloped straight line and comparatively small oscillations.
In order to better appreciate the oscillations, we thus subtracted a linear correction factor from the raw data.
This reveals the profile of $g_\mathrm{dc}(\psi),\ \psi \in [\psi_1,\psi_2]$, present in the reflector surface.

\begin{figure}[htbp]
	\includegraphics[width=0.5\linewidth]{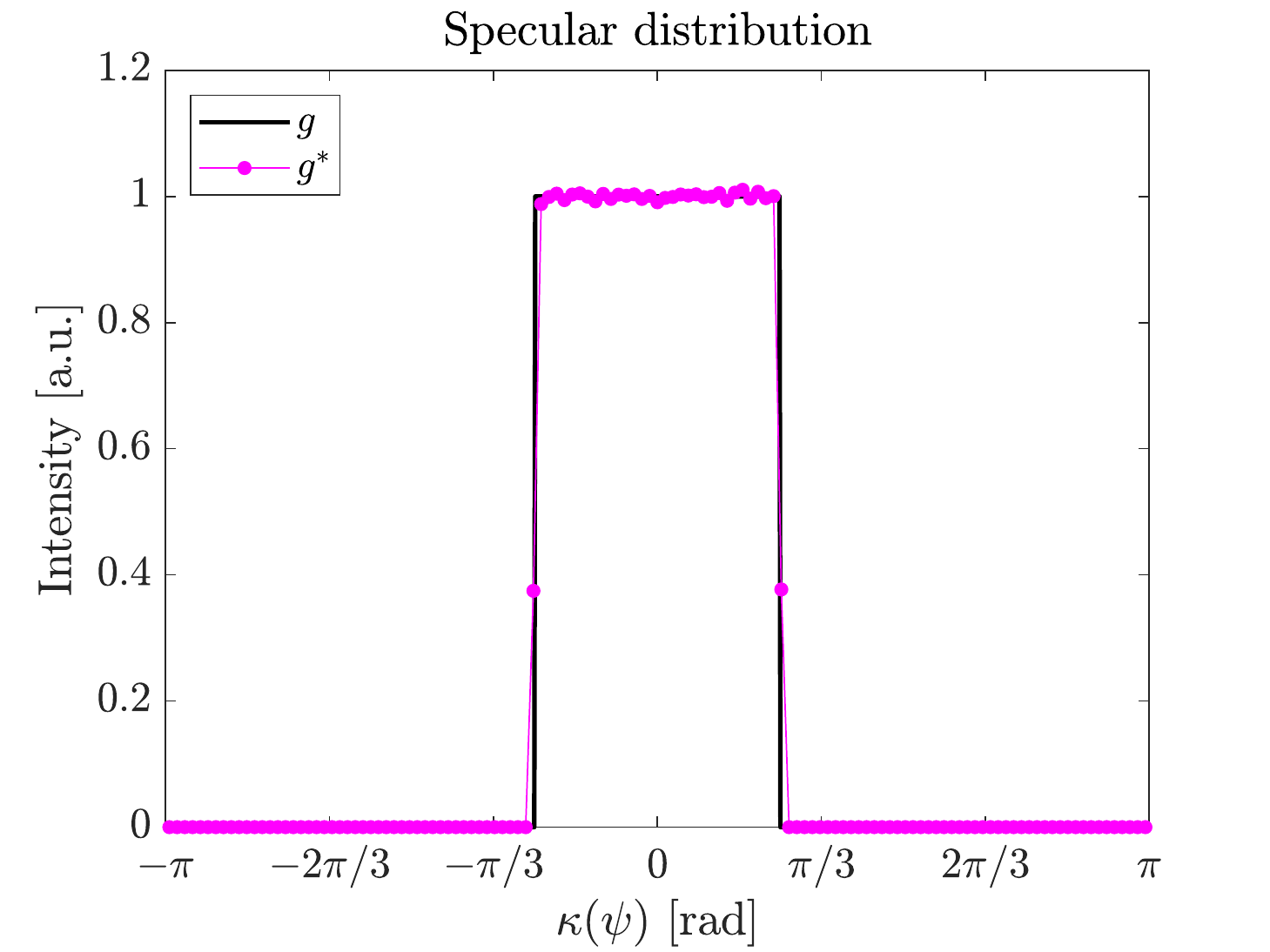}%
	\includegraphics[width=0.5\linewidth]{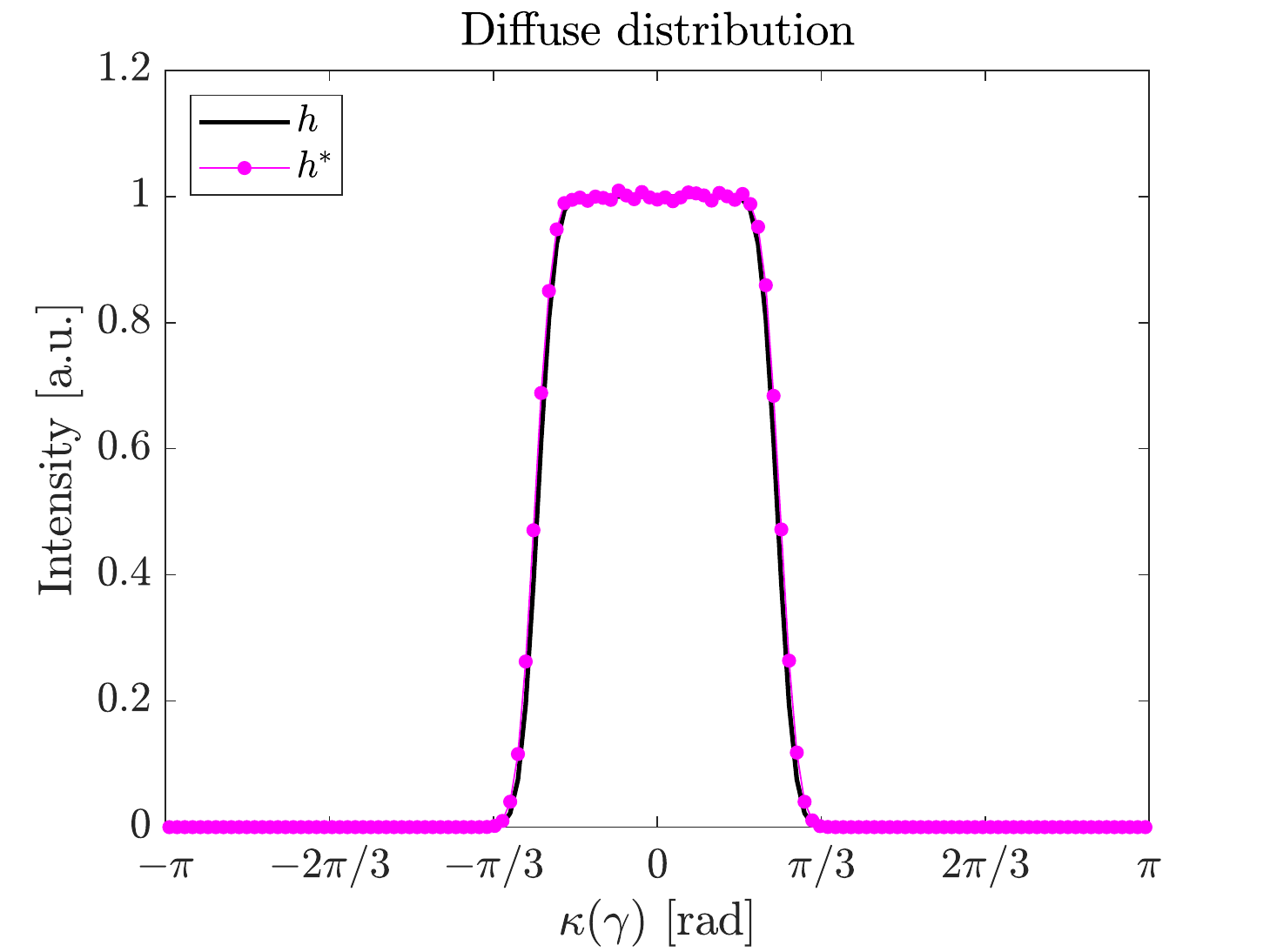}
	\captionsetup{width=\linewidth}
	\caption{Raytraced distributions; example \#2 with $g$ from Fig.~\ref{fig:example_2-1} and $m_\mathrm{conv}$; $10^6$ rays.}
	\label{fig:example_2-g-RT}
\end{figure}

\begin{figure}[htbp]
	\includegraphics[width=0.5\linewidth]{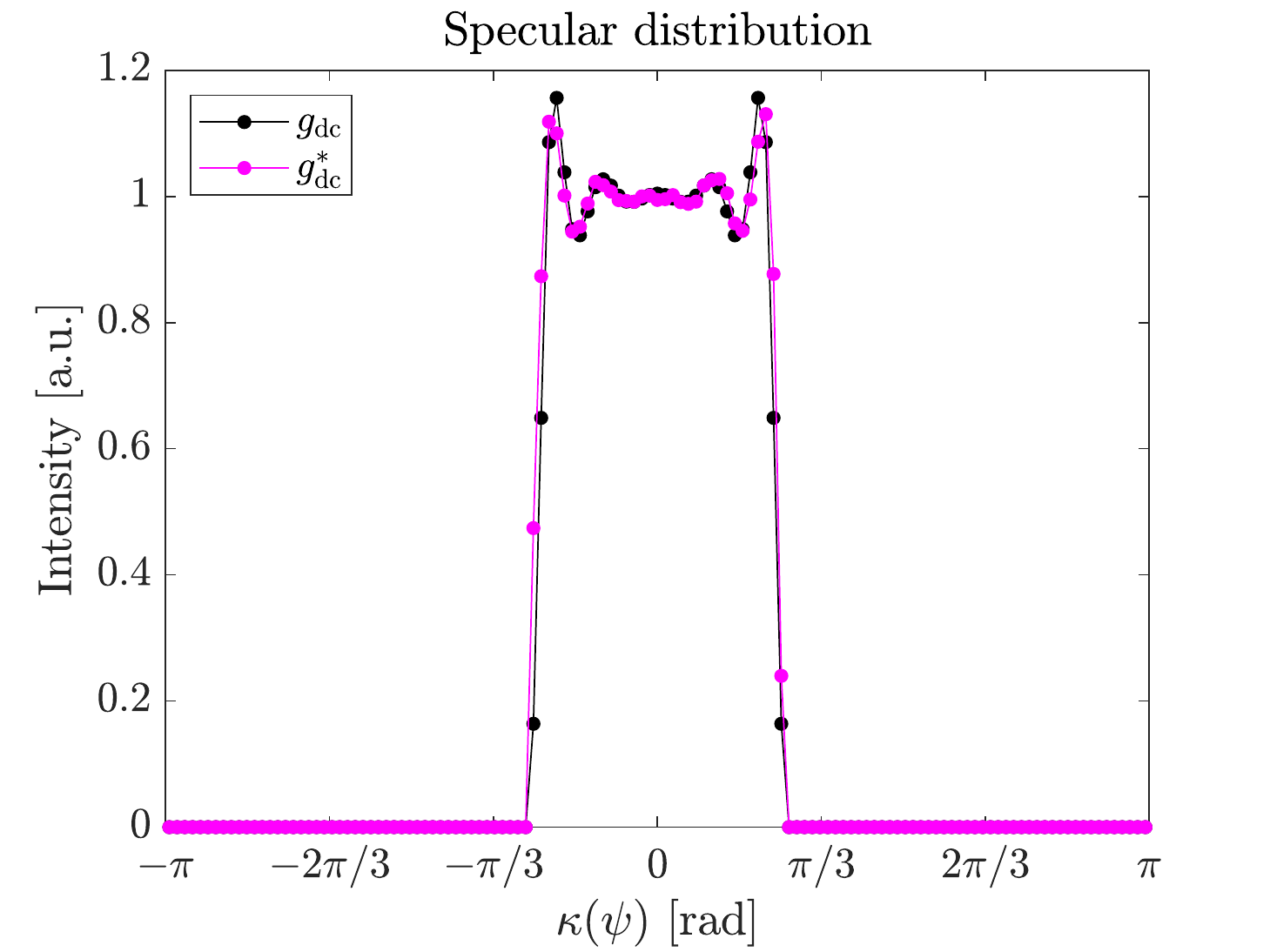}%
	\includegraphics[width=0.5\linewidth]{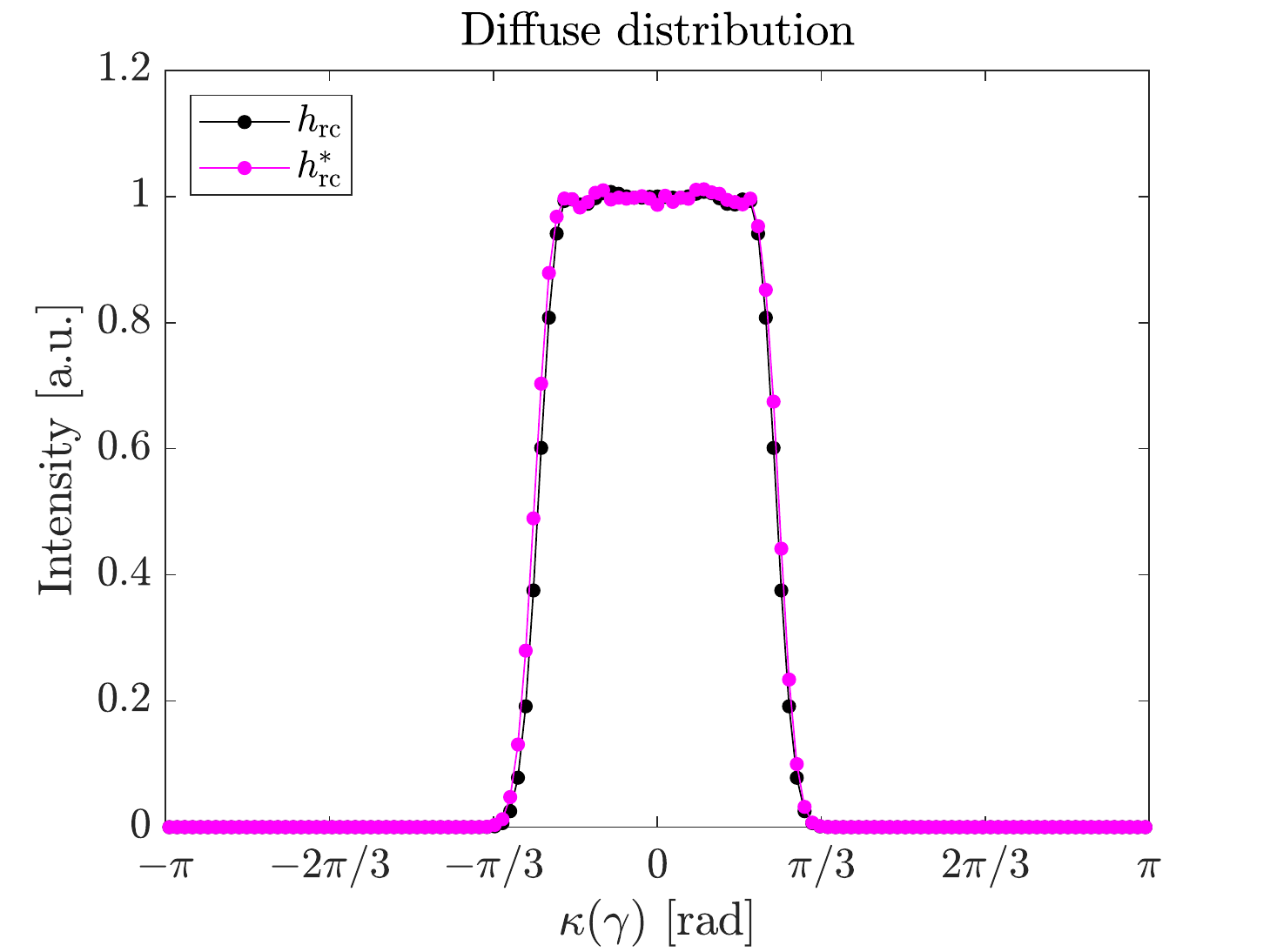}
	\captionsetup{width=\linewidth}
	\caption{Raytraced distributions; example \#2 with $g_\mathrm{dc}$ from Fig.~\ref{fig:example_2-1} and $m_\mathrm{conv}$; $10^6$ rays.}
	\label{fig:example_2-g_deconv-RT}
\end{figure}

\begin{figure}[htbp]
	\includegraphics[width=0.5\linewidth]{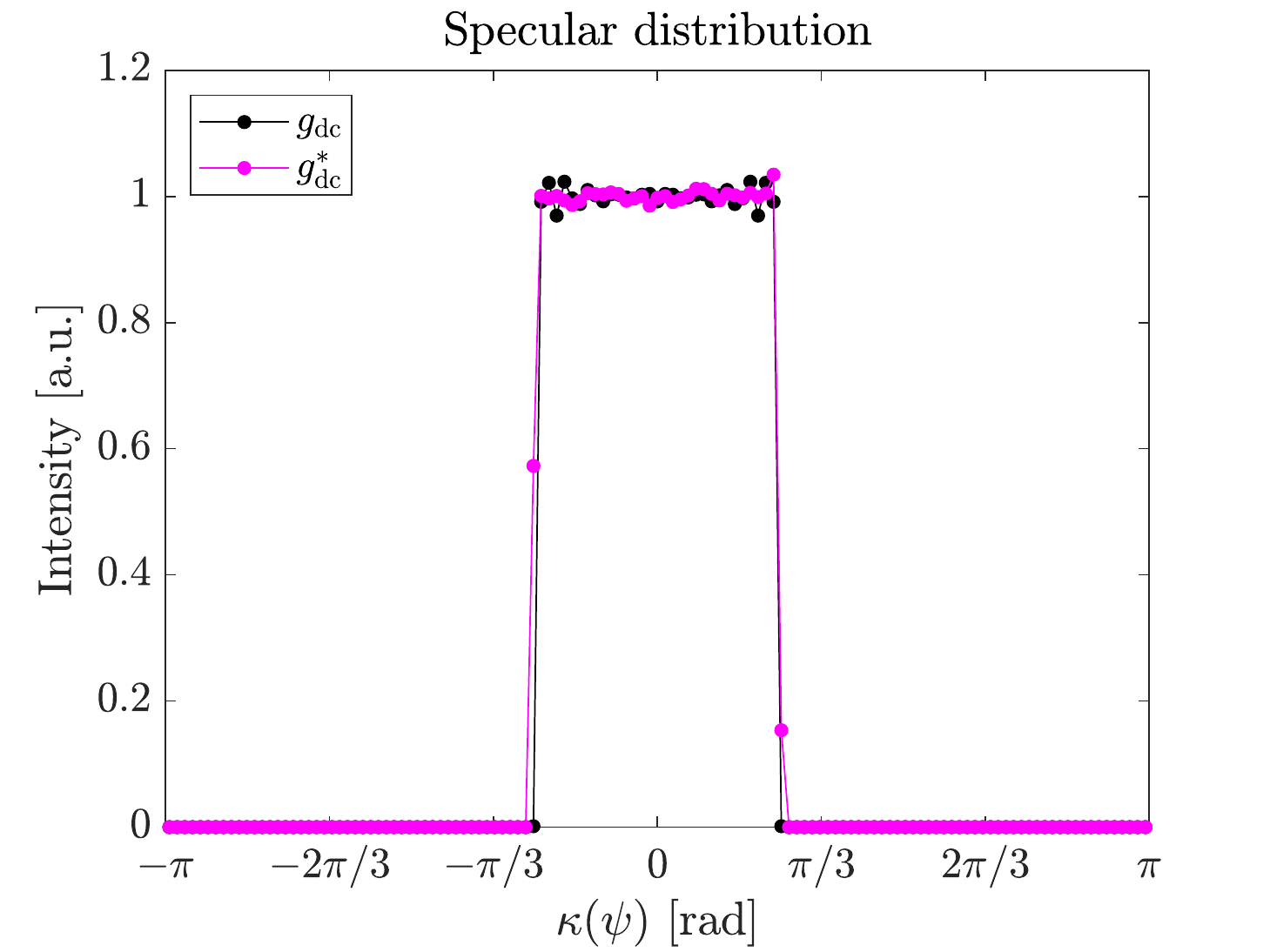}%
	\includegraphics[width=0.5\linewidth]{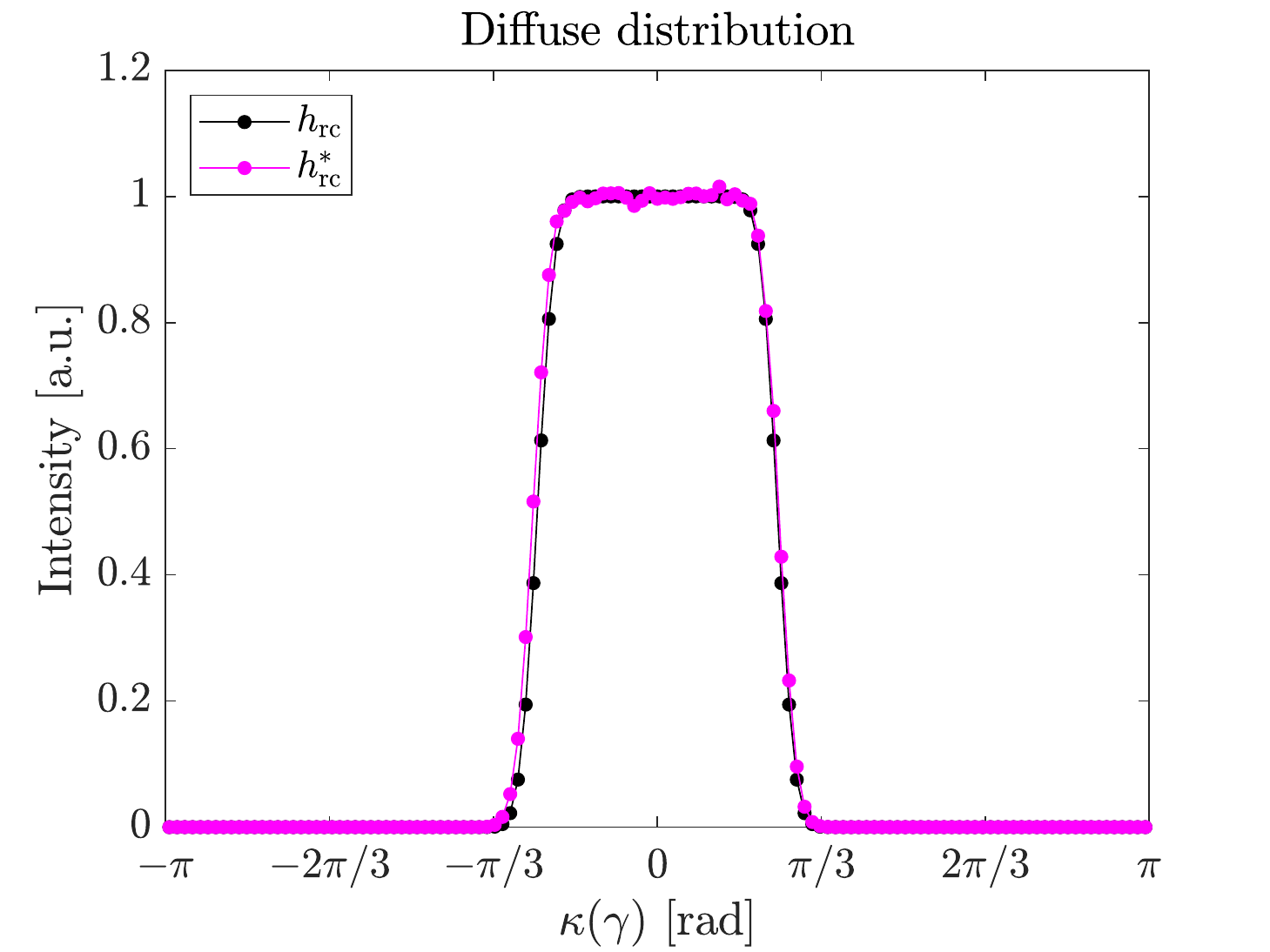}
	\captionsetup{width=\linewidth}
	\caption{Raytraced distributions; example \#2 with $g_\mathrm{dc}$ from Fig.~\ref{fig:example_2-g_deconv_10k} and $m_\mathrm{conv}$; $10^6$ rays.}
	\label{fig:example_2-g_deconv-RT_10k}
\end{figure}

\noindent Moving on to raytracing, the results are shown in Figs.~\ref{fig:example_2-g-RT}, \ref{fig:example_2-g_deconv-RT} and \ref{fig:example_2-g_deconv-RT_10k} with $g$ and $g_\mathrm{dc}$ from Fig.~\ref{fig:example_2-1} (10 iterations) and $g_\mathrm{dc}$ from Fig.~\ref{fig:example_2-g_deconv_10k} ($10^4$ iterations), respectively.
In all these cases, we used a total of $10^6$ rays, and the $m_\mathrm{conv}$ reflectors.
It is clear that the scheme works well, and the issues we see were explained when discussing the previous example.
That is, binning and numerical errors due to discretisation.
A slight asymmetry appears in the results for the reflectors designed using the deconvolved $g_\mathrm{dc}$ distributions.
This can also be seen from the slope in Fig.~\ref{fig:example_2-relfs-radii}, so the cause appears to be somewhere in the numerical computation of the reflectors, presumably due to integration from left to right.
The discrepency is very minor, so an attempt to correct it has not been made, as it is clear that the model predicts the scattered distribution very well, and that the resulting reflector can be validated using raytracing.

\clearpage
\subsection{Example \#3: Lorentzian Scattering Function}
The rotationally symmetric example we have chosen differs from the previous examples in two major ways.
The first is that we no longer know the exact $g$, but rather we prescribe an exact $h$, and to quantify the accuracy of the results, we shall use $\varepsilon(h_\mathrm{rc},h^*_\mathrm{rc})$.
The second difference is that the scattering function is a Lorentzian (also known as a Cauchy distribution).
This is significant for two reasons.
The first is that machined mirrors often exhibit this type of bidirectional reflectance distribution function (BRDF) \cite[Ch.~4]{Stover2012Optical}.
The second is that the tails fall to zero at a significantly lower rate, meaning more large-angle scattering will occur when compared to the Gau{\ss}ians we have used thus far.
As such, we increase the relevance of the method whilst testing the limits of our model.
The example is outlined in the box below, where
\begin{equation}
	\mathcal{L}(\theta;\sigma) = \frac{1}{\pi\sigma}\Bigg(\frac{\sigma^2}{\theta^2+\sigma^2}\Bigg),
\end{equation}
is a Lorentzian with a full width at half maximum (FWHM) of $2\sigma$; $\sigma$ is often denoted $\gamma$ in literature, but not here for obvious reasons.
Note that we again truncated the values of $p$ outside of $[\alpha_1, \alpha_2]$ and renormalised, such that $\uint_{\alpha_1}^{\alpha_2} p(\alpha) \dd \alpha = 1$.

\vspace{5pt}
\noindent \adjustbox{varwidth=\linewidth,scale=0.966}{%
\begin{mdframed}
	\textbf{Example \#3: Lorentzian Scattering Function}
\begin{align*}
	&\text{$\varphi$-range:} 					&[\varphi_1, \varphi_2] 					&= [\pi/4,3\pi/4-0.34]\\
	&\text{$\gamma$-range:} 					&[\kappa(\gamma_1), \kappa(\gamma_2)] 		&= [1.015,2.222]\\
	&\text{$\alpha$-range:} 					&[\alpha_1, \alpha_2] 						&= [-\pi,\pi]\\
	&\text{Source distribution:} 				&f(\varphi)									&= \begin{cases} 1, \ \varphi \in [\varphi_1, \varphi_2]\\ 0, \text{ otherwise} \end{cases}\\
	&\text{Diffuse target distribution:} 		&h(\gamma) 									&= \begin{cases} \sin^4(4\gamma)-\cos(3\gamma-3\pi/5), \ \gamma \in [\gamma_1, \gamma_2]\\ 0, \text{ otherwise} \end{cases}\\
	&\text{Surface scattering function:} 		&p(\alpha) 									&= \mathcal{L}(\alpha;5^{\circ})\\
	&\text{Boundary condition:} 				&u(\varphi_1)								&= 1
\end{align*}
\end{mdframed}
}
\vspace{5pt}

\begin{figure}[htbp]
	\includegraphics[width=0.5\linewidth]{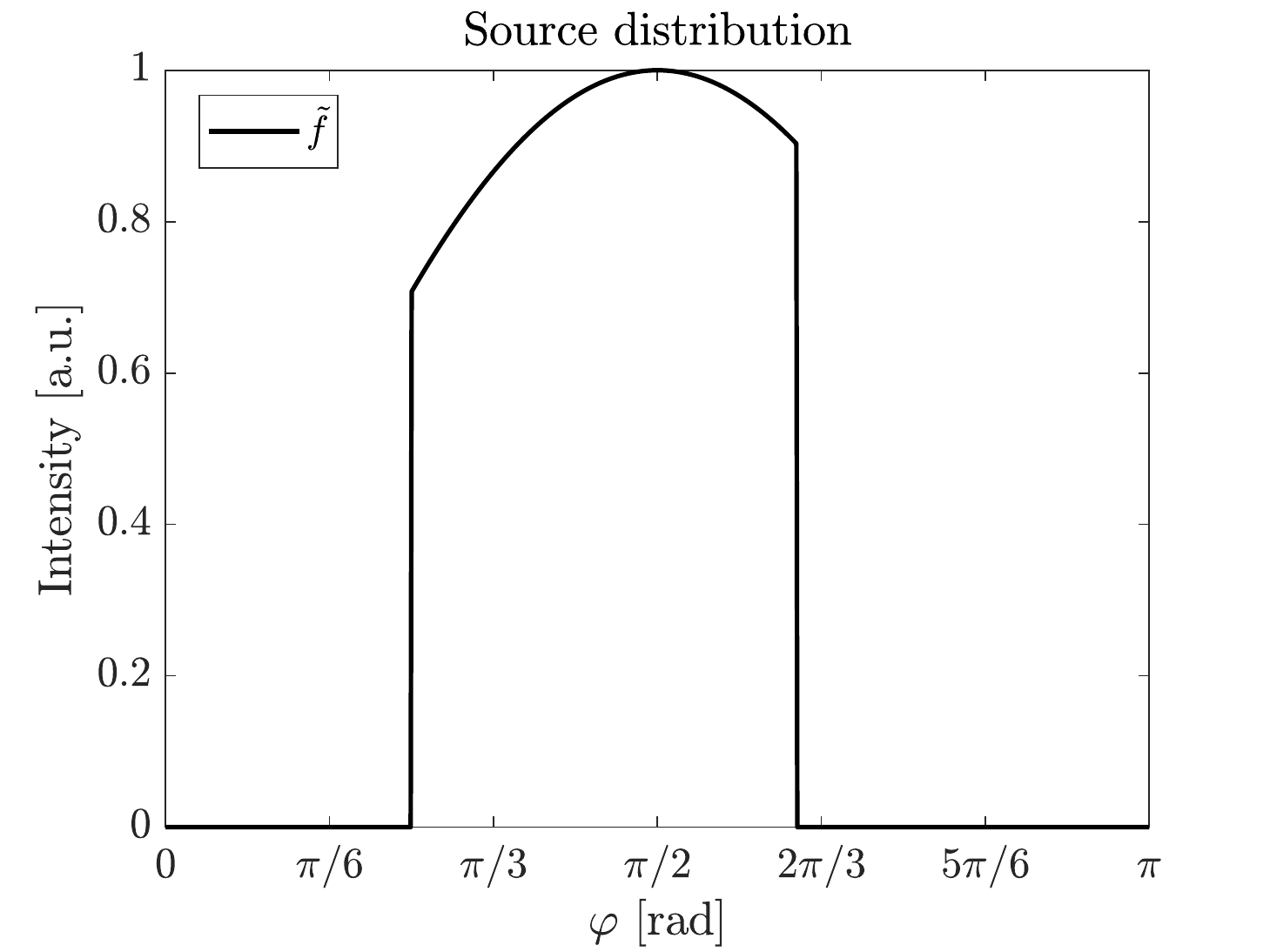}%
	\includegraphics[width=0.5\linewidth]{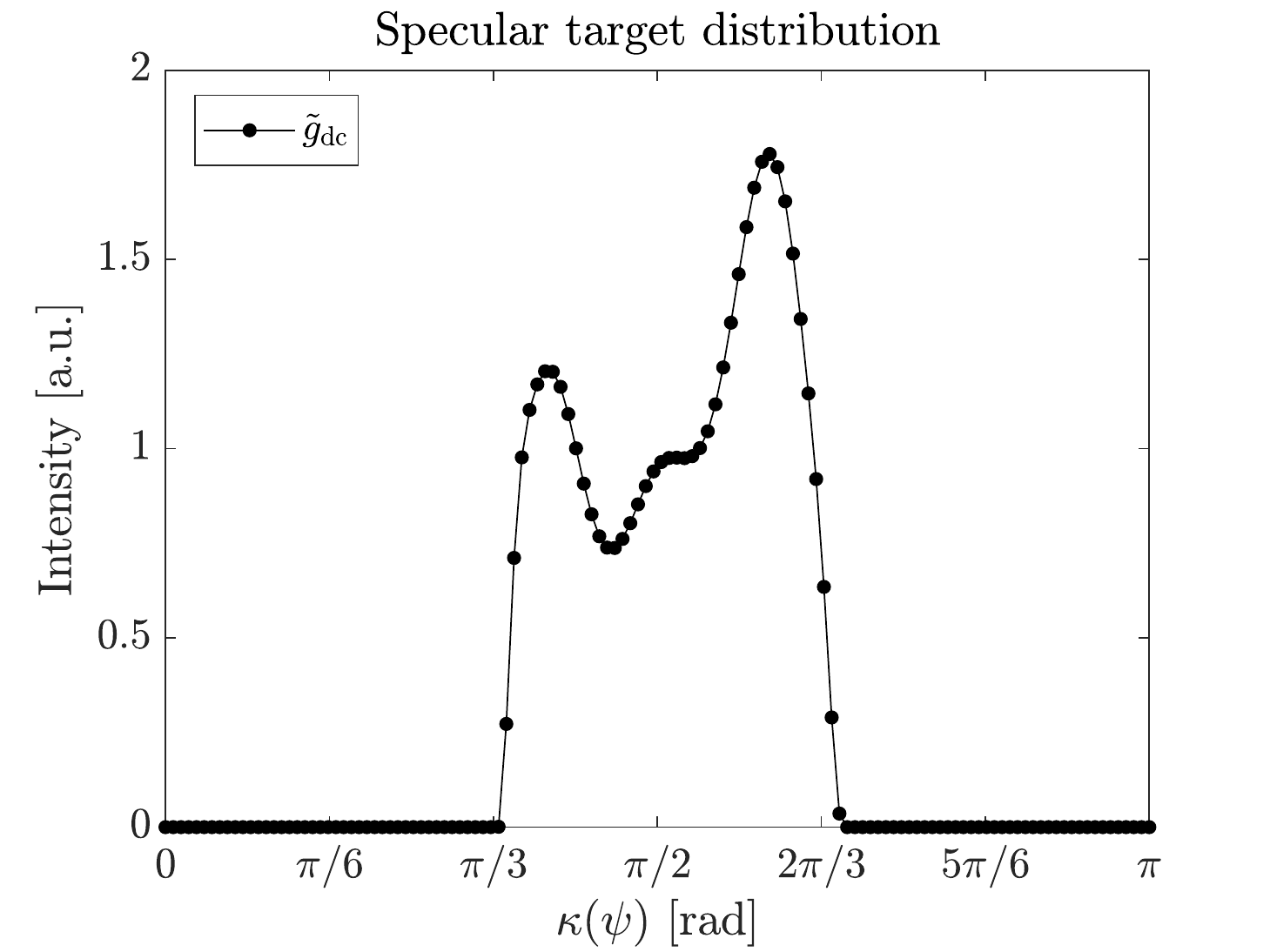}\\[5pt]
	\includegraphics[width=0.5\linewidth]{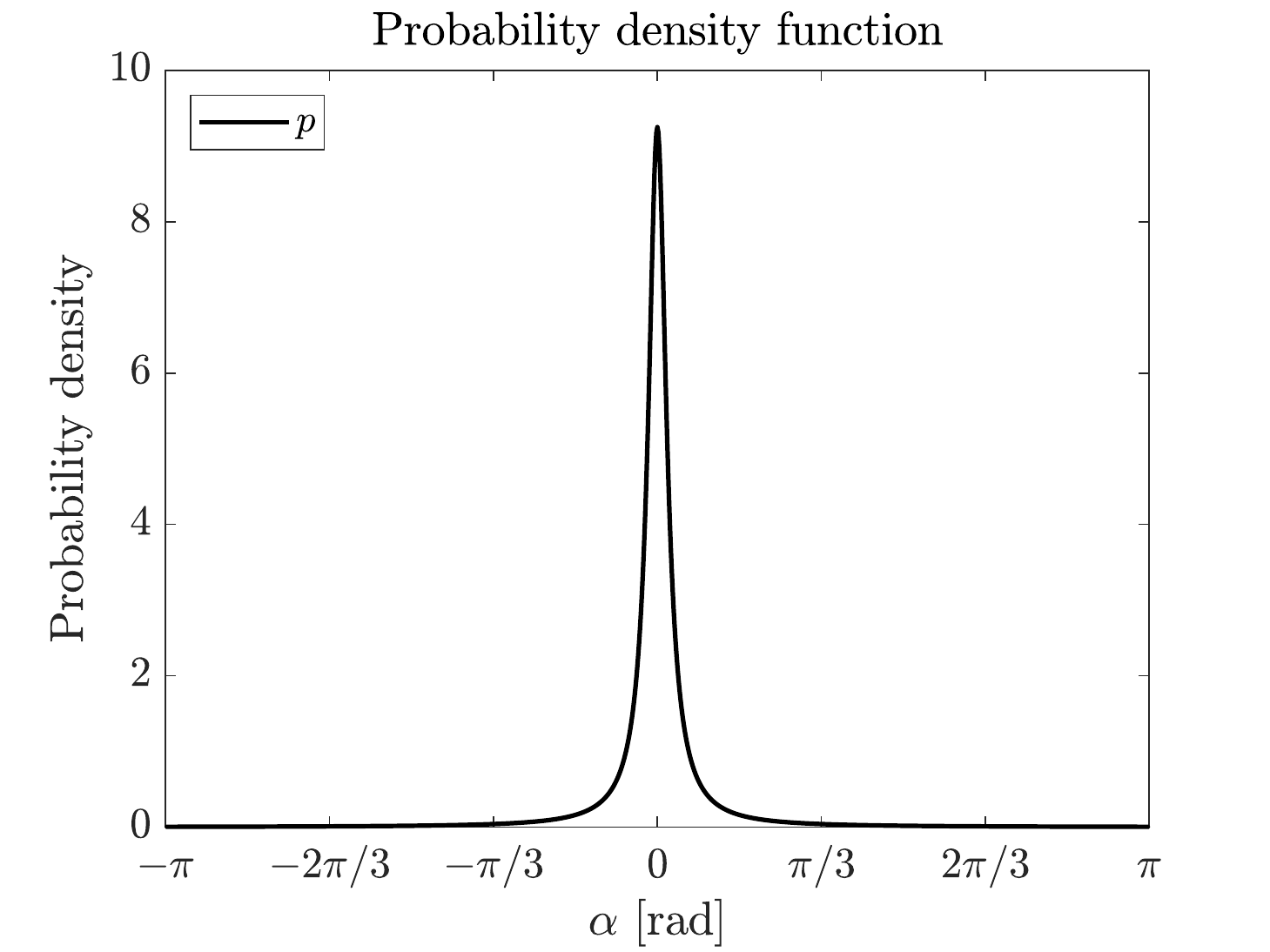}%
	\includegraphics[width=0.5\linewidth]{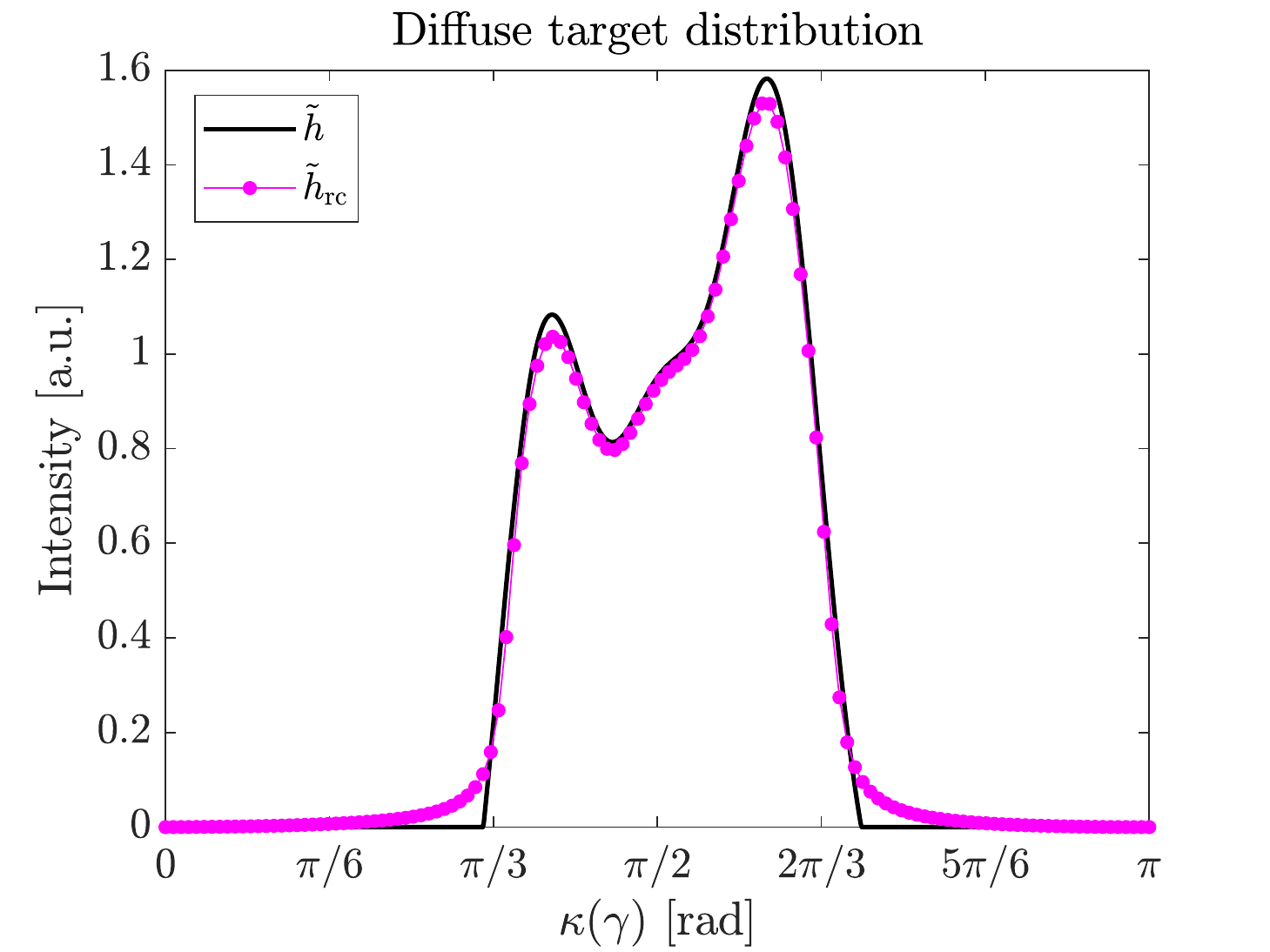}
	\captionsetup{width=\linewidth}
	\caption{Initial distributions in Example \#3; 128 sample points.}
	\label{fig:example_3-1}
\end{figure}

\begin{figure}[htbp]
	\centering
	\includegraphics[width=0.75\linewidth]{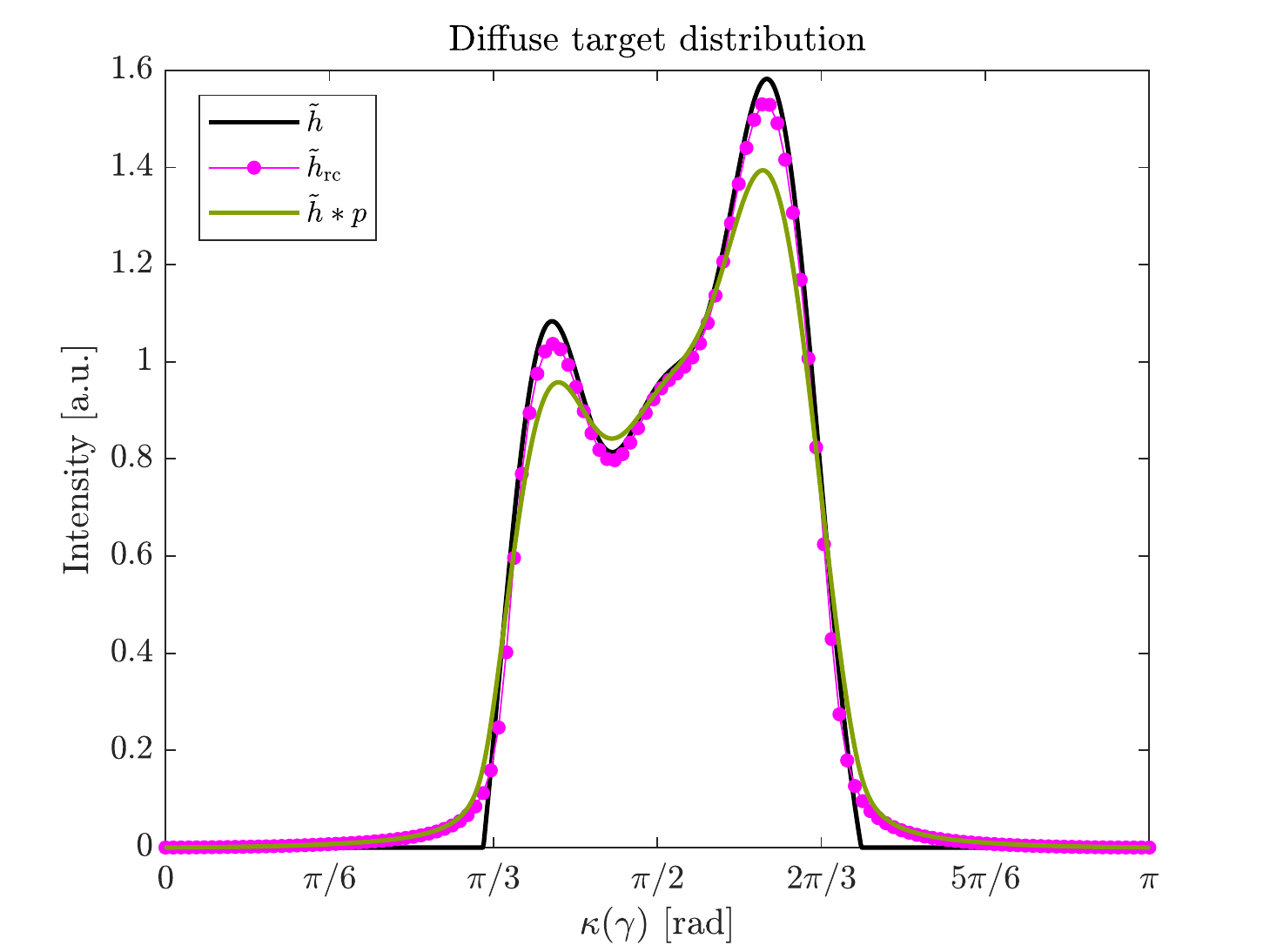}
	\captionsetup{width=0.95\linewidth}
	\caption{The prescribed and predicted targets in Example \#3.}
	\label{fig:example_3-h-2}
\end{figure}

\noindent The distributions are shown in Fig.~\ref{fig:example_3-1}, where we have opted to absorb the sine terms from the energy balancses, Eq.~\eqref{eq:3D_globalEnergyBalance}, into the distributions (indicated by the tilde).
The deconvolved specular target distribution $\tilde{g}_\mathrm{dc}$ was computed using the default \texttt{deconvlucy} settings of 10 iterations.
Before designing the reflectors using $g_\mathrm{dc}$, let us briefly consider what to expect from the final raytraced distributions.
In particular, since we have prescribed $h$, rather than $g$, we are no longer guaranteed that the deconvolution converges.
We can get an appreciation for this by comparing the reconvolved $\tilde{h}_\mathrm{rc} := \tilde{g}_\mathrm{dc} * p$ with our prescribed $\tilde{h}$ in Fig.~\ref{fig:example_3-1}.
This may seem like a disappointing result, but let us compare $\tilde{h}$, $\tilde{h}_\mathrm{rc}$ and $\tilde{h}*p$, where the latter would be the diffuse result if we disregarded scattering in the design procedure entirely, i.e., if we designed the reflectors using $f$ and took $h$ as $g$ in the design procedure, and raytraced the optical system using our scattering model.
All of these distributions are shown in Fig.~\ref{fig:example_3-h-2}, and the RMS error $\varepsilon(\tilde{h},\tilde{h}*p) = 0.0624$, whilst $\varepsilon(\tilde{h},\tilde{h}_\mathrm{rc}) = 0.0356$, i.e., our approach represents an improvement of approximately $40\%$.
Visually, we see that the problematic regions for $\tilde{h}_\mathrm{rc}$ are partly the peaks and partly near $\kappa(\gamma_1) \approx 1.02$ and $\kappa(\gamma_2) \approx 2.22$.
The deviation close to the peaks could perhaps be improved by increasing the number of deconvolution iterations, but the problems close to the boundaries are not solvable in our model.
This is due to an inherent ``maximum steepness'' dictated by the least steep function we are deconvolving (in this case $p$), and it is a property of (de-)convolution.

\begin{figure}[htbp]
	\centering
	\includegraphics[width=0.5\linewidth]{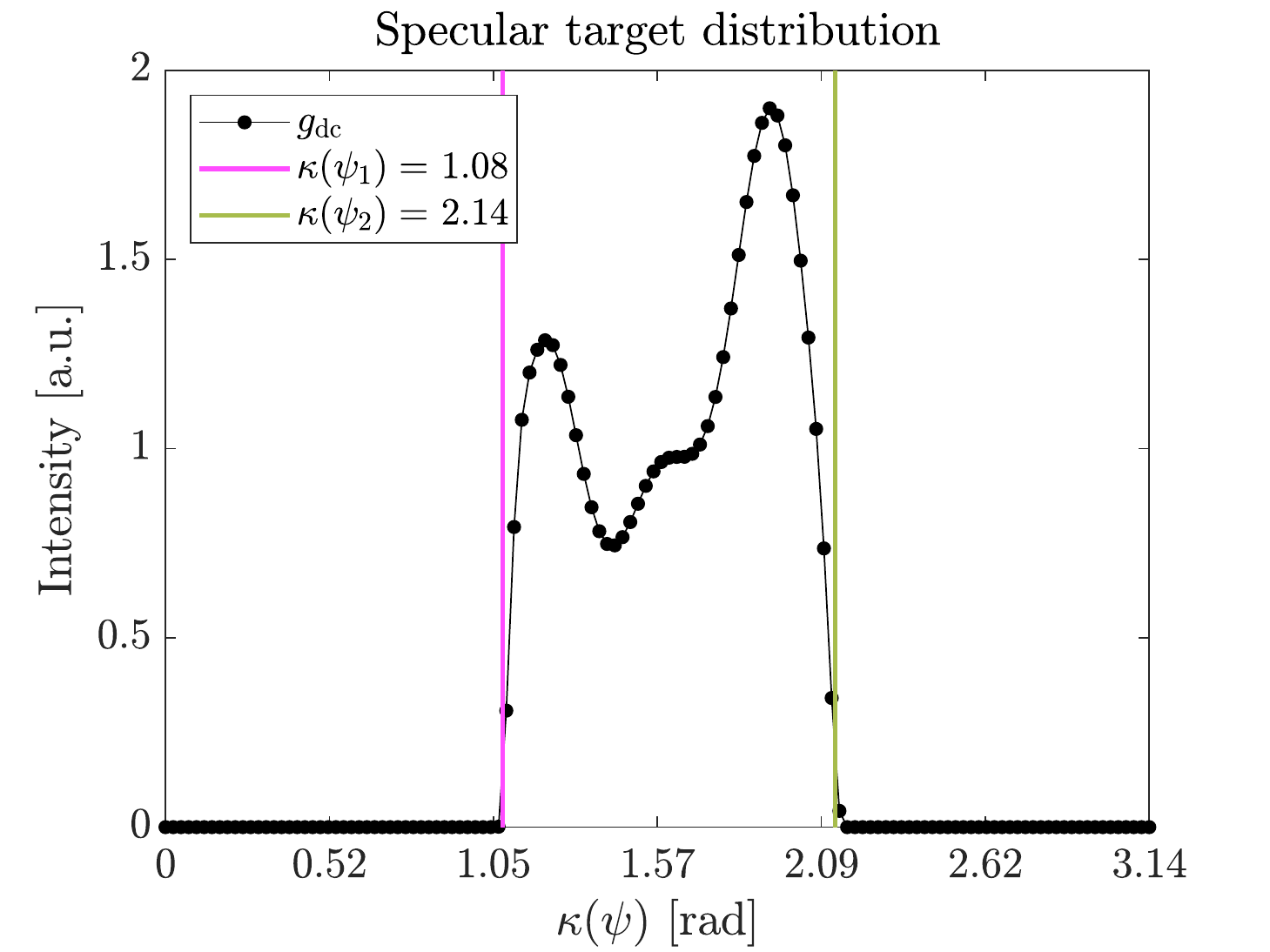}
	\captionsetup{width=0.95\linewidth}
	\caption{The $\kappa(\psi)$-boundaries used as the support of $g_\mathrm{dc}$ in Example \#3.\\[15pt]}
	\label{fig:example_3_g-psiLims}
	\centering
	\begin{minipage}{0.5\linewidth}
		\centering
		\includegraphics[width=\linewidth]{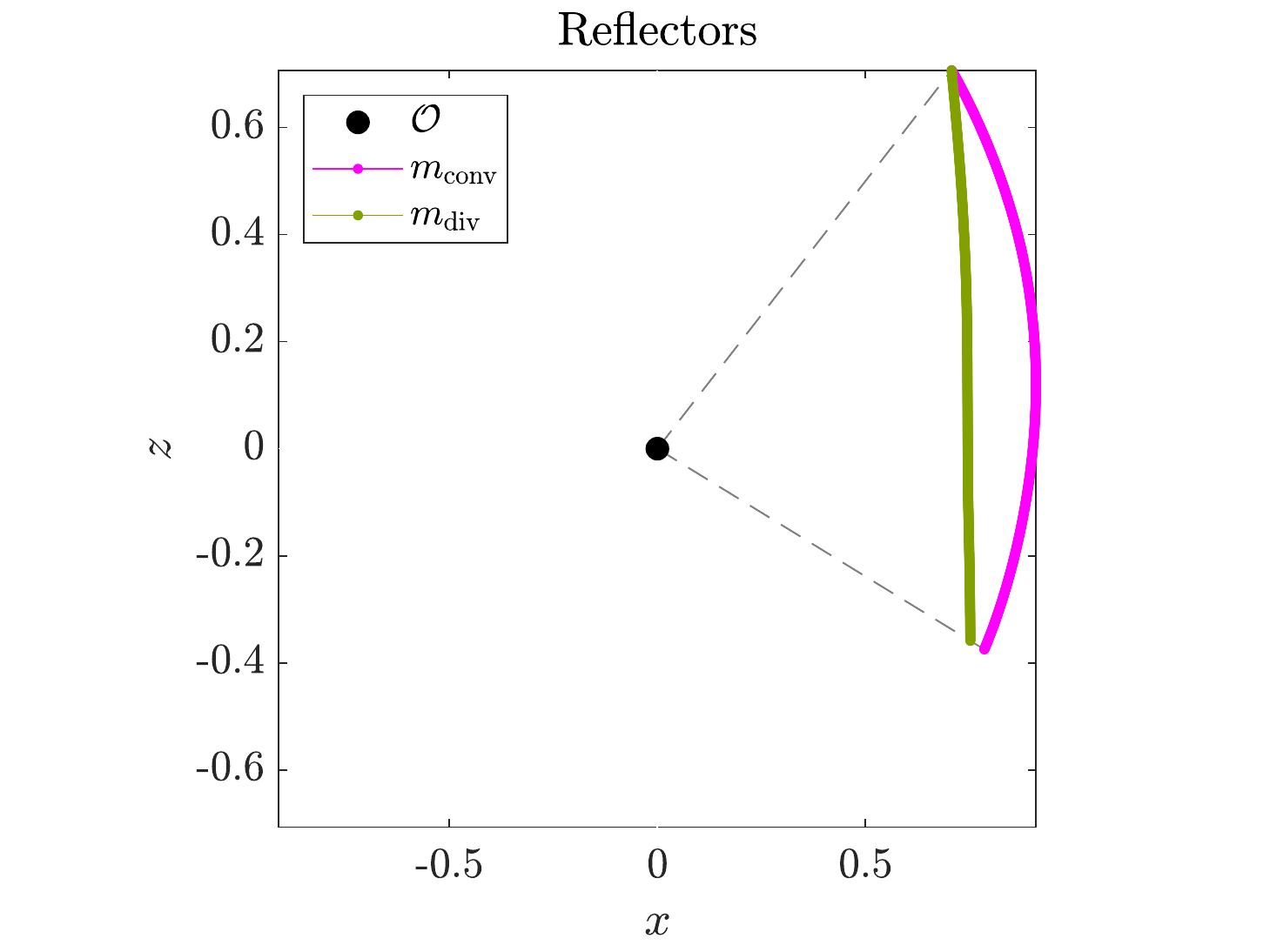}
		\captionsetup{width=0.95\linewidth}
		\caption{The reflectors computed using $g_\mathrm{dc}$ in Fig.~\ref{fig:example_3-1}; 1024 sample points.}
		\label{fig:example_3-refl-1}
	\end{minipage}\hfill
	\begin{minipage}{0.5\linewidth}
		\centering
		\includegraphics[width=\linewidth]{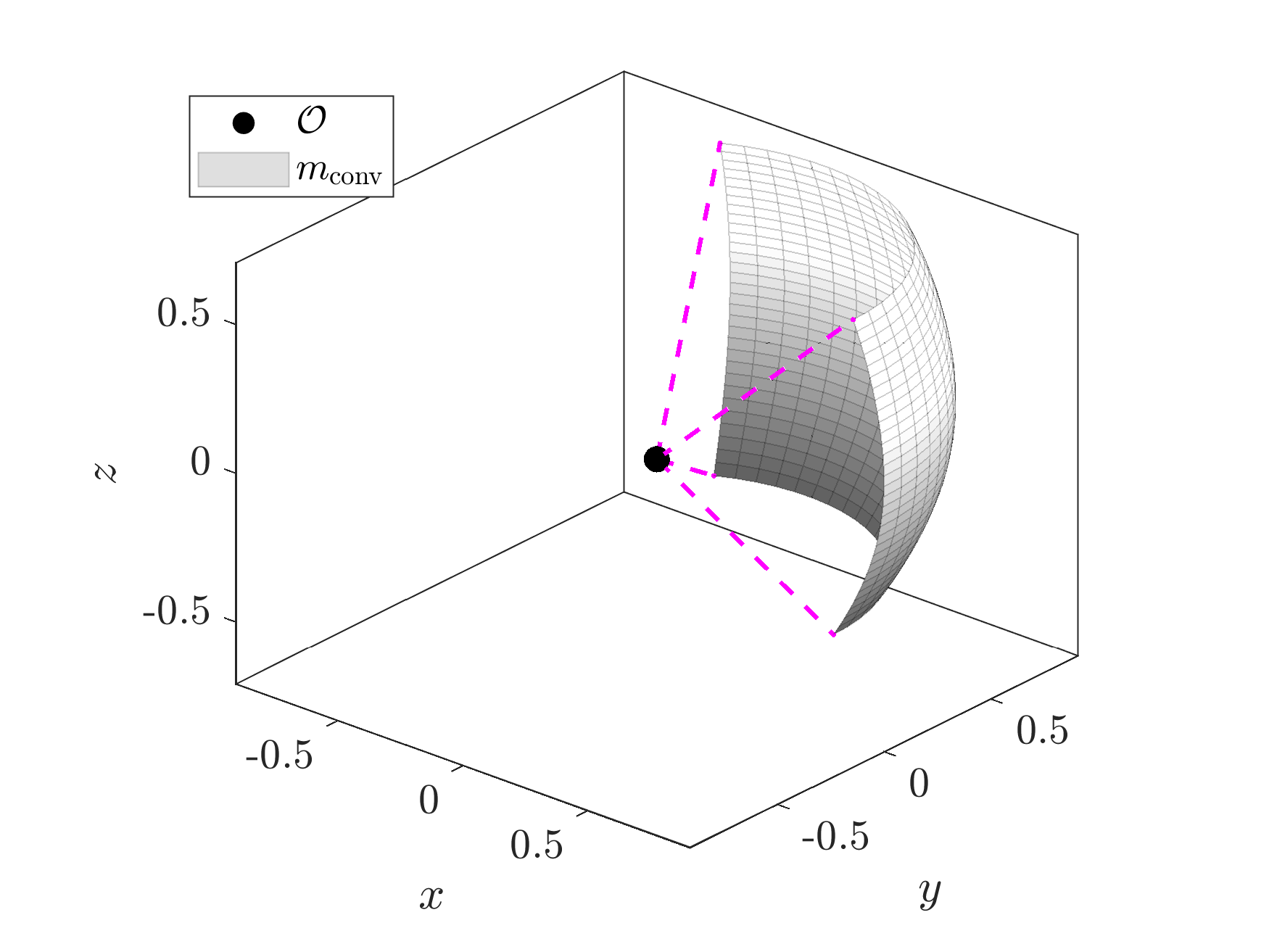}
		\captionsetup{width=0.95\linewidth}
		\caption{A three-dimensional version of the $m_\mathrm{conv}$ reflector in Fig.~\ref{fig:example_3-refl-1}.}
		\label{fig:example_3-refl-2}
	\end{minipage}
\end{figure}

\begin{figure}[htbp]
	\includegraphics[width=0.5\linewidth]{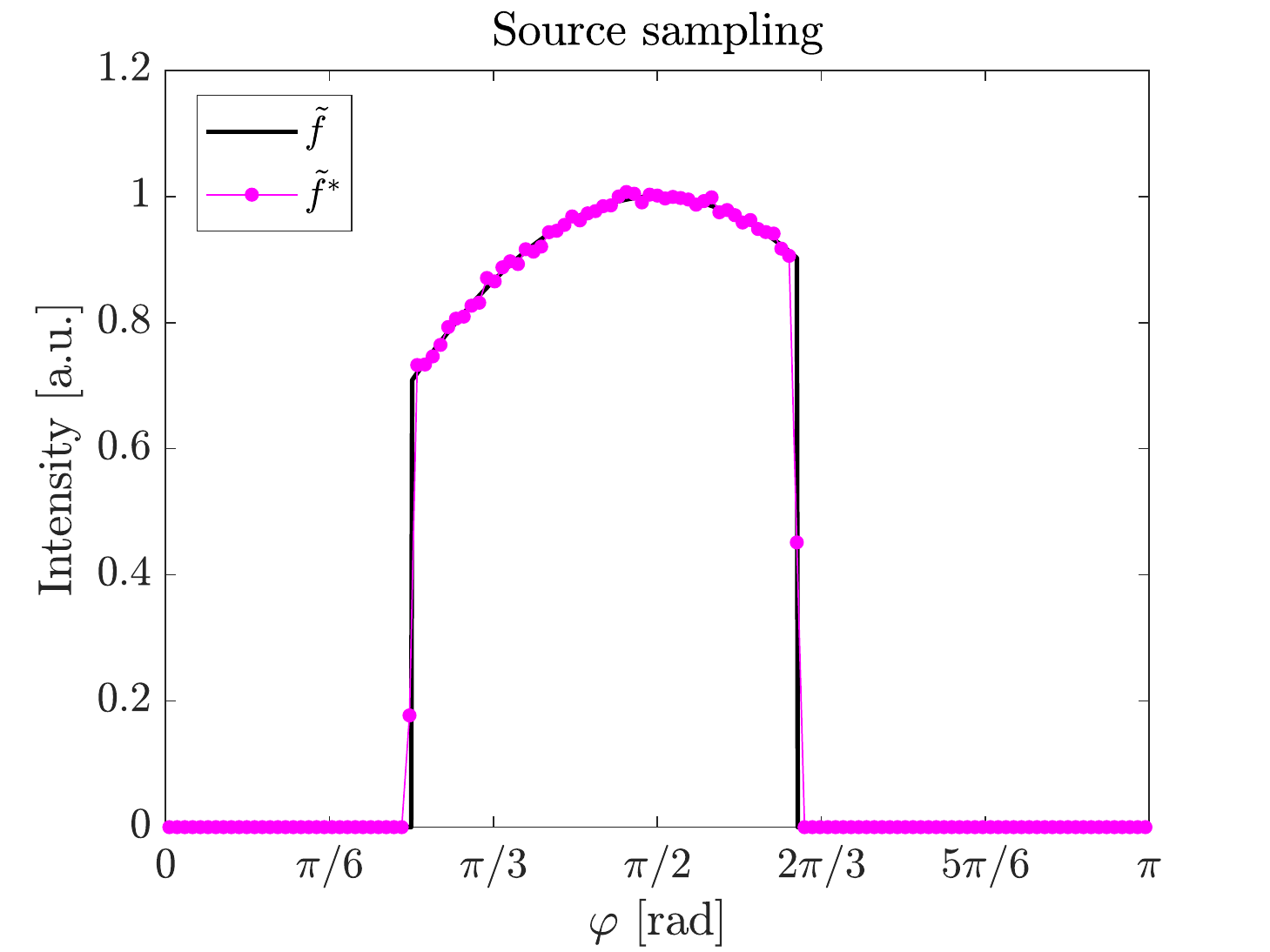}%
	\includegraphics[width=0.5\linewidth]{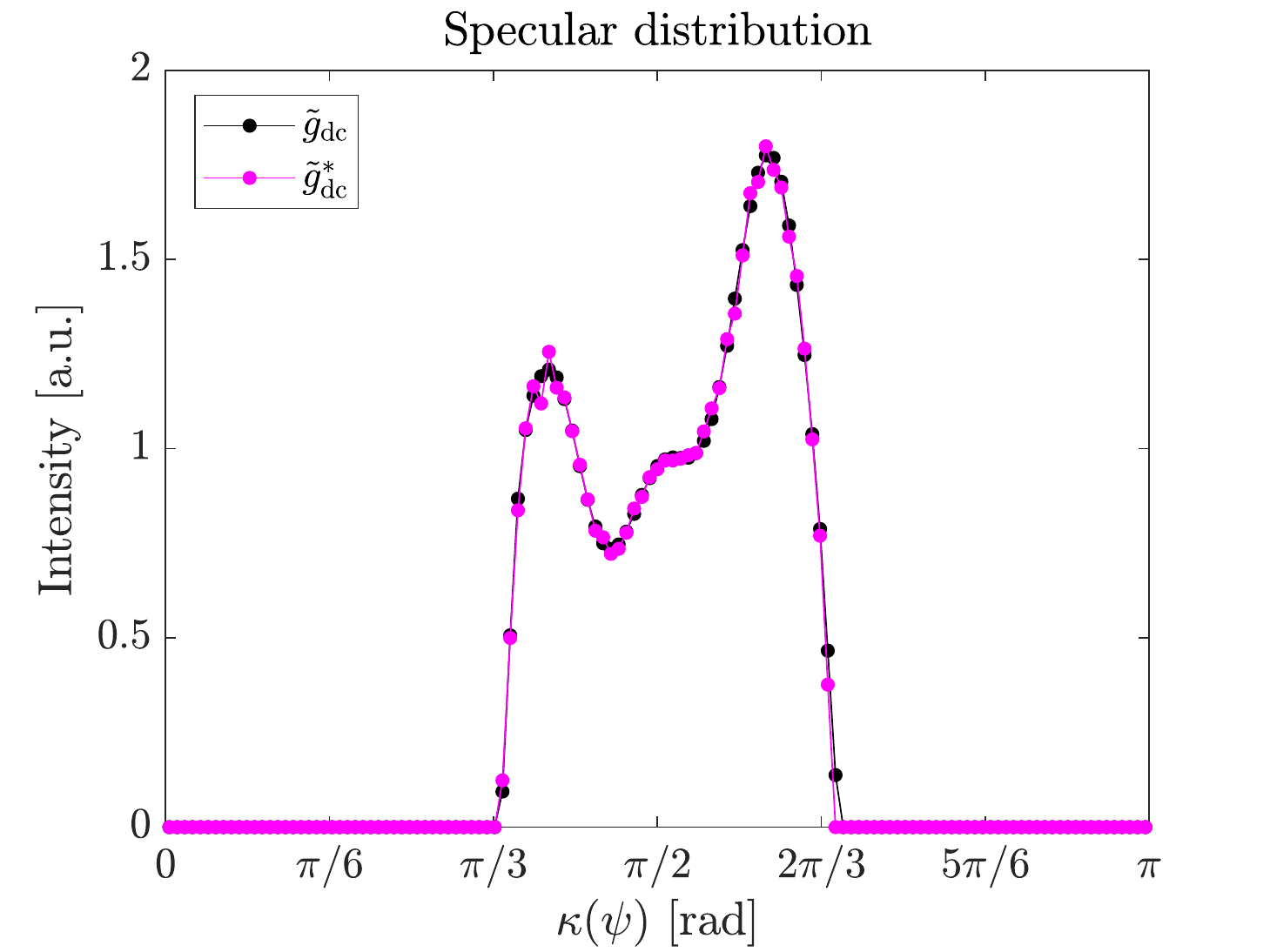}\\[5pt]
	\includegraphics[width=0.5\linewidth]{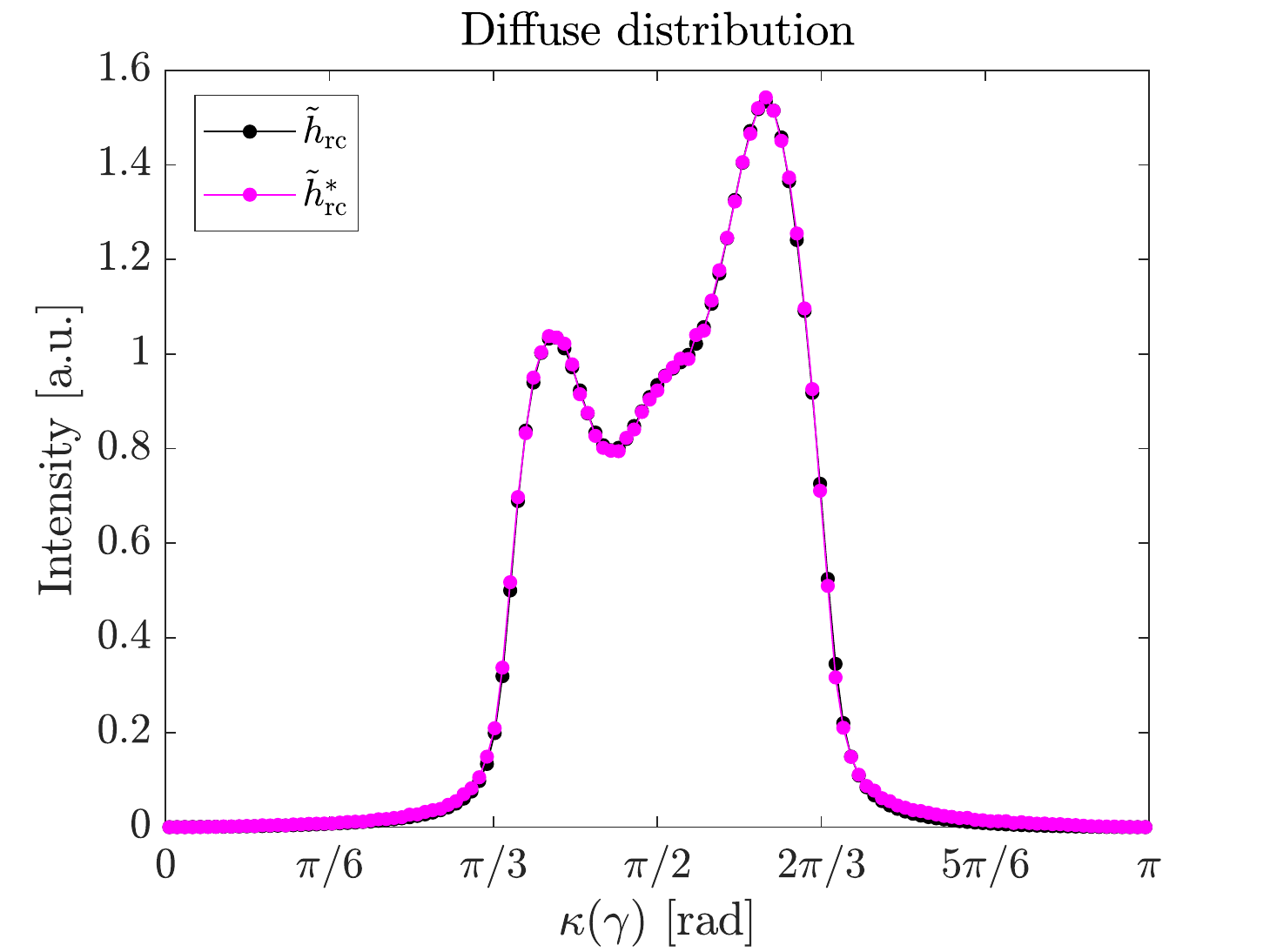}%
	\includegraphics[width=0.5\linewidth]{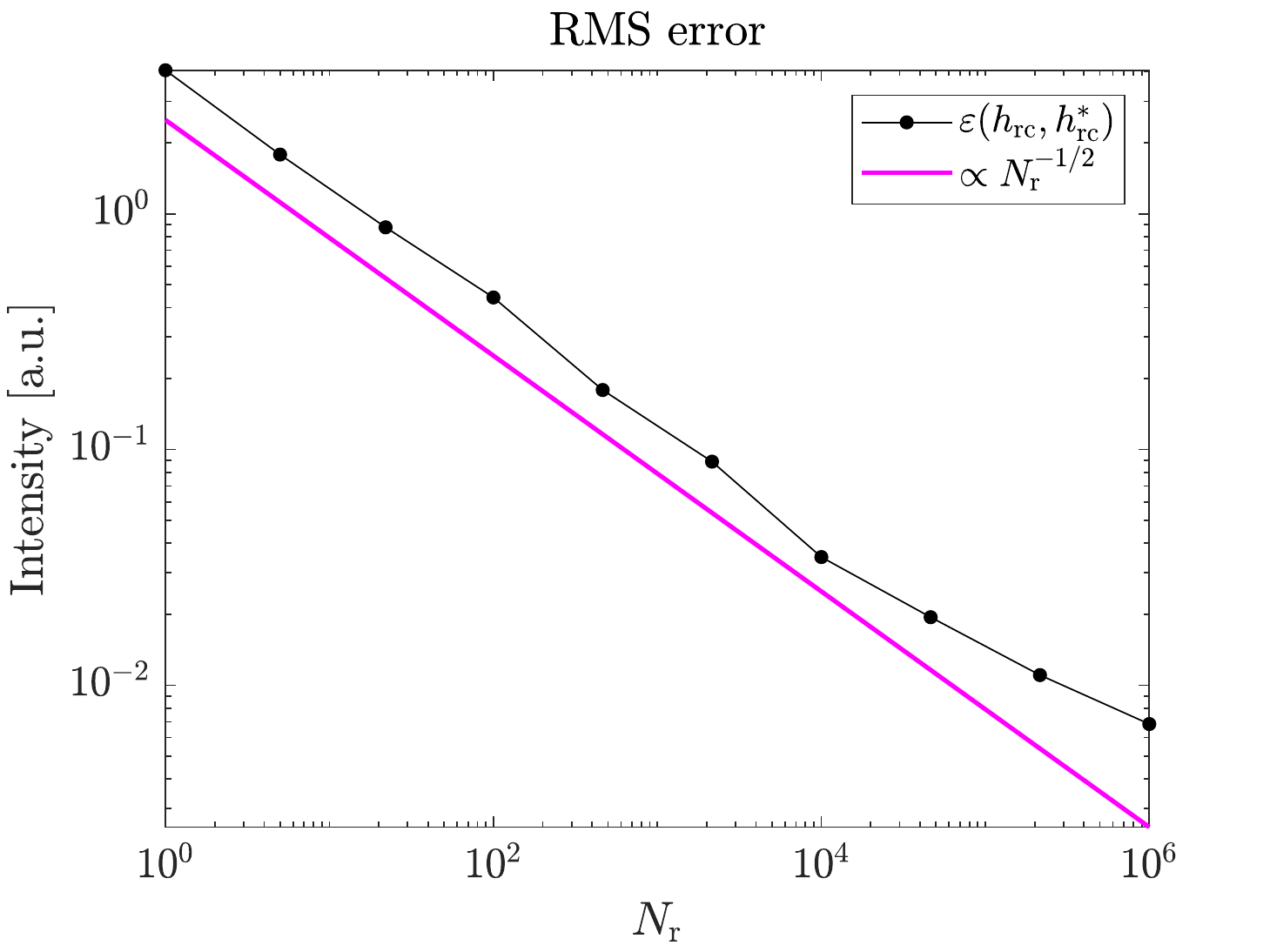}
	\captionsetup{width=\linewidth}
	\caption{Raytraced distributions; example \#3 with $g_\mathrm{dc}$ and $m_\mathrm{conv}$; $10^6$ rays.}
	\label{fig:example_3-RT-E6}
\end{figure}

\begin{figure}[htbp]
	\includegraphics[width=0.5\linewidth]{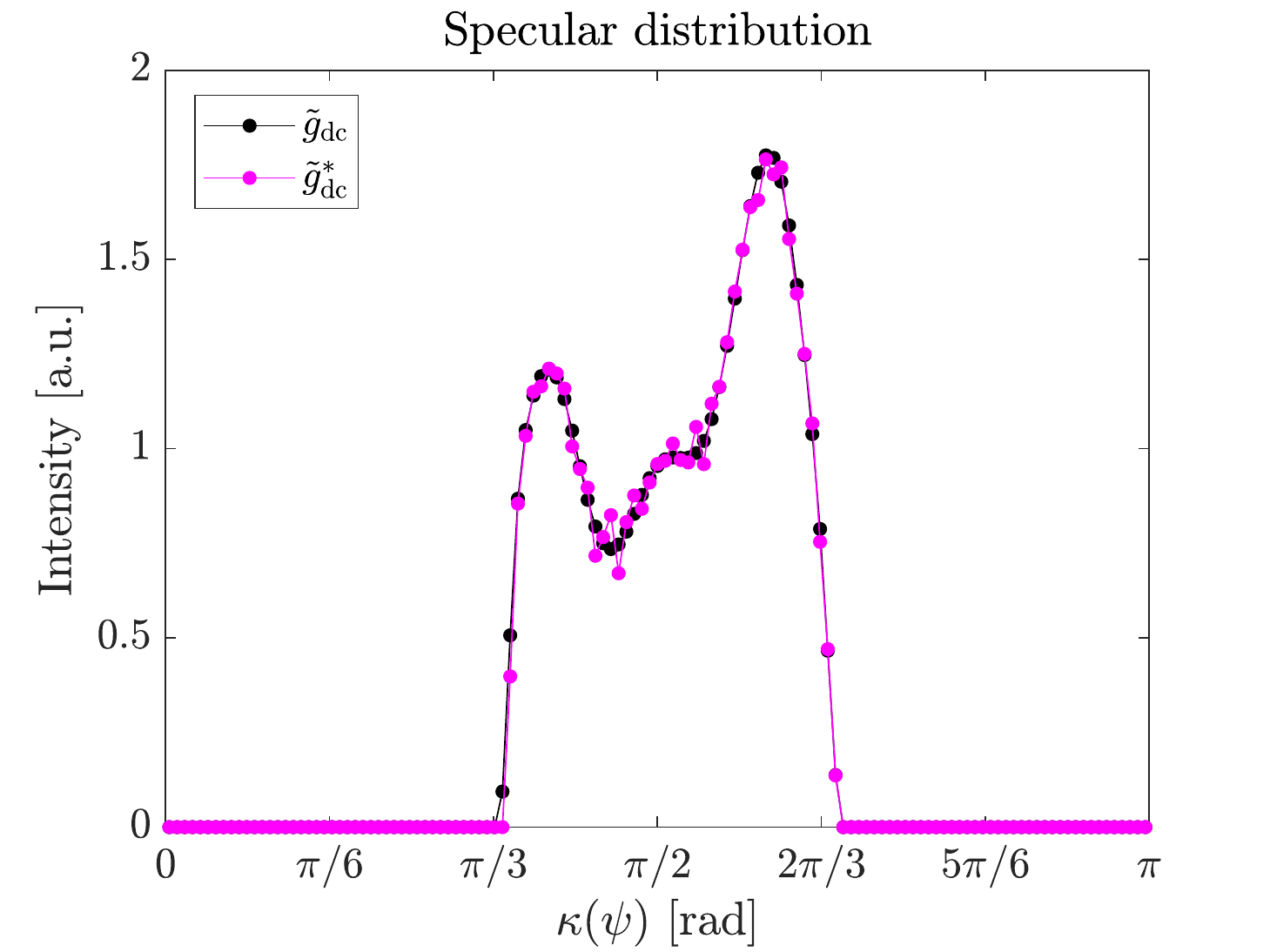}%
	\includegraphics[width=0.5\linewidth]{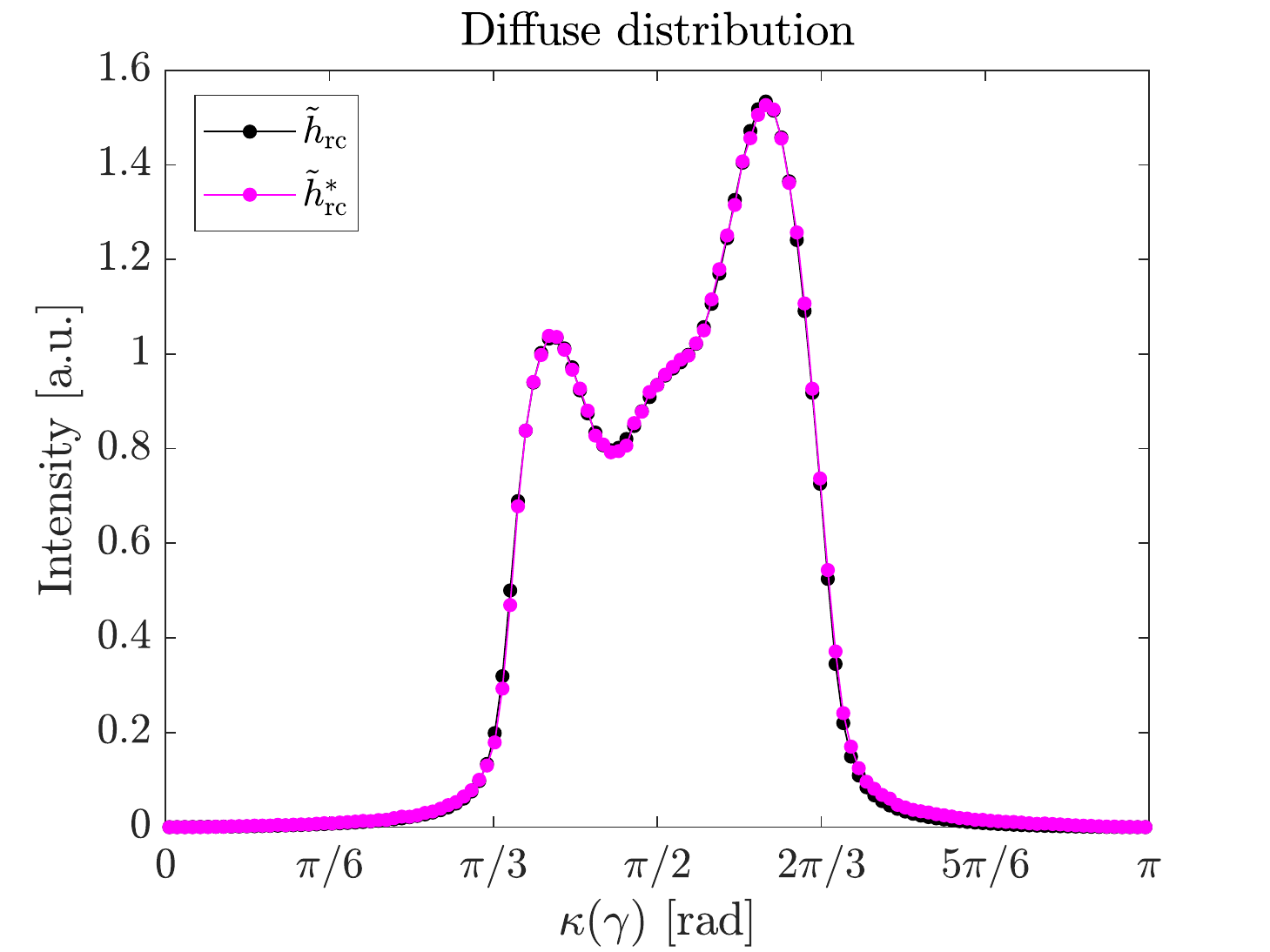}
	\captionsetup{width=\linewidth}
	\caption{Raytraced distributions; example \#3 with $g_\mathrm{dc}$ and $m_\mathrm{div}$; $10^6$ rays.}
	\label{fig:example_3-RT-mConv-E6}
\end{figure}

\noindent We are now ready to design the reflectors.
By fixing $\eta = 0.001$ and locating the two values of $\kappa(\psi)$ where $g_\mathrm{dc}(\psi) = \eta$, we found $\kappa(\psi_1) = 1.08$ and $\kappa(\psi_2) = 2.14$, see Fig.~\ref{fig:example_3_g-psiLims}.
The reflectors computed using this specular target $g_\mathrm{dc}$ and $f$ is shown in Figs.~\ref{fig:example_3-refl-1} and \ref{fig:example_3-refl-2}, where the latter is a three-dimensional version of the $m_\mathrm{conv}$ reflector.
The raytraced distributions are shown in Fig.~\ref{fig:example_3-RT-E6}, where we see that the source sampling $\tilde{f}^*$ is correct, as is the resulting diffuse distribution $\tilde{h}^*_\mathrm{rc}$.
As for the intermediate specular target distribution, we see some deviations from the target $\tilde{g}_\mathrm{dc}$, especially near the first peak.
These deviations are presumably due to difficulties solving the relevant IVPs that give the reflector radius function $u$, and they are likely the reason why our RMS error convergence slows down after approximately $10^5$ rays.
For the sake of completeness, we also raytraced the $m_\mathrm{div}$ reflector in Fig.~\ref{fig:example_3-RT-mConv-E6}.
Here, we only show $\tilde{g}_\mathrm{dc}$, $\tilde{g}_\mathrm{dc}^*$, $\tilde{h}_\mathrm{rc}$ and $\tilde{h}_\mathrm{rc}^*$, since the source sampling is identical.
It is clear that the diffuse distribution is still very closely achieved, whilst the specular distribution deviates in more obvious ways than with the $m_\mathrm{conv}$ reflector.
This highlights the non-triviality of designing specular reflectors, rather than any apparent flaw in our model of scattering, and it could perhaps be improved by increasing the sampling frequency of the distributions, or altering the number of samples in the reflectors themselves.
This has not been attempted, since we are mostly interested in validating our model for scattering, and indeed, even with these deviations from the specular target distribution, the effect of scattering smoothes them out greatly.

\clearpage
\section{Conclusions}\label{sec:conclusion}
\noindent We have presented a novel modelling approach to include surface scattering in the design of reflectors as part of optical systems.
The approach is inspired by concepts from optimal mass transport theory, and it relies on energy conservation.
In the case of isotropic in-plane scattering and cylindrical or rotational symmetry, the forward prediction reduces to a convolution integral between a probability density function and a specular target function.
By prescribing a desired diffuse target distribution, and the scattering probability density function, one can solve for the specular target distribution using deconvolution methods from literature, and then compute the reflectors using purely specular design procedures.
As such, including the effects of scattering can be considered a pre-processing step, and all the mature specular reflector design procedures remain essential.
This gives the optical designer a greater ability to use scattering to their advantage, or mitigate it in applications where it is undesirable.

In the future, we would like to treat scattering functions that depend on the incident angle, thus increasing the realism of the model significantly.
In addition, we would like to extend the model to treat three-dimensional freeform reflectors.\\[10pt]

\noindent\textsc{\textbf{Funding:}} This work was partially supported by the Dutch Research Council (\textit{Dutch:} Nederlandse Organisatie voor Wetenschappelijk Onderzoek (NWO)) through grant P15-36.\\

\noindent\textsc{\textbf{Disclosures:}} The authors declare no conflicts of interest.\\

\noindent\textsc{\textbf{Data availability:}} Data underlying the results presented in this paper are not publicly available at this time but may be obtained from the authors upon reasonable request.\\

\pagestyle{ref}
\addcontentsline{toc}{section}{References}
\printbibliography

\end{document}